\documentclass[acmsmall]{acmart}
\def\formal{}
 
\usepackage[ruled,linesnumbered]{algorithm2e}
\usepackage{xcolor}
\usepackage{tikz-cd}
\usepackage{algpseudocode} 
\usepackage{collectbox} 
\usepackage{multirow}
\usepackage{bbold}
\usepackage{cutwin}
\usepackage{lipsum}
\usepackage{diagbox}
\usepackage{makecell}
\usepackage{amsmath}
\usepackage{ebproof}
\usepackage{wrapfig}
\usepackage{threeparttable}
\usepackage{graphics}
\usepackage{epsfig}
\usepackage{setspace}
\usepackage{xspace}
\usepackage{extarrows}
\usepackage{tabu}
\usepackage{subfig}
\usepackage{color}
\usepackage{xcolor}
\usepackage{booktabs}
\definecolor{keywordcolor}{rgb}{0.8,0.1,0.5}
\usepackage{listings}
\usepackage{tikz}
\newtheorem{theorem}{Theorem}[section]

\newcommand*{\circled}[1]{\lower.7ex\hbox{\tikz\draw (0pt, 0pt)%
    circle (.4em) node {\makebox[1em][c]{\small #1}};}}
\definecolor{dkgreen}{rgb}{0,0.6,0}
\definecolor{gray}{rgb}{0.5,0.5,0.5}
\definecolor{mauve}{rgb}{0.58,0,0.82}

\usepackage{enumitem}
\usepackage{eucal} 
\newcommand{\var}[1]{{\color[rgb]{0.7,0,0}#1}}
\newcommand{\bluec}[1]{{\color[rgb]{0,0,0.7}#1}}
\newcommand{\quanti}[1]{{\color[rgb]{0, 0.4,0}#1}}

\newcommand{\mainname}{\textit{AutoLifter}\xspace}
\newcommand{\misfactor}{mismatch factor\xspace}
\newcommand{\Misfactor}{Mismatch Factor\xspace}

\newcommand{\acomp}[1]{\bluec{a\;\!@}\textit{#1}}
\newcommand{\ccomp}[1]{\quanti{c\;\!@}\textit{#1}}
\newcommand{\ignore}[1]{}
\newcommand{\cat}{\mbox{$\hspace{0.5mm}+ \hspace{-1mm} + \hspace{0.5mm}$}}
\newtheorem{assumption}[theorem]{Assumption}

\newcommand{\targetname}{\textit{orig}\xspace}
\newcommand {\spl} {\;\mbox{\raisebox{.18em}{\tiny $\bigtriangleup$}}\;}

\makeatletter
\newcommand*\bigcdot{\mathpalette\bigcdot@{.5}}
\newcommand*\bigcdot@[2]{\mathbin{\vcenter{\hbox{\scalebox{#2}{$\m@th#1\bullet$}}}}}
\makeatother

\newcommand{\jrydel}[1]{}
\newcommand{\jrymod}[2]{#2}
\newcommand{\jryadd}[1]{#1}

\newcommand{\smalltitle}[1]{{\smallskip \noindent \bf  {#1}.\ }}

 \newcommand{\variable}[1]{\textit{#1}}

\ifx\formal\undefined
  
\else
   
\fi

\newcommand{\key}[1]{{\bf #1}}

\newcommand{\bb}{\begin{array}{lllll}}
\newcommand{\ee}{\end{array}}

\newcommand{\m}[1]{\mbox{\it #1}}

\usepackage{booktabs}   

\acmJournal{PACMPL}
\acmVolume{1}
\acmNumber{} 
\acmArticle{1}
\acmYear{2020}
\acmMonth{1}
\acmDOI{} 
\startPage{1}

\setcopyright{none}

\title[Decomposition-Based Synthesis for Applying D\&C-Like Algorithmic Paradigms]{Decomposition-Based Synthesis for Applying Divide-and-Conquer-Like Algorithmic Paradigms}

\author{Ruyi Ji}
\affiliation{
  \streetaddress{Key Lab of High Confidence Software Technologies, Ministry of Education, School of Computer Science}
  \institution{Peking University}
  \city{Beijing}
  \country{China}            
}
\email{jiruyi910387714@pku.edu.cn}          

\author{Yuwei Zhao}
\affiliation{
  \streetaddress{Key Lab of High Confidence Software Technologies, Ministry of Education, School of Computer Science}
  \institution{Peking University}
  \city{Beijing}
  \country{China}            
}
\email{zhaoyuwei@stu.pku.edu.cn}   

\author{Yingfei Xiong}
\authornote{Corresponding author}
\affiliation{
  \streetaddress{Key Lab of High Confidence Software Technologies, Ministry of Education, School of Computer Science}
  \institution{Peking University}
  \city{Beijing}
  \country{China}            
}
\email{xiongyf@pku.edu.cn}          

\author{Di Wang}
\affiliation{
  \streetaddress{Key Lab of High Confidence Software Technologies, Ministry of Education, School of Computer Science}
  \institution{Peking University}
  \city{Beijing}
  \country{China}            
}
\email{wangdi95@pku.edu.cn}    

\author{Lu Zhang}
\affiliation{
  \streetaddress{Key Lab of High Confidence Software Technologies, Ministry of Education, School of Computer Science}
  \institution{Peking University}
  \city{Beijing} 
  \country{China}            
}
\email{zhanglucs@pku.edu.cn}          

\author{Zhenjiang Hu}
\affiliation{
  \streetaddress{Key Lab of High Confidence Software Technologies, Ministry of Education, School of Computer Science}
  \institution{Peking University}
  \city{Beijing} 
  \country{China}            
}
\email{huzj@pku.edu.cn}           

\begin{abstract} 
\jrydel{Algorithmic paradigms such as divide-and-conquer (D\&C) are proposed to guide developers to design efficient algorithms, but applying them to optimize existing programs is difficult. Therefore, many research efforts have been devoted to the automatic application of algorithmic paradigms. However, most existing approaches to this problem are based on deductive methods and thus put significant restrictions on how the original program is implemented. To overcome this limitation, we study the automatic application of paradigms as an \m{oracle-guided inductive synthesis} problem, where the synthesizer only invokes the original program as a black-box oracle or uses a given verifier to verify the correctness of candidate programs. Such a synthesizer puts no restriction on the original program and thus overcomes the limitation of deductive approaches.}

\indent \jryadd{
Algorithmic paradigms such as divide-and-conquer (D\&C) are proposed to guide developers in designing efficient algorithms, but it can still be difficult to apply algorithmic paradigms to practical tasks. To ease the usage of paradigms, many research efforts have been devoted to the automatic application of algorithmic paradigms. However, most existing approaches to this problem rely on syntax-based program transformations and thus put significant restrictions on the original program.
}

\jryadd{
In this paper, we study the automatic application of D\&C and several similar paradigms, denoted as D\&C-like algorithmic paradigms, and aim to remove the restrictions from syntax-based transformations. To achieve this goal, we propose an efficient synthesizer, named \mainname, which does not depend on syntax-based transformations. 
Specifically, the main challenge of applying algorithmic paradigms is from the large scale of the synthesized programs, and \mainname addresses this challenge by applying two novel decomposition methods that do not depend on the syntax of the input program, \m{component elimination} and \m{variable elimination}, to soundly divide the whole problem into simpler subtasks, each synthesizing a sub-program of the final program and being tractable with existing synthesizers.
}
  
  We evaluate \mainname on 96 programming tasks related to 6 different algorithmic paradigms. \mainname solves 82/96 tasks with an average time cost of 20.17 seconds, significantly outperforming existing approaches.

\end{abstract}


\begin{document}

\maketitle


\section{Introduction} \label{section:introduction}
\jrydel{Efficiency is a major pursuit in practical software development, and designing suitable algorithms is a fundamental way to achieve efficiency. To reduce the difficulty of algorithm design, researchers have proposed many \textit{algorithmic paradigms}~\cite{Mehlhorn1984}, such as divide-and-conquer (D\&C), dynamic programming, greedy, and incrementalization. However, an algorithmic paradigm only tells the general principles of an algorithm class, and implementing these principles for a specific problem is still difficult. For example, D\&C only suggests recursively dividing the problem into sub-problems and combining the solutions to the sub-problems into the solution to the original problem. However, 
how to combine the solutions for a concrete task is totally unknown and up to the developer to discover.}

\jryadd{\noindent Efficiency is a major pursuit in practical software development, and designing suitable algorithms is a fundamental way to achieve efficiency. To reduce the difficulty of algorithm design, researchers have proposed many \textit{algorithmic paradigms}~\cite{Mehlhorn1984} to summarize patterns of efficient algorithms. For example, the paradigm of divide-and-conquer (D\&C)~\cite{DBLP:journals/ppl/Cole95} suggests recursively dividing a possibly complex problem into sub-problems and then combining the solutions for the sub-problems into the solution for the original problem.

This paper focuses on a specific class of algorithmic paradigms that share a similar idea with D\&C, denoted as \textit{D\&C-like algorithmic paradigms}. These paradigms prescribe a recursive structure of transforming the original problem into sub-problems and aim to build up the final results step-by-step through the given recursive structure. Besides D\&C, such paradigms also include (but not limited to) incrementalization~\cite{acar2005self}, single-pass~\cite{DBLP:reference/db/Schweikardt18a}, segment trees~\cite{DBLP:conf/innovations/LauR21}, and three greedy paradigms for longest segment problems~\cite{DBLP:journals/scp/Zantema92}.

Applying D\&C-like paradigms in practice is difficult. Although these paradigms prescribe the recursive structure to build up results, how to efficiently calculate the results in each step can be significantly different among different tasks. For example, although the paradigm of D\&C suggests combining the solutions for the sub-problems, how to combine these solutions in a concrete task is totally unknown and up to the developer to discover.
}

\jrydel{
To reduce the burden on the user, many research efforts have been devoted to the automatic application of algorithmic paradigms. These approaches take a possibly inefficient original program as input, and their goal is to generate a semantically equivalent program with guaranteed efficiency by applying a specific algorithmic paradigm. Typical such approaches cover the automatic application of D\&C~\cite{DBLP:conf/pldi/MoritaMMHT07, DBLP:conf/sosp/RaychevMM15, toronto21}, dynamic programming~\cite{DBLP:conf/ijcai/LinML19}, single-pass~\cite{DBLP:conf/oopsla/PuBS11}, and incrementalization~\cite{acar2005self}.}

\jryadd{
To reduce the burden on the user, many research efforts were devoted to automatically applying individual D\&C-like paradigms, such as applying D\&C~\cite{DBLP:conf/pldi/MoritaMMHT07, DBLP:conf/sosp/RaychevMM15, toronto21}, applying single-pass~\cite{DBLP:conf/oopsla/PuBS11}, and applying incrementalization~\cite{acar2005self}. The approaches proposed by these studies take a possibly inefficient program as input. Then, they apply their respective paradigm to the original program and aim to generate a semantically equivalent program with guaranteed efficiency.
}

 
\jrymod{Applying algorithmic paradigms is difficult because optimized programs are usually complex. To cope with this challenge, most existing approaches are deductive in the sense that the source code of the original program is available to the system, and deductive program transformations are used to solve or simplify the task.}{However, the existing approaches put non-trivial restrictions on the input program, which are not easy to satisfy. Automatically applying algorithmic paradigms is difficult because optimized programs are usually complex. To cope with this challenge, most existing approaches use syntax-based program transformations. Specifically, they access the source code of the original program, transform the source code into certain forms using pre-defined program transformations, and thus simplify or even directly solve the task.} However, to ensure a successful application of program transformations, these approaches put strict restrictions on the original program, leading to a significant limitation on usage. For example, approaches for D\&C~\cite{DBLP:conf/pldi/FarzanN17, toronto21, DBLP:conf/sosp/RaychevMM15, DBLP:conf/pldi/MoritaMMHT07} require the original program to be implemented in another paradigm, namely \textit{single-pass}~\cite{DBLP:reference/db/Schweikardt18a}. An approach for incrementalization~\cite{acar2005self} requires the execution of the original program to be affected little by possible changes in the input, otherwise, the resulting program may not speed up, or even slow down the computation. Satisfying these requirements is typically difficult in practice. For example, in our dataset, applying single-pass already requires implementing 40.54\%-58.62\% of the code needed for applying D\&C (Section \ref{section:rq2}).

\jryadd{
In this paper, we aim to overcome the limitation of existing approaches and propose a more general approach for applying D\&C-like paradigms that does not depend on syntax-based transformations. 
To achieve this goal, we explore another direction for addressing the scalability challenge
: by decomposition.
Specifically, we aim to decompose the application task into a sequence of subtasks, each corresponding to a sub-program of the original synthesis target, and solve these subtasks one by one using existing synthesizers that rely little on the source code, e.g., inductive synthesizers. Such a procedure will put little restriction on the original program if the decomposition can be accomplished without accessing the source code.

However, decomposing a synthesis task is in general difficult. In most cases, there exist mutual dependencies among different sub-programs of the synthesis target, making it impossible to derive precise specifications for independently synthesizing individual sub-programs. 
Our idea is to use approximate specifications in some subtasks when the precise specifications are intractable. The key point here is that, although there may be a difference between an approximation and its respective precise specification, the whole approach will still be sound if we use the original specification in the last step, and be effective 
when the difference is small enough.
}

\smallskip

\jryadd{
Following the above decomposition-based idea, we propose a novel synthesizer named \mainname for applying D\&C-like paradigms. To support applying different D\&C-like paradigms, we design \mainname on a novel class of synthesis problems, named \m{lifting problems}, which we propose to capture the core task of applying D\&C-like paradigms. 
We reduce the applications of various D\&C-like paradigms to lifting problems, and \mainname can be instantiated as synthesizers for applying these paradigms.
}

\jryadd{
\mainname decomposes lifting problems using two novel decomposition methods, namely \textit{component elimination} and \textit{variable elimination}, which decompose through a tuple-output structure and a function-composition structure in the specification of lifting problems, respectively. Both methods break dependency among sub-programs by producing approximate specifications in some subtasks. Consequently, during the decomposition, these methods may generate problematic subtasks without any valid solution, affecting the performance of \mainname. We investigate the effect of these approximations and provide both empirical and theoretical results showing that these approximations are precise enough to ensure the effectiveness of \mainname.
}

\jrymod{\textit{The third contribution of this paper is a thorough evaluation of \mainname}.}{We conduct a thorough evaluation to verify the effectiveness of \mainname in applying D\&C-like paradigms.} Specifically, we instantiate \mainname as $6$ inductive synthesizers, each for applying a D\&C-like paradigm, including D\&C, single-pass, segment trees, and the three greedy paradigms for the longest segment problem. We construct a dataset of $96$ tasks for applying these paradigms, collected from existing datasets~\cite{DBLP:conf/pldi/FarzanN17,toronto21,DBLP:conf/oopsla/PuBS11}, existing publications on formalizing algorithms~\cite{note1989,DBLP:journals/scp/Zantema92}, and an online contest platform for competitive programming (\url{codeforces.com}). We compare \mainname with existing approaches on these tasks, and our evaluation results demonstrate the effectiveness of \mainname.
\begin{itemize}
    \item \mainname solves $82$ out of $96$ tasks with an average time cost of $20.01$ seconds, significantly outperforming existing synthesizers that can be applied to lifting problems. 
    {Among solved tasks, the largest result includes 157 AST nodes and is found by \mainname in $100.0$ seconds.}
    \item \mainname outperforms a specialized synthesizer for applying single-pass, and when compared with an existing synthesizer for applying D\&C programs, \mainname can offer competitive or even better performance while putting less restriction on the original program.
\end{itemize}
\jrydel{Besides, we conduct a case study using two tasks in our dataset, which shows the advantage of inductive synthesis and the ability of \mainname to solve tasks difficult for human programmers.}

\smallskip 

To sum up, this paper makes the following main contributions. 
\begin{itemize}
    \item \jrymod{We introduce a novel class of synthesis problems named lifting problems (Section \ref{section:problem}) and reduce the application of various algorithmic paradigms to lifting problems (Section \ref{section:application}).}{We introduce a novel class of synthesis problems, named lifting problems, for capturing the key task of applying D\&C-like paradigms (Section \ref{section:problem}), and reduce the application tasks of various D\&C-like algorithmic paradigms to lifting problems (Section \ref{section:application}).}
    \item We propose an efficient approach named \mainname for solving lifting problems (Section \ref{section:approach}), which decomposes lifting problems into subtasks tractable by existing inductive synthesizers \jryadd{with two novel decomposition methods, \textit{component elimination} and \textit{variable elimination}}.
    \item We implement \mainname (Section \ref{section:implementation}) and evaluate it on a dataset of 96 related tasks (Section \ref{section:evaluation}). The results demonstrate the advantage of \mainname compared with existing approaches.
\end{itemize}

\section{Overview}\label{sec:simpleOverview} 

In this section, we give an overview of our approach. {Starting from an example task for calculating the second minimum of lists (Section \ref{subsection:example1}), we discuss the synthesis task (Section \ref{subsection:moti-problem-challenge}), the limitation of existing approaches (Section \ref{subsection:moti-limitation}), and the synthesis procedure of \mainname (Section \ref{subsection:moti-overview}). 

For simplicity, we focus on applying the D\&C paradigm in this section. The full definition of lifting problems can be found in Section \ref{section:problem}.} 

\subsection{Example: Divide-and-Conquer for Second Minimum} \label{subsection:example1}

\begin{figure*}
  \hfill 
  \begin{minipage}{0.3\textwidth}
    \begin{figure}[H]
      \begin{lstlisting}
if len(xs) <= 1: return INF;
return sorted(xs)[1]; 
          \end{lstlisting}
          \vspace{-1em}
          \caption{Second minimum.}
          \label{fig:smin}
    \end{figure}
    \end{minipage}
    \hfill
  \begin{minipage}{0.4\textwidth}
    \begin{figure}[H]
    \centering
    \includegraphics[width=0.65\linewidth]{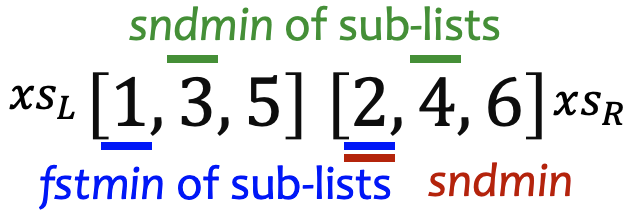}
    \vspace{-0.6em}
    \captionof{figure}{An example of calculating \m{sndmin}.}
    \label{fig:sndmin-case}
    \end{figure}
    \end{minipage}
  \hfill \ 
  \vspace{-1.5em}
\end{figure*}
 
Let $\textit{sndmin}$ be a function returning the second-smallest value in an integer list. A natural implementation of \m{sndmin} (Figure~\ref{fig:smin}, in Python-like syntax) first sorts the input list ascendingly and then returns the second element of the sorted list. {Given a list of length $n$, this program takes $O(n\log{n})$ time to calculate the second minimum, being inefficient}.

To optimize this natural implementation, let us consider manually applying D\&C, a paradigm widely used for optimization. In general, D\&C decomposes a task into simpler subtasks of the same type and calculates by combining the results of subtasks. For the \m{sndmin} task, a standard procedure of D\&C is to divide the input list $\m{xs}$ into two halves $\m{xs}_L$ and $\m{xs}_R$, recursively calculates $\textit{sndmin}\ \m{xs}_L$ and $\textit{sndmin}\ \m{xs}_R$, and then combines them into $\textit{sndmin}\ \m{xs}$. In this procedure, a combinator $\textit{comb}$ satisfying the formula below is required, where $\m{xs}_L \cat \m{xs}_R$ represents the concatenation of two lists.
$$
\textit{sndmin}\ (\m{xs}_L \cat \m{xs}_R) = \textit{comb}\ (\textit{sndmin}\ \m{xs}_L, \textit{sndmin}\ \m{xs}_R) 
$$

\begin{figure*}

\hfill 
\begin{minipage}{0.5\textwidth}
\begin{figure}[H]
\small
\vspace{0.7em}
\fbox{\parbox{\textwidth}{
  \vspace{-0.5em}
\begin{align*}
  &\m{aux}\ \m{xs} = \m{min}\ \m{xs} \\
  &\m{comb}\ ((\m{smin}_L, \m{aux}_L), (\m{smin}_R, \m{aux}_R)) = \\
  &\quad \textbf{let } \m{csmin} = \min(\m{smin}_L, \m{smin}_R, \max(\m{aux}_L, \m{aux}_R))\textbf{ in} \\
  &\quad \quad \textbf{let } \m{caux} = \min(\m{aux}_L, \m{aux}_R) \textbf{ in} \\
  &\quad\quad\quad (\m{csmin}, \m{caux}) 
\end{align*}}}
\vspace{-0.5em}
\caption{\m{aux} and \m{comb} for {\it sndmin}.} \label{fig:auxcomb}
\end{figure}
  \end{minipage}
\quad \quad 
\begin{minipage}{0.42\textwidth}
\begin{figure}[H]
  \begin{lstlisting}[morekeywords=parallel]
def dac(xs, l, r):
  if r - l <= 1:
    return (orig([xs[l]]), aux([xs[l]])) 
  mid = (l + r) // 2
  lres = dac(xs, l, mid)
  rres = dac(xs, mid, r)
  return comb(lres, rres)
return dac(xs, 0, len(xs))[0]
      \end{lstlisting}
    \vspace{-1.4em}
    \caption{A divide-and-conquer template on lists.}
    \label{fig:dac}
\end{figure}
\end{minipage}
  \hfill \ 
\end{figure*}

However, such a combinator does not exist because the second minimum of the whole list may not be the second minimum of any of the two sub-list. In the example in Figure \ref{fig:sndmin-case}, the second minimums of the sub-lists are 3 and 4, respectively, but the second minimum of the whole list is 2. 
To solve this problem, a standard way is to extend the original program \m{sndmin} with a program $\m{aux}$ (denoted as an \textit{auxiliary program}) specifying necessary auxiliary values to make a valid combinator $\m{comb}$ exist, as shown below.
\begin{equation}
\begin{aligned}
&\textit{sndmin}'\ (\m{xs}_L \cat \m{xs}_R) = \textit{comb}\ (\textit{sndmin}'\ \m{xs}_L, \textit{sndmin}'\ \m{xs}_R) \\ 
& \qquad \key{where}\  \textit{sndmin}'\ \m{xs} \triangleq (\m{sndmin}\ \m{xs}, \m{aux}\ \m{xs}) 
\end{aligned}\label{formula:dac-sndmin}
\end{equation}

In this example, a valid auxiliary value is the first minimum of each sub-list. The corresponding $(\m{aux}, \m{comb})$ is shown in Figure~\ref{fig:auxcomb}, written in a syntax related to our synthesizer (Section \ref{subsection:moti-problem-challenge}). 
A D\&C program can be obtained by filling these two programs into a template (Figure~\ref{fig:dac}), where \targetname stands for the original program \m{sndmin} (Figure \ref{fig:smin}). In this template, function \m{dac} deals with the sub-list in range $[l, r)$ of the input array \m{xs} and calculates the expected result (second minimum here) and the auxiliary value of this sub-list. When the sub-list contains only one element, the original program and the \m{aux} are applied directly. Otherwise, \m{dac} is recursively invoked on the two halves of the sub-list, and the results are combined by \m{comb}. Note that although \m{aux} is applied only to singleton lists in this template, it is defined for all lists to guide the design of the \m{comb}. 

The time complexity of the resulting D\&C program is $O(n)$ {on a list of length $n$ }when \m{comb} runs in $O(1)$ time, and both \m{orig} and \m{aux} run in $O(1)$ time on singleton lists. This complexity can be further reduced to $O(n/p)$ on $p \leq n/\log n$ processors with proper parallelization. 

\smallskip
As demonstrated in the above procedure, applying D\&C is non-trivial. Although the template in Figure~\ref{fig:dac} is standard for D\&C programs on lists, we still need to find an auxiliary program $\m{aux}$ specifying proper auxiliary values and a corresponding combinator \m{comb}. These programs are observably more complex than the original program in Figure \ref{fig:smin}.


\subsection{Problem and Challenge} \label{subsection:moti-problem-challenge}
Motivated by the difficulty in manual optimization, we study the automatic application of D\&C. Concretely, given the original program \m{sndmin} (Figure \ref{fig:smin}) as the input, we aim to automatically synthesize proper $\m{aux}$ and $\m{comb}$ to fill the D\&C template (Figure \ref{fig:dac}), and meanwhile ensures both the correctness and the efficiency of the resulting D\&C program.


\begin{figure*} \small
  \begin{centering}
      \subfloat[The program space $\mathcal L_{\textit{aux}}^{\textit{ex}}$ of $\m{aux}$.]{
        \begin{tabular}{cccl} 
          \toprule 
          Start symbol & $S$ & $\rightarrow$ & $N_{\mathbb Z}\ |\ (S, S)$\\
          Integer expr & $N_{\mathbb Z}$& $\rightarrow$ & $N_{\mathbb Z} + N_{\mathbb Z}\ |\ \m{min}\ N_{\mathbb L}$ \\
          & & $|$ & $\m{max}\ N_{\mathbb L}\ \ \ |\ \m{sum}\ N_{\mathbb L}$ \\
          List expr & $N_{\mathbb L}$& $\rightarrow$ & $\text{Input}$  \\
          \bottomrule
          \vspace{-0.5em}
          \label{fig:aux-space}
  \end{tabular}
      }
      \hfill
      \subfloat[The program space $\mathcal L_{\textit{comb}}^{\textit{ex}}$ of $\m{comb}$.]{
      \begin{tabular}{cccl} 
        \toprule 
        Start symbol & $S$ & $\rightarrow$ & $N_{\mathbb Z}\ |\ (S, S)$\\
        Integer expr & $N_{\mathbb Z}$& $\rightarrow$ & $\text{Inputs}\hspace{1.4em} |\ \min(N_{\mathbb Z}, N_{\mathbb Z})$ \\
        & & $|$ & $N_{\mathbb Z} + N_{\mathbb Z}\ |\ \max (N_{\mathbb Z}, N_{\mathbb Z})$ \\
        \bottomrule
        \vspace{0.7em}
          \label{fig:comb-space}
      \end{tabular}
      }
      \vspace{-0.1em}
    \caption{A solution space for synthesizing a D\&C program of \m{sndmin}, where the output of $\m{aux}$ can be a tuple of integers, representing the usage of multiple auxiliary values (Figure \ref{fig:aux-space}), and the output of $\m{comb}$ can also be a tuple since \m{comb} usually needs to calculate multiple values (Figure \ref{fig:comb-space}).} \label{fig:solution-space}
  \end{centering}
  \vspace{-0.3em}
\end{figure*}

\begin{itemize}
  \item \textbf{(Correctness)} \jrymod{To ensure the resulting program correctly calculates the second minimum, we use Formula~\ref{formula:dac-sndmin} as the specification for synthesizing \m{aux} and \m{comb}.}{The resulting D\&C program should be semantically equivalent to the original program, that is, it should correctly calculate the second minimum of the input list. To ensure this point, we use Formula~\ref{formula:dac-sndmin} as the specification for synthesizing $\m{aux}$ and $\m{comb}$. At this time, by filling the synthesized $\m{aux}$ and $\m{comb}$ to the D\&C template (Figure \ref{fig:dac}), the resulting program must be correct.}
  \item \textbf{(Efficiency)} To ensure an efficient D\&C program that runs in $O(n/p)$-time in parallel, we apply the SyGuS framework~\cite{DBLP:conf/fmcad/AlurBJMRSSSTU13} and constrain the space of solutions to include only $\m{comb}$ that runs in $O(1)$ time and \m{aux} that runs in $O(1)$ time on singleton lists. 
  
  In this section, we use a toy solution space (Figure \ref{fig:solution-space}), which is simplified from the one in our implementation (Section \ref{section:implementation}), \jryadd{to illustrate the main idea of our approach}. \jryadd{This solution space satisfies the constraint above, that is, every \m{comb} in $\mathcal L_{\textit{comb}}^{\textit{ex}}$ runs in $O(1)$ time and every \m{aux} in $\mathcal L_{\textit{aux}}^{\textit{ex}}$ runs in $O(1)$ time on singleton lists.} One can verify that any possible solution $(\m{aux}, \m{comb})$ in this toy space can lead to an efficient D\&C program.

\end{itemize}


The synthesis task here is challenging because we need to synthesize two interrelated programs from a relational specification, and meanwhile {the total size of these two programs can be large in real-world algorithmic problems (up to $157$ AST nodes in our dataset)}. General program synthesis approaches that handle relational specifications such as enumerative synthesis~\cite{DBLP:conf/fmcad/AlurBJMRSSSTU13} and relational synthesis~\cite{DBLP:journals/pacmpl/0001WD18} do not scale up to solve most of the problems in our dataset. On the other hand, most other scalable synthesis approaches~\cite{DBLP:conf/iclr/BalogGBNT17,DBLP:conf/pldi/OseraZ15,DBLP:conf/pldi/FeserCD15,DBLP:conf/icse/RolimSDPGGSH17,DBLP:journals/pacmpl/MiltnerNBCD22,DBLP:journals/pacmpl/JiXXH21} work only for synthesizing a single program and require obtaining input-output examples. They cannot 
work for synthesizing two programs from a relational specification.

\subsection{An Existing Approach and Its Limitation} \label{subsection:moti-limitation}
\textit{Parsynt}~\cite{DBLP:conf/pldi/FarzanN17,toronto21} is a state-of-the-art synthesizer for D\&C. It solves the scalability challenge using a \jrymod{deductive}{syntax-based} program transformation system specifically designed for D\&C. Specifically, \m{Parsynt} applies its transformation system to the source code of the original program to directly derive \m{aux}. After \m{aux} is derived, only \m{comb} is unknown and can be synthesized using existing synthesizers. In this procedure, \m{Parsynt} will derive the full definition of \m{aux} to help synthesize \m{comb}, though \m{aux} is invoked only on singleton lists in the D\&C template (Figure \ref{fig:dac}).


\begin{wrapfigure}[]{r}{0.4\textwidth}
  \vspace{-1.1em}
  \begin{lstlisting}
fstmin, sndmin = INF, INF 
for v in xs:
  sndmin = min(sndmin, max(fstmin, v))
  fstmin = min(fstmin, v)
return sndmin
  \end{lstlisting}
  \vspace{-1.5em}
\caption{A single-pass program for \textit{sndmin}.}
\vspace{-1em}
\label{fig:smin-single-pass}
\end{wrapfigure}
The \jrymod{deductive}{syntax-based} transformation system in \m{Parsynt} puts a strict restriction on the original program, that is, the original program must be implemented as a single-pass program that enumerates each element in the input list only once. Figure \ref{fig:smin-single-pass} shows a single-pass implementation of \m{sndmin}, which is formed by a loop visiting each element in the input list $\m{xs}$ only once. This program takes the first minimum as an auxiliary value and updates the second minimum using the property that, each time a new element is visited, the new second minimum must be the medium value among the previous first minimum, the previous second minimum, and the new element. 

\jryadd{Although any functions that can be implemented as D\&C can also be implemented as single-pass after introducing enough auxiliary values\footnote{By the second list-homomorphism theorem~\cite{DBLP:journals/jfp/Gibbons96a}, any function that can be implemented as D\&C with a set of auxiliary values can also be implemented as single-pass with the same set of auxiliary values.}, the single-pass restriction of \m{Parsynt} still leads to significant burdens on the user from two aspects.
\begin{itemize}
  \item Similar to D\&C, implementing single-pass programs is difficult because many functions cannot be implemented as single-pass unless auxiliary values are introduced. In the above example of \m{sndmin} in Figure \ref{fig:smin-single-pass}, the user has to introduce the first minimum as an auxiliary value, which is already the auxiliary value required by D\&C. In the dataset we used for evaluation, the auxiliary values required by single-pass already account for $40.54\%$-$58.62\%$ of the auxiliary values required by D\&C (Section \ref{section:rq2}).
  \item Implementing single-pass programs is error-prone. The dataset used by \citet{toronto21} contains two bugs introduced when the authors manually implemented the original programs into single-pass. These bugs have been confirmed by the authors.
\end{itemize}
}

\subsection{\mainname on the Second Minimum Example} \label{subsection:moti-overview}
\jrydel{To remove the requirement on single-pass original programs, our approach \mainname solves this problem from the aspect of inductive synthesis. It never directly accesses the source code of the original program, and instead, it invokes the original program with some input to obtain the corresponding output or invokes a given verifier to verify the correctness of a candidate program. In this way, \mainname can accept any implementation of the original program that is supported by the verifier, such as the implementation in Figure~\ref{fig:smin}. 

{
  However, the scalability challenge resurfaces under the inductive setting. We can no longer extract $\m{aux}$ using deductive transformations (as \textit{Parsynt} does) because such transformations cannot be applied without accessing the source code of the original implementation. In this paper, we explore another direction to cope with the scalability challenge by answering the following question.
}}

\jryadd{To remove the requirement on single-pass original programs, we aim to solve the synthesis task without using syntax-based program transformations. 
Instead, we explore a decomposition-based approach to cope with the scalability challenge by answering the following question.
}

\smallskip

{\hfill
\begin{minipage}{0.7\textwidth}
  \fbox{\parbox{\textwidth}{\emph{Is it possible to derive a specification that involves only a sub-program of the synthesis target $(\m{aux}, \m{comb})$, such as \m{aux} only?}}}
\end{minipage}\hfill
}

\smallskip 

Our answer is positive. We propose two decomposition methods, named \emph{component elimination} and \emph{variable elimination}, to derive specifications for sub-programs of the synthesis target. By applying these methods, we can first synthesize a sub-program of the synthesis target using the derived specification and then synthesize the remainder with the help of the obtained sub-program.
In this way, we greatly reduce the scale of the program to be synthesized in each step. \jryadd{Given the difficulty of deriving a precise specification for a sub-program, we derive approximate specifications. At the end of this section we would discuss why approximate specifications do not affect the soundness of our approach.}

\begin{figure*}
  \centering
  \includegraphics[width=\linewidth]{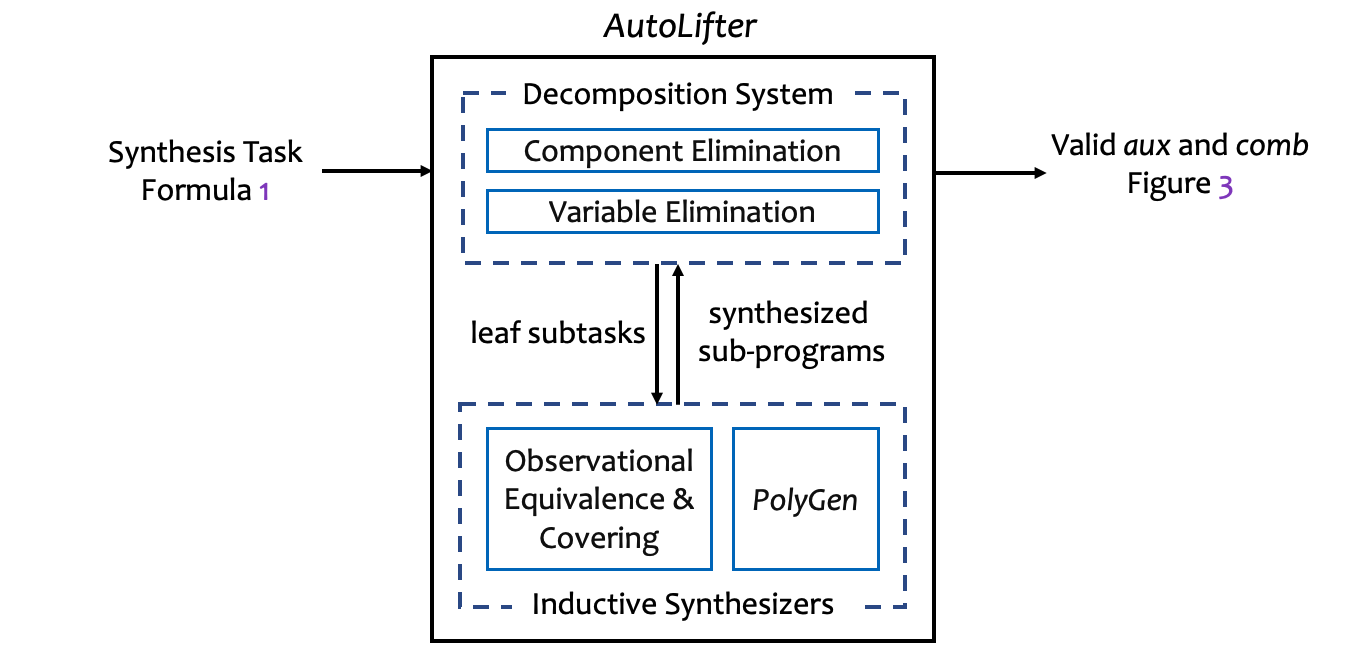}
  
  \vspace{-0.5em}
  \captionof{figure}{The workflow of \mainname.} 
  \label{fig:autolifter-framework}
  \end{figure*}

Figure \ref{fig:autolifter-framework} shows the workflow of \mainname, which synthesizes through an interaction between a decomposition system and two inductive synthesizers. Given a synthesis task, \jrymod{the decomposition methods are applied to decompose the task into}{the decomposition methods are iteratively applied to decompose the task and in this process, we would obtain} a series of \textit{leaf} subtasks (i.e., subtasks that cannot be further decomposed), each with a smaller scale and a simpler form. \mainname solves these subtasks one by one using inductive synthesizers and collects the results for generating subsequent subtasks and constructing the final result.




\smallskip

\jryadd{\noindent \textbf{Component elimination}. In the original specification (Formula \ref{formula:dac-sndmin}), the output of {\m{comb}} is a pair of two components, corresponding to the expected output of \m{sndmin} and the auxiliary values defined by {\m{aux}}, respectively. Accordingly, we can synthesize two sub-programs, denoted as $\m{comb}_1$ and $\m{comb}_2$, each for calculating an output component, and then constructs $\m{comb}$ as follows.
$$
\m{comb}~\m{in} \triangleq (\m{comb}_1~\m{in}, \m{comb}_2~\m{in})
$$

A natural idea here is to synthesize the two sub-programs individually~\cite{DBLP:conf/pldi/OseraZ15}. However, this method does not work here because the two sub-programs are both dependent on \m{aux}. Specifically, we could have the specifications below for the two sub-programs, where for clarity, we use \bluec{blue} to denote the original program, \var{red} to denote unknown programs to be synthesized, and \quanti{green} to denote universally quantified variables that range over all integer lists.

\begin{gather}
  \begin{aligned}
  &\bluec{\textit{sndmin}}\ (\quanti{\m{xs}_L} \cat \quanti{\m{xs}_R}) = \var{\textit{comb}_1}\ (\textit{sndmin}'\ \quanti{\m{xs}_L}, \textit{sndmin}'\ \quanti{\m{xs}_R}), \key{where}\  \textit{sndmin}'\ \m{xs} \triangleq (\bluec{\m{sndmin}}\ \m{xs}, \var{\m{aux}}\ \m{xs}) 
  \end{aligned}\label{formula:ce-1} \\
  \begin{aligned}
  &\var{\m{aux}}\ (\quanti{\m{xs}_L} \cat \quanti{\m{xs}_R}) = \var{\textit{comb}_2}\ (\textit{sndmin}'\ \quanti{\m{xs}_L}, \textit{sndmin}'\ \quanti{\m{xs}_R}), \key{where}\  \textit{sndmin}'\ \m{xs} \triangleq (\bluec{\m{sndmin}}\ \m{xs}, \var{\m{aux}}\ \m{xs}) 
  \end{aligned}\label{formula:ce-2}
\end{gather}

\noindent Both  specifications involve the same unknown program $\m{aux}$. If we synthesize from these specifications individually, we may get two incompatible results that use different $\m{aux}$.

\smallskip 
To solve this problem, we analyze the requirement on $\m{aux}$ put by each specification.
\begin{itemize}
  \item In Formula \ref{formula:ce-1}, $\m{aux}$ needs to provide enough information for calculating the second minimum.
  \item In Formula \ref{formula:ce-2}, $\m{aux}$ needs to ensure that the auxiliary values provide enough information for calculating themselves.
\end{itemize}

Our observation here is that, for the first requirement, it does not matter if $\m{aux}$ provides more information than necessary. Consequently, these two requirements can be satisfied in order. We can first find $\m{aux}$ to satisfy the first requirement and then if necessary, add more auxiliary values to satisfy the second requirement. At this time, the first requirement will still be satisfied because more information is provided. 

Following this idea, we design our first decomposition method \textit{component elimination}. 
The decomposition procedure is shown below.
\begin{itemize}
  \item The first subtask is the same as Formula \ref{formula:ce-1}. It targets finding those auxiliary values necessary for the second minimum and the corresponding combinator. Here, one possible result is to take $\m{aux}$ as $\m{min}$ and take $\m{comb}_1$ as the $\m{csmin}$ expression in Figure \ref{fig:auxcomb}.
  \item Then, the second subtask is shown below. It aims to expand the found auxiliary value to satisfy the second requirement. In this specification, the new auxiliary program $\m{aux}'$ denotes the new auxiliary values needed to calculate auxiliary values themselves. 
  \begin{gather}
  \bb
  &\m{aux}~(\quanti{\m{xs}_L} \cat \quanti{\m{xs}_R}) = \var{\m{comb}_2}~(\m{sndmin}'~\quanti{\m{xs}_L}, \m{sndmin}'~\quanti{\m{xs}_R}) \\
  &\textbf{where }\m{sndmin}'~\m{xs} \triangleq (\bluec{\m{sndmin}}~\m{xs}, \m{aux}~\m{xs}) \\
  &\hspace{4.8em}\m{aux}~\m{xs} \triangleq (\bluec{\m{min}}~\m{xs}, \var{\m{aux}'}~\m{xs})
  \ee \label{formula:ce-2-simplified}
  \end{gather}
  In the \m{sndmin} example, no other information is needed for calculating the auxiliary value of the first minimum. One possible result here is to take $\m{aux}'$ as an empty program (i.e., returns an empty tuple) and take $\m{comb}_2$ as the $\m{caux}$ expression in Figure \ref{fig:auxcomb}. 

  We shall discuss another example where new auxiliary values are needed in Section \ref{section:approach}.
  \item By merging the above results of subtasks, we can obtain the intended solution in Figure \ref{fig:auxcomb}.
\end{itemize}

\mainname will further decompose both subtasks to achieve efficient synthesis. 
\begin{itemize}
  \item The first subtask (Formula \ref{formula:ce-1}) will be decomposed by another decomposition method, \textit{variable elimination}, which will be introduced later.
  \item The second subtask (Formula \ref{formula:ce-2-simplified}) will be recursively decomposed by component elimination. Specifically, this subtask is similar to the original task (Formula \ref{formula:dac-sndmin}) in form. Both tasks are about finding new auxiliary values to calculate the output of a known function (together with the auxiliary values themselves). Therefore, this subtask can still be decomposed similarly. 
\end{itemize}


}

\smalltitle{Variable elimination} The first task generated by component elimination is still challenging because it involves two unknown programs, $\m{comb}_1$ and $\m{aux}$.
Our second decomposition method \m{variable elimination} decomposes this task by deriving a subtask involving only $\m{aux}$. In other words, this method eliminates a program variable (i.e., $\m{comb}_1$) from the specification.

To derive a specification only for \m{aux}, we first revisit the fundamental reason why \m{aux} is needed. Let us consider two pairs of lists, $(\m{xs}_L, \m{xs}_R)$ in Figure \ref{fig:sndmin-case} and another pair $(\m{xs}_L', \m{xs}_R')$.
\begin{equation}\label{formula:example-list-2}
  \m{xs}_L\ [1, 3, 5], [2, 4, 6]\ \m{xs}_R \qquad \m{xs}'_L\ [1, 3, 5], [1, 4, 6]\ \m{xs}'_R
\end{equation}
Although the second minimums of $\m{xs}_L$ and $\m{xs}_R$ ($3$ and $4$) are the same as their counterparts of $(\m{xs}_L', \m{xs}_R')$, the second minimum of the combined list $\m{xs}_L \cat \m{xs}_R$ (which is $2$) differs from that of $\m{xs}_L' \cat \m{xs}_R'$ (which is $1$). Consequently, if $\m{aux}$ is not involved, a conflict will emerge after substituting these two list pairs into the specification of $\m{comb}_1$ (Formula \ref{formula:ce-1}). Specifically, $(\m{xs}_L, \m{xs}_R)$ requires $\m{comb}_1$ to output $1$ from input $(3, 4)$ but $(\m{xs}_L', \m{xs}_R')$ requires $\m{comb}_1$ to output $2$ from the same input. Such a $\m{comb}_1$ does not exist because it must produce the same output from the same input.

\smallskip

\jryadd{
The above analysis indicates that a necessary condition on \m{aux} is to ensure that a function exists for $\m{comb}_1$ to implement. When the expected outputs of $\m{comb}_1$ (i.e., the second minimums of the combined lists) are different, the inputs (i.e., the second minimums or the auxiliary values on the two halves) must also be different. Our method \m{variable elimination} takes this necessary condition as the specification for synthesizing $\m{aux}$ and thus separates the synthesis of $\m{aux}$ and $\m{comb}_1$. This method decomposes the first task of component elimination (Formula \ref{formula:ce-1}) as follows.
\begin{itemize}
\item The first subtask is shown below. It targets finding auxiliary values such that two inputs of $\m{comb}_1$ must be different when their respective outputs differ.
\begin{equation*}
  \begin{aligned}
  \bluec{\m{sndmin}}\ (\quanti{xs_L} \cat \quanti{xs_R}) \neq \bluec{\m{sndmin}}\ (\quanti{xs'_L} \cat \quanti{xs'_R}) \hspace{16em}&\\
\rightarrow (\m{sndmin}'\ \quanti{xs_L}, \m{sndmin}'\ \quanti{xs_R}) \neq (\m{sndmin}'\ \quanti{xs_L'}, \m{sndmin}'\ \quanti{xs_R'})& \\
\textbf{where}\ \m{sndmin}' \m{xs} \triangleq (\bluec{\m{sndmin}}\ \m{xs}, \var{\m{aux}}\ \m{xs})&
  \end{aligned}
\end{equation*}

\ignore{
\begin{equation*}
  \begin{aligned}
  {\m{sndmin}'}\ ({xs_L} \cat {xs_R}) \neq {\m{sndmin}'}\ ({xs'_L} \cat {xs'_R}) \hspace{16em}&\\
\rightarrow (\m{sndmin}'\ {xs_L}, \m{sndmin}'\ {xs_R}) \neq (\m{sndmin}'\ {xs_L'}, \m{sndmin}'\ {xs_R'})& \\
\textbf{where}\ \m{sndmin}' \m{xs} \triangleq ({\m{sndmin}}\ \m{xs}, {\m{aux}}\ \m{xs})&
  \end{aligned}
\end{equation*}}
For clarity, we transform this specification into the following equivalent form to make the constraint on \m{aux} clear. One possible result here takes $\m{aux}$ as $\m{min}$.

\smallskip
\hspace{-2.7em}
\noindent\fbox{\parbox{\textwidth}{ 
\begin{equation}\label{formula:ve-1}
  \begin{aligned}
  (\bluec{\m{sndmin}}\ \quanti{xs_L}, \bluec{\m{sndmin}}\ \quanti{xs_R}) = (\bluec{\m{sndmin}}\ \quanti{xs_L'}, \bluec{\m{sndmin}}\ \quanti{xs'_R})& \\ 
  \wedge\ \bluec{\m{sndmin}}\ (\quanti{xs_L} \cat \quanti{xs_R}) \neq \bluec{\m{sndmin}}\ (\quanti{xs'_L} \cat \quanti{xs'_R})& \\
  \rightarrow (\var{\m{aux}}\ \quanti{xs_L}, \var{\m{aux}}\ \quanti{xs_R})& \neq (\var{\m{aux}}\ \quanti{xs_L'}, \var{\m{aux}}\ \quanti{xs_R'})
  \end{aligned}
  \end{equation}
}}
\smallskip 

This subtask will be used for synthesis without further decomposition; in other words, it is a leaf subtask of the decomposition. In this section, we include leaf subtasks in framed boxes to distinguish them from the other tasks.

\smallskip

\item Then, the second subtask is shown below. It aims to synthesize a corresponding $\m{comb}_1$ that calculates the second minimum using the auxiliary value found in the first subtask.

\smallskip
\hspace{-2.7em}
\noindent \fbox{\parbox{\textwidth}{
\begin{gather}
  \bluec{\m{sndmin}}\ (\quanti{\m{xs}_L} \cat \quanti{\m{xs}_R}) = \var{\m{comb}_1}\ \big((\bluec{\m{sndmin}}\ \quanti{\m{xs}_L}, \bluec{\m{min}}\ \quanti{\m{xs}_L}), (\bluec{\m{sndmin}}\ \quanti{\m{xs}_R}, {\bluec{\m{min}}}\ \quanti{\m{xs}_R})\big) \label{formula:ve-2}
\end{gather}
}}
\smallskip 

One possible result here takes ${\m{comb}_1}$ as the $\m{csmin}$ expression in Figure \ref{fig:auxcomb}.

\item By merging the results of the above subtasks, we can obtain a valid solution to the original task (Formula \ref{formula:ce-1}).
\end{itemize}
}

\jryadd{
\smalltitle{Synthesis from leaf tasks} After applying the above two decomposition methods, the original synthesis task (Formula \ref{formula:dac-sndmin}) is decomposed into two series of leaf tasks, one for sub-programs of $\m{aux}$ (represented by Formula \ref{formula:ve-1}) and the other for sub-programs of $\m{comb}$ (represented by Formula \ref{formula:ve-2}). We solve these leaf tasks following the framework of counter-example guided inductive synthesis (CEGIS)~\cite{DBLP:journals/sttt/Solar-Lezama13}. In CEGIS, the synthesizers focus on satisfying a set of examples (i.e., instances of the quantified variables $\m{xs}_L$, $\m{xs}_R$, $\m{xs}'_L$, and $\m{xs}_R'$) instead of the full specification, and a verifier verifies the correctness of the program synthesized from examples and provides new counter-examples when it is incorrect. 

Among the leaf tasks, the task for sub-programs of $\m{comb}$ (e.g., Formula \ref{formula:ve-2}) is already in the input-output form, where input-output examples are available. For example, under example $(\m{xs}_L, \m{xs}_R) \triangleq ([1,3,5], [2,4,6])$, Formula \ref{formula:ve-2} requires sub-program $\m{comb}_1$ to output $2$ from input $((3, 1), (4, 2))$. As a result, these tasks can be solved by existing inductive synthesizers that rely on input-output examples. We use a state-of-the-art synthesizer $\m{PolyGen}$~\cite{DBLP:journals/pacmpl/JiXXH21} in our implementation. 

In contrast, the task for sub-programs of $\m{aux}$ (e.g., Formula \ref{formula:ve-1}) is still relational. It involves the outputs of \m{aux} on four different inputs $(\m{xs}_L$, $\m{xs}_R$, $\m{xs}_L', \m{xs}_R')$, making input-output examples unavailable. Fortunately, a domain property here is that the size of $\m{aux}$ is usually much smaller than $\m{comb}$. Specifically, since $\m{aux}$ will only be invoked on singleton lists in the resulting program (Figure \ref{fig:dac}), it does not need to be efficient and thus can be synthesized compactly using high-level list operators. Using this property, we solve the leaf tasks for $\m{aux}$ using \m{observational equivalence}~\cite{DBLP:conf/fmcad/AlurBJMRSSSTU13}, a general enumeration-based synthesizer, and also proposes a specialized pruning method, named \textit{observational covering}, to speed up the synthesis for the cases requiring multiple auxiliary values. The details on this synthesizer can be found in Section \ref{subsection:inductive}.
}

\smallskip

\noindent \textbf{Notes}. 
There are two points worth noting in the synthesis procedure.
\begin{itemize}
  \item Although the full definition of \m{aux} is not used in the resulting D\&C program, it reduces the difficulty of synthesizing $\m{comb}$. As we can see, after the full \m{aux} is synthesized, the subsequent subtasks for $\m{comb}$ are no longer relational and can be solved easily.
  \item Neither \m{PolyGen} nor OE can be directly applied to the original problem (Formula \ref{formula:dac-sndmin}) without the decomposition. For \m{PolyGen}, neither input-output examples of $\m{aux}$ nor those of $\m{comb}$ can be extracted from Formula \ref{formula:dac-sndmin}; and for OE, 
  the target $\m{comb}$ is too large to be efficiently synthesized by enumeration.
\end{itemize}

\smallskip 

\noindent \textbf{Properties of \mainname.} 
The decomposition of \mainname is \m{sound} in the sense that any solution constructed from valid sub-programs for the leaf subtasks must satisfy the full specification. This is because, in each decomposition, the second subtask is always obtained by putting the result of the first subtask into the original specification before the decomposition. Therefore, the original specification must be satisfied when the second subtask is solved successfully.

\jrydel{
However, the decomposition methods used by \mainname is \emph{not complete}, possibly decomposing a realizable task (i.e., a task whose valid solution exists) into unrealizable subtasks.
This is because, in both decomposition methods, the specification derived for the first subtask is weaker than the original one, and thus it is possible to synthesize a program in the first subtask that makes the second subtask unrealizable. For example, the subtask for $\m{aux}$ (Formula \ref{formula:ve-1}) only ensures that a function exists for $\m{comb}_1$ to implement but does not ensure that such an implementation exists in the program space ($\mathcal L_{\textit{comb}}^{\textit{ex}}$, Figure \ref{fig:comb-space}). One can verify that $(\m{min}\ \m{xs}) + (\m{min}\ \m{xs})$ is also valid for this subtask, but a corresponding combinator does not exist in $\mathcal L_{\textit{comb}}^{\textit{ex}}$.

To deal with incompleteness, one possible way is to combine the decomposition system with a backtracking mechanism that guides \mainname to synthesize alternative \m{aux} once the previous \m{aux} leads to unrealizable subtask. However, we found such backtracking is not needed because this theoretical incompleteness of \mainname never happens in our evaluation; in other words, \mainname never generates unrealizable subtasks from realizable tasks in our dataset. In the remainder of this subsection, we shall intuitively discuss why this happens on this \m{sndmin} example.} 

\jryadd{
However, both decomposition methods in \mainname are approximate, possibly decomposing a realizable task (i.e., a task whose valid solution exists) into unrealizable subtasks. This is because, both decomposition methods use approximate specifications in their first subtask, and thus it is possible to synthesize a sub-program from the first subtask that can never form a valid solution, making the corresponding second subtask unrealizable. For example, the subtask for $\m{aux}$ (Formula \ref{formula:ve-1}) only ensures that a function exists for $\m{comb}_1$ to implement but does not ensure that such a program exists in the program space ($\mathcal L_{\textit{comb}}^{\textit{ex}}$, Figure \ref{fig:comb-space}). One can verify that $(\m{min}\ \m{xs}) + (\m{min}\ \m{xs})$ is also valid for this subtask, but a corresponding combinator does not exist in $\mathcal L_{\textit{comb}}^{\textit{ex}}$.

There are two possible strategies for using such approximate decomposition methods in practice.
\begin{itemize}
  \item (Greedy strategy) In each decomposition, consider only the first sub-program synthesized from the first subtask and then focus only on the corresponding second subtask.
  \item (Backtracking strategy) Each time an unrealizable subtask is met, backtrack to the previous decomposition step and try other valid sub-programs to the first subtask.
\end{itemize}
Both strategies are effective only when the approximation is precise enough to ensure that unrealizable subtasks are seldom generated. Otherwise, the greedy strategy will be frequently stuck into an unrealizable subtask, significantly harming the effectiveness; and the backtracking strategy will frequently roll back and switch to other search branches, significantly harming the efficiency. 

Fortunately, our evaluation results suggest that our decomposition methods are precise enough: they never generate any unrealizable subtask from realizable tasks in our dataset. In the remainder of this section, we shall intuitively discuss why this happens on the \m{sndmin} example.
}

\smallskip  

\jrymod{The key to solving the \m{sndmin} task is to ensure that $\m{aux}$ is exactly synthesized as $\m{min}~\m{xs}$ from the first subtask (Formula \ref{formula:ve-1})}{To ensure that no unrealizable subtask is generated when solving the \m{sndmin} task, the key is to ensure that $\m{aux}$ is exactly synthesized as $\m{min}~\m{xs}$ from the first subtask (Formula \ref{formula:ve-1})}, given that the subsequent two subtasks (Formulas \ref{formula:ve-2} and \ref{formula:ce-2-simplified}) are both determined by this result. \mainname achieves this through a combined effect between the enumeration-based synthesizer (mainly OE) and the program space ($\mathcal L_{\textit{aux}}^{\textit{ex}}$, Figure \ref{fig:aux-space}).

Programs in $\mathcal L_{\textit{aux}}^{\textit{ex}}$ can be divided into two categories. The first includes programs \textit{derived} by the intended solution $\m{min}\ \m{xs}$, for example, by including more auxiliary values (e.g., $(\m{min}\ \m{xs}, \m{max}\ \m{xs})$) or performing some arithmetic operations (e.g.,$(\m{min}\ \m{xs}) + (\m{min}\ \m{xs})$). Although many programs in this category satisfy the specification (Formula \ref{formula:ve-1}) as well, the pinciple of \emph{Occam's razor}~\cite{DBLP:journals/ipl/BlumerEHW87,DBLP:journals/pacmpl/JiXXH21} applies here: the intended solution $\m{min}\ \m{xs}$ is the smallest in this category. Since \mainname synthesizes $\m{aux}$ by enumerating programs from small to large, it prefers smaller programs and thus can successfully find $\m{min}\ \m{xs}$ from those unnecessarily complex programs.

The second category includes the remaining programs not related to $\m{min}\ \m{xs}$. The specification for $\m{aux}$ (Formula \ref{formula:ve-1}) is strong enough to exclude all programs in this category because of a property of these functions, which we name as the \m{compressing} property. {As a side effect of ensuring an efficient D\&C program, program space $\mathcal L_{\textit{aux}}^{\textit{ex}}$ includes only programs mapping a list (whose size is unbounded) to a constant-sized tuple of integers. Such programs \textbf{compress} a large input space to a much smaller output space\footnote{Here we assume the integer range is fixed for simplicity. The effect of the integer range on the compressing property will be discussed in Section \ref{section:discussion}.} and thus frequently output the same on different inputs. 
Consequently, an incorrect program in $\mathcal L_{\textit{aux}}^{\textit{ex}}$ can hardly satisfy the specification (Formula \ref{formula:ve-1}) because this specification requires \m{aux} to generate different outputs on a series of input pairs. 
For example, $\m{sum}\ \m{xs}$ and $\m{max}\ \m{xs}$ are two candidates in $\mathcal L_{\textit{aux}}^{\textit{ex}}$ that are not related to $\m{min}\ \m{xs}$. Both of them are rejected by Formula \ref{formula:ve-1}, and the corresponding counter-examples are listed in Table \ref{table:counterexample}.

Note that the specification for $\m{aux}$ (Formula \ref{formula:ve-1}) may be weak without the compressing property because it only requires $\m{aux}$ to output differently on some pairs of inputs. It accepts all programs that seldom output the same, such as the identity program $\m{id}\ \m{xs} \triangleq \m{xs}$. 
}

\smallskip 

The above two factors will be revisited formally in Section \ref{subsection:properties}.
\begin{itemize}
\item First, we prove that the probability for \mainname to \jrymod{be incomplete}{generate an unrealizable subtask} converges to $0$ under a probabilistic model where the semantics of programs are modeled as independent random functions with the compressing property (Theorem \ref{theorem:completeness}).
\item Second, we prove that \mainname can always find a minimal auxiliary program (i.e., no strict sub-program of the synthesized auxiliary program is valid), helping avoid unnecessarily complex solutions when the dependency among semantics is considered (Theorem \ref{theorem:minimal}).
\end{itemize}
  \begin{table*}
    \renewcommand\arraystretch{1.2}
    \caption{Counter-examples of $\m{sum}\ \m{xs}$ and $\m{max}\ \m{xs}$ for Formula \ref{formula:ve-1}.}
    \vspace{-0.5em}
      \small
      \begin{tabular}{|c|c|c|c|}
        \Xhline{1pt}
        Program & $(\m{xs}_L, \m{xs}_R)$ & $(\m{xs}_L', \m{xs}_R')$ & Simplified Specification \\
        \Xhline{1pt} 
        $\m{sum}\ \m{xs}$ & $([0, 2], [0, 1, 2])$ & $([0, 2], [1, 1, 1])$ & $(2, 1) = (2, 1) \wedge 0 \neq 1 \rightarrow (2, 3) \neq (2, 3)$\\
        \hline
        $\m{max}\ \m{xs}$ & $([0, 2], [0, 2])$ & $([0, 2], [1, 2])$ & $(2, 2) = (2, 2) \wedge 0 \neq 1 \rightarrow (2, 2) \neq (2, 2)$\\
        \Xhline{1pt}
      \end{tabular}
    \label{table:counterexample} 
  \end{table*}
\section{Lifting Problem} \label{section:problem}

Section \ref{sec:simpleOverview} shows how \mainname works for applying the D\&C paradigm. In this section, we show how to capture 
\jrymod{the application tasks of similar algorithmic paradigms}{the application tasks of D\&C-like paradigms}
uniformly as \textit{lifting problems}, a novel class of synthesis tasks considered by \mainname. 

\subsection{Example: Incrementalization for Second Minimum}\label{subsection:incre-example}

We use the paradigm of incrementalization as an example. Suppose now a series of changes are going to be applied to a list, each time a new integer will be appended, and the task is to determine the second minimum of the new list after each change. The incrementalization paradigm suggests computing some auxiliary values such that the new result after each change can be incrementally calculated from the previous one. In other words, we need to find a program \m{aux} for specifying auxiliary values and a combinator $\m{comb}$ for quickly updating the result, as shown below.
\begin{equation}
  \textit{sndmin}'\ (\m{append}\ \m{xs}\ v) = \textit{comb}\ (v, \textit{sndmin}'\ \m{xs}), \key{where}\ \textit{sndmin}'\ \m{xs} \triangleq (\m{sndmin}\ \m{xs}, \m{aux}\ \m{xs}) \label{formula:incre}
\end{equation}

\begin{wrapfigure}[]{r}{0.45\textwidth}
  \jrydel{\vspace{-1.3em}}
  \begin{minipage}{0.40\textwidth}
  \fbox{\parbox{\textwidth}{
    \vspace{-0.7em} 
    \small
  \begin{align*}
    &\m{aux}\ \m{xs} = \m{min}\ \m{xs} \\
    &\m{comb}\ (v, (\m{smin}_\textit{pre}, \m{aux}_\textit{pre})) = \\
    &\quad \textbf{let } \m{csmin} = \min(\m{smin}_{\textit{pre}}, \max(\m{aux}_{\textit{pre}}, v))\textbf{ in} \\
    &\quad \quad \textbf{let } \m{caux} = \min(\m{aux}_{\textit{pre}}, v) \textbf{ in} \\
    &\quad\quad\quad (\m{csmin}, \m{caux}) 
  \end{align*}}
  \vspace{-0.5em}
  }
\end{minipage}
\vspace{-0.7em}
\caption{\m{aux} and \m{comb} for incrementalization}
\vspace{-1.5em}
\label{fig:auxcomb4incre}
\end{wrapfigure}
A valid solution to this specification is shown in Figure~\ref{fig:auxcomb4incre}. Similar to the D\&C case, the auxiliary value is still the minimum element of the list, and the combinator updates both the second minimum and the first minimum with the newly appended integer $v$. This program takes $O(1)$ time for each update, but it is not easy to write since a proper auxiliary value is required.


\smallskip

{We can see that the above example task of applying incrementalization has many commonalities compared with the previous example task of applying D\&C to \m{sndmin} (Section \ref{subsection:example1}).}
\begin{itemize}
  \item In both tasks, a list is created from existing lists via an operator  ($\textit{xs}_L \cat \textit{xs}_R$ and $\textit{append}\ \textit{xs}\ v$) and the output of an original program ($\textit{sndmin}$) on the created list is calculated. 
  \item Both tasks aim at finding (1) a program $\textit{aux}$ (denoted as an \textit{auxiliary program}) for specifying auxiliary values, and (2) a corresponding combinator $\textit{comb}$ for calculating the outputs on the created list from those on the existing lists.
\end{itemize}
We denote such a problem as a \textit{lifting problem}. As we shall demonstrate later, the auxiliary program and the original program form a homomorphism that preserves a given operation; in other words, the auxiliary program \emph{lifts} the original program to be a homomorphism. 

\subsection{Lifting Problem} \label{subsection:lifting-problem}

\noindent \textbf{Notations}. In this paper, we regard a type as a set of values of the type and use the two terms interchangeably. To distinguish between types and values, we use uppercase letters such as $A, B$ to denote types, and lowercase letters (or words) such as $a, \m{xs}, \m{func}$ to denote values and functions. Particularly, we use overline letters such as $\overline{a}$ to denote values in the form of tuples. 

To operate types and functions, we use $T^n$ to denote the $n$-arity product $T \times \dots \times T$ of type $T$, $\textit{func}_1 \spl \textit{func}_2$ to apply two functions to the same value, $\m{func}_1 \times \m{func}_2$ to apply two functions to the two components in a pair, and $\textit{func}^n$ to apply a function to each component in an $n$-tuple.
\begin{gather*}
(\textit{func}_1 \spl \textit{func}_2)\ x \triangleq (\textit{func}_1\ x, \textit{func}_2\ x)\qquad (\textit{func}_1 \times \textit{func}_2)\ (x_1, x_2) \triangleq (\textit{func}_1\ x_1, \textit{func}_2\ x_2)\\
\textit{func}^n\ (x_1, \dots, x_n) \triangleq (\textit{func}\ x_1, \dots, \textit{func}\ x_n)
\end{gather*}

\noindent \textbf{Problem definition}. Given an original program \m{orig} over some data-structure type and an operator that creates an instance $a$ from some other instances $a_1, \ldots, a_n$ of the data structure, a \emph{lifting problem} is to find an auxiliary program and a combinator such that $\m{orig}\ a$ can be calculated from $\m{orig}\ a_1, \ldots, \m{orig}\ a_n$. Formally, a lifting problem is defined as follows.


\begin{definition}[Lifting Problem]\label{def:lifting} A lifting problem is specified by the following components.
  \begin{itemize}
    \item An original program $\textit{orig}$, whose input type is denoted as $A$. 
    \item An operator $\textit{op}$ with input type $C \times A^n$ and output type $A$ for some type $C$ and arity $n$. It constructs an $A$-element from $n$ existing $A$-elements and a complementary input in $C$. 
    \item {Two domain-specific languages $\mathcal L_{\textit{aux}}$ and $\mathcal L_{\textit{comb}}$, each specified by a grammar and the corresponding interpretations (i.e., semantics), defining the spaces of candidate programs.}
  \end{itemize}
The task of a lifting problem is to find an auxiliary program $\m{aux} \in \mathcal L_{\textit{aux}}$ and a combinator $\m{comb} \in \mathcal L_{\textit{comb}}$ such that the formula below is satisfied for any $c \in C$ and $\overline{a} \in A^n$: 
\begin{equation}
(\bluec{\m{orig}} \spl \var{\m{aux}})\ (\bluec{\m{op}}\ (\quanti{c}, \quanti{\overline{a}})) = \var{\m{comb}}\ (\quanti{c}, (\bluec{\m{orig}} \spl \var{\m{aux}})^n\ \quanti{\overline{a}}) \label{formula:lifting-definition}
\end{equation}
Following the notations in Section \ref{sec:simpleOverview}, we mark the known programs given in the synthesis task (e.g., original program \m{orig} and operator \m{op}) as \bluec{blue}, the unknown programs to be synthesized as \var{red}, and those universally quantified values as \quanti{green}. Furthermore, we shall also omit the range of a universally quantified value (such as $\forall\, \overline{a} \in A^n$ and $\forall c \in C$ here) if it is clear from the context.
\end{definition}

\begin{example} The specification of a lifting problem can be transformed into the following equivalent form to better correspond to the previous synthesis tasks (Formulas \ref{formula:dac-sndmin} and \ref{formula:incre}). 
$$
\m{orig}'\ \big(\bluec{\m{op}}\ (\quanti{c}, (\quanti{a_1, \dots, a_n}))\big) = \var{comb}\ (\quanti{c}, (\m{orig}'\ \quanti{a_1}, \dots, \m{orig}'\ \quanti{a_n})), \textbf{where }\m{orig}'\ x  \triangleq (\bluec{\m{orig}}\ x, \var{aux}\ x)
$$
Table \ref{table:correspondence} associates the concepts in a lifting problem with the two previous tasks, where \texttt{Unit} is a singleton type, and $()$ is the only element in \texttt{Unit} that provides no information. 
  \begin{table*}
    \renewcommand\arraystretch{1.15}
    \caption{The correspondence between the lifting problem and previous synthesis tasks.}
    \label{table:correspondence} 
    \begin{spacing}{1}
        \small
        \begin{tabular}{|c|c|c|c|c|c|c|}
            \Xhline{1pt}
            Paradigm & Specification  &  \targetname & $A$ & $\m{op}$ & $n$ & $C$ \\
            \Xhline{1pt}
             D\&C & Formula \ref{formula:dac-sndmin} & \multirow{3}{*}{\m{sndmin}} & \multirow{3}{*}{\texttt{List}}  & $\m{op}\ ((), (\m{xs}_L, \m{xs}_R)) \triangleq \m{xs}_L \cat \m{xs}_R$ & 2 & \texttt{Unit} \\
            \cline{1-2} \cline{5-7}
            \makecell{incrementalization \\ \jryadd{(list append)} } & Formula \ref{formula:incre} & & & $\m{op}\ (c, (\m{xs})) \triangleq \m{append}\ \m{xs}\ c$ & 1& \texttt{Int} \\
            \Xhline{1pt}
        \end{tabular}
    \end{spacing}
\end{table*}
\end{example}

A lifting problem is defined as a syntax-guided synthesis (SyGuS) problem~\cite{DBLP:conf/fmcad/AlurBJMRSSSTU13}, where the two languages $\mathcal L_{\textit{aux}}$ and $\mathcal L_{\textit{comb}}$ can be used to control the complexity of the generated program. We have seen the case of D\&C in Section \ref{subsection:moti-problem-challenge}, and the incremental program generated by solving the respective lifting problem (Formula \ref{formula:incre}) must run in $O(1)$ time per change if $\mathcal L_{\textit{comb}}$ includes only programs running in constant time.

In this paper, we assume that $\mathcal L_{\textit{aux}}$ and $\mathcal L_{\textit{comb}}$ are implicitly given and denote a lifting problem as $\mathsf{LP}(\targetname, \m{op})$. Besides, we assume that $\mathcal L_{\textit{aux}}$ contains a constant function $\textit{null}$ mapping anything to the unit constant $()$, corresponding to the case where no auxiliary value is required. 

\smallskip

\begin{wrapfigure}[]{r}{0.3\textwidth}
  \vspace{-1.7em}
  \begin{center}
    \includegraphics[width=0.3\textwidth]{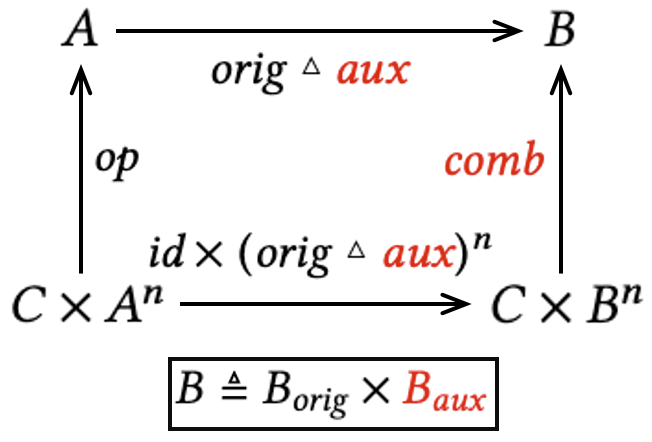}
  \end{center}
  \vspace{-1em}
\end{wrapfigure}
\noindent \textbf{Meaning}. A lifting problem has a clear algebraic meaning about synthesizing a \m{homomorphism}. For clarity, we draw the commutative diagram of its specification on the right, where each arrow represents a function application and the two paths from the lower-left to the upper-right result in the same function, $\m{id}$ is the identity function, and $B$ denotes the output type of $\m{orig} \spl \m{aux}$. 
This diagram shows that $\m{orig} \spl \m{aux}$ is a homomorphism mapping from $A$ to $B$, where the operator $\m{op}$ (related to $A$) is preserved as the combinator $\m{comb}$ (related to $B$).

In the sense of program optimization, the lifting problem is about eliminating the construction of $A$-elements. In its specification (Formula \ref{formula:lifting-definition}), the left-hand side explicitly constructs an $A$-element via $\m{op}$ and immediately consumes it via $\m{orig}$; in contrast, the right-hand side avoids this construction by directly calculating from existing results, via the synthesized combinator $\m{comb}$. This intuition matches a general optimization strategy, named \textit{fusion}~\cite{PePr96}, which suggests that a program is efficient if there is no unnecessary intermediate data structure produced and consumed during the computation.

\jrydel{
\smalltitle{Applying to Algorithmic Paradigms} As mentioned before, the application of many algorithmic paradigms can be reduced to lifting problems. The reductions for paradigms other than D\&C and incrementalization can be found in Section~\ref{section:application}. {Given a reduction, every synthesizer for lifting problems can be instantiated as a synthesizer for applying the corresponding paradigm. In practice, to apply a certain algorithmic paradigm, the end user only needs to pick up the corresponding instantiated synthesizer and provide the original program, and then the instantiated synthesizer will automatically generate a semantically equivalent program in the target paradigm.}}

\jryadd{
\smalltitle{Applying to D\&C-like algorithmic paradigms} Besides D\&C and incrementalization, there are many other algorithmic paradigms sharing the idea of building up the final results step-by-step through a prescribed recursive structure. We denote these paradigms as D\&C-like paradigms, and other such paradigms include single-pass~\cite{DBLP:reference/db/Schweikardt18a}, segment trees~\cite{DBLP:conf/innovations/LauR21}, and three greedy paradigms for longest segment problems~\cite{DBLP:journals/scp/Zantema92}. The application of these paradigms can also be reduced to lifting problems, as shall be discussed in Section \ref{section:application}.

Given a reduction from the application of a certain D\&C-like paradigm to lifting problems, any synthesizer for lifting problems can be instantiated as a synthesizer for applying the respective paradigm. In practice, to obtain an efficient algorithm for a specific task, the user needs only to select an available D\&C-like algorithmic paradigm, pick up the corresponding instantiated synthesizer, and provide the original program. Then, the instantiated synthesizer will automatically generate a semantically equivalent program in the target paradigm. 

We shall discuss how to select an algorithmic paradigm in Section \ref{section:discussion}.
}

\section{Approach} \label{section:approach}
In this section, we shall illustrate \mainname in detail with a more complex example related to a classic problem, \textit{maximum segment sum (mss)}~\cite{Bird89}. This section is organized as follows. Section \ref{subsection:example2} introduces the \m{mss} example, Section \ref{subsection:decomposition} discusses the decomposition methods, Section \ref{subsection:inductive} shows how to solve the leaf subtasks via inductive synthesis, and Section \ref{subsection:properties} summarizes the theoretical properties of \mainname.

\subsection{Example: Divide-and-Conquer for Maximum Segment Sum} \label{subsection:example2}

\begin{figure*}
	\hfill
	\centering
	\begin{minipage}{.36\textwidth}
		\begin{figure}[H]
			\begin{lstlisting}
mss = -INF
for i in range(len(x)):
  for j in range(i, len(x)):
    mss = max(mss, sum(x[i: j+1]))
return mss
			\end{lstlisting}
		  \vspace{-1.5em}
		  \caption{Maximum segment sum}
		  \label{fig:mss}
		  \end{figure}
	\end{minipage}
	\hfill \hfill
	\begin{minipage}{.36\textwidth}
		\centering
		\begin{figure}[H]
		\includegraphics[width=1\linewidth]{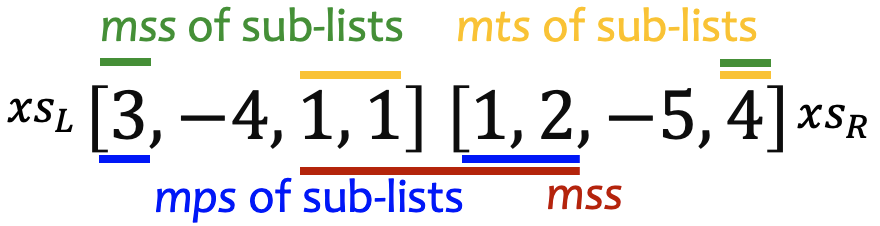}
		\vspace{-2em}
		\captionof{figure}{An example of calculating \m{mss}.}
		\vspace{-2em}
		\label{fig:mss-case}
		\end{figure}
	  \begin{figure}[H]
	  \includegraphics[width=1\linewidth]{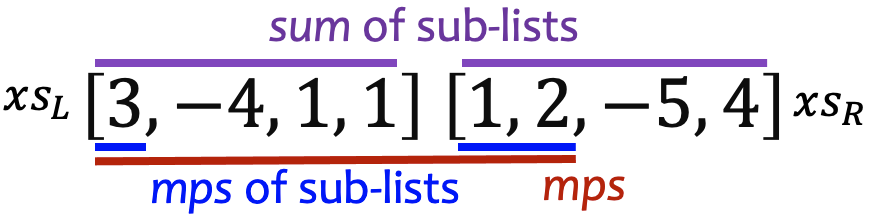}
	  \vspace{-2em}
	  \captionof{figure}{An example of calculating \m{mps}.}
	  \label{fig:mps-case}
	  \end{figure}
	\end{minipage}%
	\hfill \ 
	\end{figure*}

  Given a list of integers, we can create many contiguous subsequences, called segments. For each segment, we can add up integers within the segment to get the segment sum. The {\it mss} problem is to find, for a given list, the greatest sum we can get among all segments.
  A natural implementation of $\m{mss}$ (Figure \ref{fig:mss}) enumerates all segments, calculates their segment sums, and returns the maximum. This program runs in $O(n^3)$ time on a list of length $n$ and thus is quite inefficient.
  
  D\&C can be applied to optimize this natural implementation. However, similar to the second minimum example, if we divide the input list into two halves, the \m{mss} of the whole list cannot be calculated from those of the two halves. In the case shown in Figure \ref{fig:mss-case}, the segment with the maximum sum of the left half is the prefix list $[3]$, that of the right half is the prefix list $[1, 2]$, but the segment with the maximum sum of the whole list (i.e.,$[1, 1, 1, 2]$) is a concatenation of a tail-segment of the left half and a prefix of the right half. 
  
  To resolve the issue exposed by Figure \ref{fig:mss-case}, we can take the maximum prefix sum (\m{mps}) and the maximum tail-segment sum (\m{mts}) as auxiliary values so that the $\m{mss}$ of the whole list in Figure \ref{fig:mss-case} can be produced by adding up the \m{mts} of the left half and the \m{mps} of the right half. However, the problem is not completely solved yet. These auxiliary values should also be calculated during D\&C, and the same issue shall occur again: no corresponding combinator exists unless new auxiliary values are introduced. Figure~\ref{fig:mps-case} demonstrates such a case, where the \m{mps} of the whole list covers the full left half, and its sum cannot be produced using only \m{mps}, \m{mts}, and \m{mss} of the two halves. Here, we can introduce the sum of integers in the list as a supplementary auxiliary value to enable the calculation of $\m{mps}$ and $\m{mts}$. In this way, the $\m{mps}$ of the whole list in Figure \ref{fig:mps-case} can be produced by adding up the sum of the left half and the $\m{mps}$ of the right half.
  
  \begin{figure*}
	\begin{minipage}{.44\textwidth}
	  \begin{figure}[H]
		{\small
		\begin{tabular}{cccl} 
		  \toprule 
		  Start symbol & $S$ & $\rightarrow$ & $N_{\mathbb Z}\ |\ (S, S)$\\
		  Integer expr & $N_{\mathbb Z}$& $\rightarrow$ & $N_{\mathbb Z} + N_{\mathbb Z}\ |\ \m{min}\ N_{\mathbb L}$ \\
		  & & $|$ & $\m{max}\ N_{\mathbb L}\ \ \ |\ \m{sum}\ N_{\mathbb L}$ \\
		  & & $|$ & $\m{mps}\ \; \! N_{\mathbb L}\ \ \ |\ \m{mts}\ \; \!N_{\mathbb L}$ \\
		  List expr & $N_{\mathbb L}$& $\rightarrow$ & $\text{Input}$  \\
		  \bottomrule
  \end{tabular}}\vspace{-0.8em}
  \begin{centering}
  \begin{lstlisting}[frame=none]
	mps xs = max([sum(xs[:i+1])
					 for i in range(lens(xs))])
	mts xs = max([sum(xs[i:]) 
					 for i in range(lens(xs))])
  \end{lstlisting}
  \end{centering}
  \vspace{-0.5em}
  \caption{The extended program $\mathcal L_{\textit{aux}}^{\textit{mss}}$ of $\m{aux}$ for the \m{mss} example, where semantics of $\m{mps}$ and $\m{mts}$ are explained using a Python-like syntax.}
  \label{fig:aux-space-extended}
	  \end{figure}
	\end{minipage}
  \hfill
  \begin{minipage}{0.5\textwidth}
  \begin{figure}[H]
  \small
  \centering
  \includegraphics[width=1\linewidth]{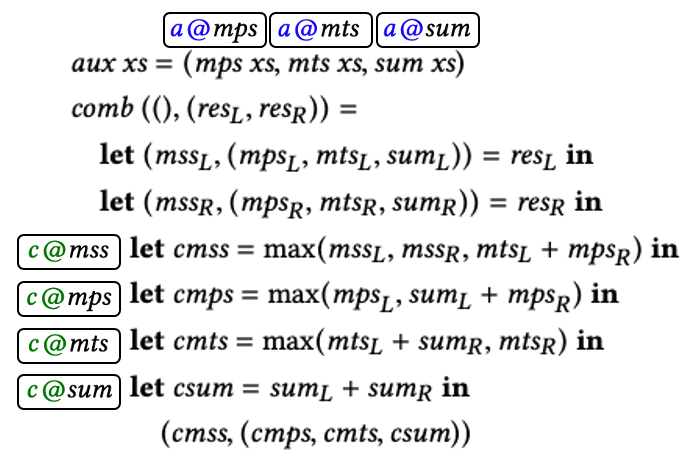}
  \vspace{-1.2em}
  \captionof{figure}{The expected \m{aux} and \m{comb} for \m{mss}.}
  \label{fig:mss-dac}
  \end{figure} 
	\end{minipage}
	\end{figure*}
  
  \smallskip
  
  {The task of applying D\&C to $\m{mss}$ can be regarded as a lifting problem $\mathsf{LP}(\m{mss}, \m{op})$ for operator $\m{op}\ ((), (\m{xs}_L, \m{xs}_R)) \triangleq \m{xs}_L \cat \m{xs}_R$. The raw specification of this lifting problem is as follows.
  \begin{gather*}
	(\bluec{\m{mss}} \spl \var{\m{aux}})\ (\quanti{\m{xs}_L} \cat \quanti{\m{xs}_R}) = \var{comb}\ ((), (\bluec{\m{mss}} \spl \var{\m{aux}})^2\ (\quanti{\m{xs}_L}, \quanti{\m{xs}_R}))
  \end{gather*}}
  \jryadd{
  which can be transformed into a more readable form as shown below.
  \begin{gather}
	\m{mss}'\ (\quanti{\m{xs}_L} \cat \quanti{\m{xs}_R}) = \var{comb}\ \big((), (\m{mss}'~\quanti{\m{xs}_L}, \m{mss}'~\quanti{xs_R})\big), \textbf{where }\m{mss}' \triangleq \bluec{\m{mss}} \spl \var{\m{aux}}
	\label{formula:mss-spec}
  \end{gather}
  }

  For simplicity, in this example, we continue using $\mathcal L_{\textit{comb}}^{\textit{ex}}$ (Figure \ref{fig:comb-space}) and extend $\mathcal L_{\textit{aux}}^{\textit{ex}}$ (Figure \ref{fig:aux-space}) by directly introducing $\m{mps}$ and $\m{mts}$ as language constructs ($\mathcal L_{\textit{aux}}^{\textit{mss}}$, Figure \ref{fig:aux-space-extended}). Note that the full languages used in our implementation are formed by more primitive constructs where, for example, $\m{mps}$ is implemented as $\m{max}\ (\m{scanl}\ (+)\ \m{xs})$ (Example \ref{example:implement-mps}, Section \ref{section:implementation}).
  
  Figure~\ref{fig:mss-dac} shows the expected solution synthesized from $\mathcal L_{\textit{aux}}^{\textit{mss}}$ and $\mathcal L_{\textit{comb}}^{\textit{ex}}$, where \m{aux} returns a 3-tuple and \m{comb} returns a 4-tuple. This solution is formed by expressions for producing components in the output tuples. We label these expressions in Figure \ref{fig:mss-dac} for later reference.

\subsection{Decomposition System} \label{subsection:decomposition}
\mainname decomposes lifting problems using two decomposition methods,
\m{component elimination} and \m{variable elimination}. \jryadd{For clarity, for each method, we shall first discuss its general idea on a compact specification and then show how to apply it to decompose lifting problems.}

\jryadd{
\subsubsection{Component Elimination} In lifting problems, the output of $\m{comb}$ is a pair of two components, corresponding to the expected output of the original program and the auxiliary values defined by \m{aux}. Our first decomposition method, component elimination, is proposed for decomposing a lifting problem into two subtasks, each involving only one component in the output pair.

\smallskip 
\noindent \textbf{General idea}. The task considered by component elimination is to calculate the output of a program from the output of another program. The general specification of this task is shown below.
\begin{align}
	\forall (\quanti{x}, \quanti{y}) \in S, (\bluec{\m{out}}~\quanti{x}, \var{g}~\quanti{x}) = \var{f}~(\bluec{in}~\quanti{y}, \var{g}~\quanti{y}) \label{formula:ce-general-full}
	\end{align}
In this specification, \m{out} and \m{in} are known functions with the same input type, $S$ is a set of input pairs, each comprising an input of \m{out} and an input of \m{in}, and $f$ and $g$ are two unknown programs to be synthesized. Intuitively, $g$ specifies some auxiliary values, and $f$ calculates the output of $\m{out}$ and the new auxiliary values from the output of $\m{in}$ and the corresponding auxiliary values.

Component elimination decomposes this task into two subtasks, each involving sub-programs of $f$ and $g$. Specifically, since $f$ is required to output a pair of two values, we can assume the target program of $f$ is formed by two sub-programs $f_1$ and $f_2$ such that $f \triangleq f_1 \spl f_2$, each for calculating one value in the output pair. 
Furthermore, if the program space of $f$ includes operators for accessing tuples, without loss of generality, we can assume $g$ is formed by two sub-programs $g_1$ and $g_2$ such that $g \triangleq g_1 \spl g_2$, where $g_1$ provides the auxiliary values for $f_1$, and $g_2$ provides the extra auxiliary values needed by $f_2$. Then, the specification is decomposed into two subtasks. 

\begin{enumerate}
	\item The first subtask synthesizes $f_1$ and $g_1$ from the specification below, where $f_1$ returns the output of $\m{out}$, and $g_1$ provides necessary auxiliary values for $f_1$.
	\begin{align}
	\forall (\quanti{x}, \quanti{y}) \in S, \bluec{\m{out}}~\quanti{x} = \var{f_1}~(\bluec{in}~\quanti{y}, \var{g_1}~\quanti{y}) \label{formula:ce-general-1}
	\end{align}
	\item If $g_1$ is \m{null}, i.e., no auxiliary value is needed, a solution for the full specification (Formula \ref{formula:ce-general-full}) is $f \triangleq f_1 \spl \m{null}$ and $g \triangleq \m{null}$. At this time, the second subtask is not needed.
	\item Otherwise, the second subtask synthesizes 
	$f_2$ and 
	$g_2$ from the specification below, where $f_2$ returns the new auxiliary values and $g_2$ extends the resulting $g_1$ of the first subtask with extra auxiliary values.
	\begin{align}
	\forall (\quanti{x}, \quanti{y}) \in S, (\bluec{g_1}~\quanti{x}, \var{g_2}~\quanti{x}) = \var{f_2}~((\m{\bluec{in}}\spl\bluec{g_1})~\quanti{y}, \var{g_2}~\quanti{y}) \label{formula:unfold-ce-2}
	\end{align}
	The second subtask has the same form as the full specification. Therefore, it can be recursively decomposed by component elimination.
	\item Using the results of subtasks, a solution for the full specification can be constructed as follows.
	$$
	f~(v_{\textit{in}}, (v_1, v_2)) \triangleq \big(f_1~(v_{\textit{in}}, v_1), f_2~((v_{\textit{in}}, v_1), v_2)\big) \qquad g~y \triangleq (g_1~y, g_2~y)
	$$
\end{enumerate}

Please note that the above decomposition is approximate. There is no guarantee that the auxiliary values found in the first subtask (i.e., the output of $g_1$) can be calculated using programs in the program spaces of $f_2$ and $g_2$. Consequently, the second subtask may be unrealizable.


\smallskip

\noindent \textbf{Usage in \mainname}. Let us first rewrite the specification of the lifting problem (Formula \ref{formula:lifting-definition}) into the following equivalent form.
\begin{equation}
(\bluec{\m{orig}}~(\bluec{\m{op}}\ (\quanti{c}, \quanti{\overline{a}})), \var{\m{aux}}~ (\bluec{\m{op}}\ (\quanti{c}, \quanti{\overline{a}}))) = (\var{\m{comb}}~(\quanti{c}~(\bluec{\m{orig}}^n~\quanti{\overline{a}}, \var{\m{aux}}^n~\quanti{\overline{a}})) \label{formula:lifting-definition-force}
\end{equation}
We can see that the above formula has mostly the same form as the general form in Formula~\ref{formula:ce-general-full} (repeated below), where the correspondence is shown in the following table. 
$$\textrm{General form: }(\bluec{\m{out}}~\quanti{x}, \var{g}~\quanti{x}) = \var{f}~(\bluec{in}~\quanti{y}, \var{g}~\quanti{y}) $$
\begin{center}
\begin{tabular}{ll|ll}
	\hline
	General Form & Lifting Problem  & General From & Current Task\\
	\hline
	$\var{f}$ & $\var{\m{comb}}$ & $\bluec{\m{in}}$ &  $\bluec{\m{orig}}^n$ \\
	$\var{g}$ & $\var{\m{aux}}$ & $\bluec{\m{out}}$ & $\bluec{\m{orig}}$\\
	\cline{3-4}
	$\quanti{x}$ & $\bluec{\m{op}}\ (\quanti{c}, \quanti{\overline{a}})$\\
	$\quanti{y}$ & $\quanti{\overline{a}}$\\
	\cline{1-2}
\end{tabular}
\end{center}
\smallskip

\noindent The two differences here are that (1) $\m{comb}$ has an extra parameter $c$ and (2) $\m{aux}$ is applied to every element of a tuple. Such differences do not affect the core idea of component elimination, and our
general discussion can be trivially extended to cover this form.

Following the decomposition procedure of component elimination, we can assume the form of $\m{comb}$ as a pair of two sub-programs $\m{comb}_1$ and $\m{comb}_2$, assume the form of $\m{aux}$ as a pair of two sub-programs $\m{aux}_1$ and $\m{aux}_2$, and then decompose the lifting problems as follows.
\begin{enumerate}
	\item The first subtask synthesizes $\m{comb}_1$ and $\m{aux}_1$ from the specification below.
	\begin{align}
	\bluec{\m{orig}}~(\bluec{\m{op}}~(\quanti{c}, \quanti{\overline{a}})) = \var{\m{comb}_1}~\big(\quanti{c}, (\bluec{\m{orig}}^n~\quanti{\overline{a}}, \var{\m{aux}_1}^n~\quanti{\overline{a}})\big) \label{formula:ce-sub1}
	\end{align}
	\item If the synthesis result of $\m{aux}_1$ is \m{null}, which means, no auxiliary value is needed to calculate the expected output of $\m{orig}$, a solution for the lifting problem is $\m{comb} \triangleq \m{comb}_1 \spl \m{null}$ and $\m{aux} \triangleq \m{null}$. At this time, the second subtask is not needed.
	\item Otherwise, the second subtask synthesizes $\m{comb}_2$ and $\m{aux}_2$ from the specification below, where $\m{comb}_1$ and $\m{aux}_1$ denote the result of the first subtask. This task can be recursively solved in the same way. 
	\begin{align}
		(\bluec{\m{aux}_1} \spl \var{\m{aux}_2})~(\bluec{\m{op}}~(\quanti{c}, \quanti{\overline{a}})) = \var{\m{comb}_2}~\big(\quanti{c}, ((\bluec{\m{orig}}\spl \bluec{\m{aux}_1})^n~\quanti{\overline{a}}, \var{\m{aux}_2}^n~\quanti{\overline{a}})\big) \label{formula:ce-sub2}
	\end{align}

\item Given the results of the two subtasks, a solution for the lifting problem can be obtained by taking $\m{aux}$ as $\m{aux}_1 \spl \m{aux}_2$ and constructing $\m{comb}$ by pairing up $\m{comb}_1$ and $\m{comb}_2$ with properly adjusting the structure of their inputs.
\end{enumerate}

\begin{example} After applying component elimination to the \m{mss} task (Formula \ref{formula:mss-spec}), the specification of the first subtask is as follows.
	\begin{align}
		\bluec{\m{mss}}\ (\quanti{\m{xs}_L} \cat \quanti{\m{xs}_R}) = \var{\m{comb}_1}\ \big((), \big((\bluec{\m{mss}}~\quanti{\m{xs}_L}, \bluec{\m{mss}}~\quanti{\m{xs}_R}), (\var{\m{aux}_1}~\quanti{\m{xs}_L}, \var{\m{aux}_1}~\quanti{\m{xs}_R})\big) \big) \label{formula:mss-ce-sub1}
		\end{align}
	
\noindent One can verify that components $\acomp{mps}, \acomp{mts}$ and $\ccomp{mss}$ in Figure \ref{fig:mss-dac} form a valid solution for this subtask. Given this solution, the specification of the second subtask is as follows. 
	\begin{equation*}
		\begin{aligned}
		&(\bluec{\m{aux}_1} \spl \var{\m{aux}_2})\ (\quanti{\m{xs}_L} \cat \quanti{\m{xs}_R}) 
		= \var{\m{comb}_2}\ \big((), 
		  \big((\bluec{\m{known}}~\quanti{\m{xs}_L}, \bluec{\m{known}}~\quanti{\m{xs}_R}), 
		  (\var{\m{aux}_2}~\quanti{\m{xs}_L}, \var{\m{aux}_2}~\quanti{\m{xs}_R})
		  \big)\big) \\
		&\qquad \textbf{where }\bluec{\m{aux}_1} \triangleq (\bluec{\m{mps}} \spl \bluec{\m{mts}}) \textrm{ and } \bluec{\m{known}} \triangleq \bluec{\m{mss}}\spl\bluec{\m{aux}_1} 
		\end{aligned}
	  \end{equation*}
This subtask can be recursively decomposed by component elimination. The first subtask of this recursive decomposition (denoted as subtask 2.1) is as follows. 
\begin{align*}
	&\bluec{\m{aux}_1}\ (\quanti{\m{xs}_L} \cat \quanti{\m{xs}_R}) = \var{\m{comb}_{2.1}}\ \big((), \big((\bluec{\m{known}}~\quanti{\m{xs}_L}, \bluec{\m{known}}~\quanti{\m{xs}_R}), (\var{\m{aux}_{2.1}}~\quanti{\m{xs}_L}, \var{\m{aux}_{2.1}}~\quanti{\m{xs}_R})\big) \big)\\
	&\qquad \textbf{where }\bluec{\m{aux}_1} \triangleq (\bluec{\m{mps}} \spl \bluec{\m{mts}}) \textrm{ and } \bluec{\m{known}} \triangleq \bluec{\m{mss}}\spl\bluec{\m{aux}_1} 
\end{align*}
One can verify that components $\acomp{sum}$, $\ccomp{mps}$, and $\ccomp{mts}$ in Figure \ref{fig:mss-dac} form a valid solution for subtask 2.1, which leads to the following subtask 2.2.
\begin{equation*}
	\begin{aligned}
	&(\bluec{\m{aux}_{2.1}} \spl \var{\m{aux}_{2.2}})\ (\quanti{\m{xs}_L} \cat \quanti{\m{xs}_R}) 
	= \var{\m{comb}_{2.2}}\ \big((), 
	  \big((\bluec{\m{known}}~\quanti{\m{xs}_L}, \bluec{\m{known}}~\quanti{\m{xs}_R}), 
	  (\var{\m{aux}_{2.2}}~\quanti{\m{xs}_L}, \var{\m{aux}_{2.2}}~\quanti{\m{xs}_R})
	  \big)\big) \\
	&\qquad \textbf{where }\bluec{\m{aux}_{2.1}} \triangleq \bluec{\m{sum}} \textrm{ and } \bluec{\m{known}} \triangleq \big(\bluec{\m{mss}}\spl(\bluec{\m{mps}} \spl \bluec{\m{mts}}) \big) \spl \bluec{\m{aux}_{2.1}}
	\end{aligned}
\end{equation*}
This subtask can still be recursively decomposed by component elimination. At this time, we shall get $\m{null}$ for the first sub-program of $\m{aux}_{2.2}$, representing that no extra auxiliary value is needed, and thus the synthesis finishes.

\end{example}
}

\jryadd{
\subsubsection{Variable Elimination} The first task of component elimination involves two unknown programs $\m{aux}_1$ and $\m{comb}_1$ that occur in the form of a composition. Our second decomposition method, variable elimination, is proposed for decomposing this task into two subtasks, each involving only one unknown program.

\smallskip 

\noindent \textbf{General idea}. Variable elimination aims to eliminate an unknown program from an input-output specification. Specifically, it considers specifications in the following form, where $f$ and $g$ are two unknown programs to be synthesized.
\begin{align}
\forall (\quanti{x}, \quanti{y}) \in S, \quanti{x} = \var{f}~(\bluec{\m{in}}~\var{g}~\quanti{y})  
\label{formula:ve-general-full}
\end{align}
This is an input-output specification for program $f$, where $S$ is a set of value pairs, each comprising an output of $f$ and a parameter for generating the input, and $\m{in}$ is a second-order program that generates an input of $f$ using the other unknown program $g$ and a parameter $y$.

Variable elimination decomposes this specification using the fact that $f$ acts as a function. To ensure such a function exists, $g$ must ensure that for any two pairs in $S$, the inputs of $f$ must be different when the expected outputs differ. Using this property, variable elimination decomposes the full specification (Formula \ref{formula:ve-general-full}) into the two sequential subtasks below, where the $g$ synthesized from the first subtask (Formula \ref{formula:ve-general-sub1}) is put into the second one (Formula \ref{formula:ve-general-sub2}), making the input and the output of $f$ no longer symbolic in the second subtask.\vspace{-0.3em}
\begin{gather}
	\forall (\quanti{x}, \quanti{y}), (\quanti{x'}, \quanti{y'}) \in S, \quanti{x} \neq \quanti{x'}  \rightarrow \bluec{\m{in}}~\var{g}~\quanti{y} \neq \bluec{\m{in}}~\var{g}~\quanti{y'} \label{formula:ve-general-sub1} \\
	\forall (\quanti{x}, \quanti{y}) \in S, \quanti{x} = \var{f}~(\bluec{\m{in}}~\bluec{g}~\quanti{y}) \label{formula:ve-general-sub2}
\end{gather}

\ignore{
\begin{minipage}{0.58\textwidth}
	\begin{gather}
		\forall (\quanti{x}, \quanti{x'}), \bluec{\m{out}}~\var{g}~\quanti{x} \neq \bluec{\m{out}}~\var{g}~\quanti{x'} \rightarrow \bluec{\m{in}}~\var{g}~\quanti{x} \neq \bluec{\m{in}}~\var{g}~\quanti{x'}\label{formula:ve-general-sub1}
	\end{gather}
\end{minipage}
\hfill
\begin{minipage}{0.37\textwidth}
	\begin{gather}
		\forall \quanti{x}, \bluec{\m{out}}~\bluec{g}~\quanti{x} = \var{f}~(\bluec{\m{in}}~\bluec{g}~\quanti{x}) \label{formula:ve-general-sub2}
	\end{gather}
\end{minipage}
\hfill
\smallskip
}

Note that the above decomposition is approximate. The resulting $g$ synthesized from the first subtask only ensures that a function exists for $f$ to implement, but there is no guarantee that this function can be implemented within the program space of $f$.

\smallskip 

\noindent \textbf{Usage in \mainname}. The first task generated by component elimination (Formula \ref{formula:ce-sub1}, repeated below) has the same form as the general form (Formula \ref{formula:ve-general-full}, repeated below). We list the correspondence in the table below.
$$
\begin{array}{cc}
	\text{General form}: & \quanti{x} = \var{f}~(\bluec{\m{in}}~\var{g}~\quanti{y})   \\
\text{Current task}: &  \bluec{\m{orig}}~(\bluec{\m{op}}~(\quanti{c}, \quanti{\overline{a}})) = \var{\m{comb}_1}~\big(\quanti{c}, (\bluec{\m{orig}}^n~\quanti{\overline{a}}, \var{\m{aux}_1}^n~\quanti{\overline{a}})\big)
\end{array}
$$
\begin{center}
\begin{tabular}{ll}
	\hline
	General Form & Current Task\\
	\hline
	$\var{f}$ & $\var{\m{comb}_1}$ \\
	$\var{g}$ & $\var{\m{aux}_1}$\\
	$\quanti{x}$ & $\bluec{\m{orig}}~(\bluec{\m{op}}~(\quanti{c}, \quanti{\overline{a}}))$\\
	$\quanti{y}$ & $(\quanti{c}, \quanti{\overline{a}})$ \\ 
	$\bluec{\m{in}}$ & $\lambda \var{\m{aux}_1}.~\lambda (\quanti{c}, \quanti{\overline{a}}).~\big(\quanti{c}, (\bluec{\m{orig}}^n~\quanti{\overline{a}}, \var{\m{aux}_1}^n~\quanti{\overline{a}})\big)$ \\
	\hline
\end{tabular}
\end{center}
\smallskip

Following the decomposition procedure of variable elimination, we can further decompose the first task of component elimination into the following two subtasks.

\begin{enumerate}
	\item The first subtask targets synthesizing $\m{aux}_1$ from the specification below.
	$$
	\bluec{\m{orig}}\ (\bluec{\m{op}}\ (\quanti{c}, \quanti{\overline{a}})) \neq \bluec{\m{orig}}\ (\bluec{\m{op}}\ (\quanti{c'}, \quanti{\overline{a'}})) \rightarrow \big(\quanti{c}, (\bluec{\m{orig}}^n~\quanti{\overline{a}}, \var{\m{aux}_1}^n\ \quanti{\overline{a}})\big) \neq  \big(\quanti{c'}, (\bluec{\m{orig}}^n~\quanti{\overline{a'}}, \var{\m{aux}_1}^n\ \quanti{\overline{a'}})\big)
	$$

	We transform this specification into the following equivalent form to clarify the constraint on $\m{aux}_1$. Following the notation in Section \ref{sec:simpleOverview}, we include leaf subtasks of decomposition within framed boxes to distinguish them from the other intermediate subtasks.
\end{enumerate}

	\noindent\fbox{\parbox{\textwidth}{
		\begin{equation}
			\begin{aligned}
			\big(\bluec{\m{orig}}^n\ \quanti{\overline{a}} = \bluec{\m{orig}}^n\ \quanti{\overline{a'}} \wedge \bluec{\m{orig}}\ (\bluec{\m{op}}\ (\quanti{c}, \quanti{\overline{a}})) \neq \bluec{\m{orig}}\ (\bluec{\m{op}}\ (\quanti{c}, \quanti{\overline{a'}}))\big)\rightarrow \var{\m{aux}_1}^n\ \quanti{\overline{a'}} \neq \var{\m{aux}_1}^n\ \quanti{\overline{a'}}
			\end{aligned}
			\label{formula:spec-aux}
		  \end{equation}
		}}
\begin{enumerate}
	\item[(2)] The second subtask targets synthesizing $\m{comb}_2$ from the specification below, where $\m{aux}_1$ here is the program obtained by solving the first subtask.
\end{enumerate}
\noindent\fbox{\parbox{\textwidth}{
\begin{equation}
	\bluec{\m{orig}}\ (\bluec{\m{op}}\ (\quanti{c}, \quanti{\overline{a}})) = \var{\m{comb}_1}\ \big(\quanti{c}, (\bluec{\m{orig}}^n~\quanti{\overline{a}}, \bluec{\m{aux}_1}^n\ \quanti{\overline{a}})\big) \label{formula:spec-comb}
	\end{equation}}}
}

\smallskip 

\noindent 
  
  \begin{example}  \label{example:decoupling} After applying variable elimination to the first subtask generated in the previous example (Formula \ref{formula:mss-ce-sub1}), the specification of the first subtask is shown below. 
\jrydel{
  \begin{align*}
	\bluec{\m{mss}}^2\ (\quanti{xs_L}, \quanti{xs_R}) = \bluec{\m{mss}}^2\ (\quanti{xs_L'}, \quanti{xs_R'}) \wedge \bluec{\m{mss}}\ (\quanti{xs_L} \cat \quanti{xs_R}) \neq \bluec{\m{mss}}\ (\quanti{xs_L'} \cat \quanti{xs_R'})& \\ \rightarrow \var{\m{aux}_1}^2\ (\quanti{xs_L}, \quanti{xs_R}) \neq&~\var{\m{aux}_1}^2\ (\quanti{xs_L'}, \quanti{xs_R'})
  \end{align*}
}
\jryadd{
\begin{align*}
  (\bluec{\m{mss}}\ \quanti{xs_L}, \bluec{\m{mss}}\ \quanti{xs_R}) = (\bluec{\m{mss}}\ \quanti{xs_L'}, \bluec{\m{mss}}\ \quanti{xs_R'}) \wedge \bluec{\m{mss}}\ (\quanti{xs_L} \cat \quanti{xs_R}) \neq \bluec{\m{mss}}\ (\quanti{xs_L'} \cat \quanti{xs_R'})& \\ \rightarrow (\var{\m{aux}_1}\ \quanti{xs_L}, \var{\m{aux}_1}\ \quanti{xs_R}) \neq(\var{\m{aux}_1}\ \quanti{xs_L'}, \var{\m{aux}_1}\ \quanti{xs_R'})&
\end{align*}

One can verify that components $\acomp{mps}$ and $\acomp{mts}$ form a valid solution to this subtask.
}
  \end{example}

\SetKwFunction{SGLP}{ComponentElimination}
\SetKwFunction{SPLP}{VariableElimination}
\SetKwFunction{IAUX}{InductiveSynthesisForAux}
\SetKwFunction{ICOMB}{InductiveSynthesisForComb}
\ignore{\begin{algorithm}[t]
	\small
	\caption{\jrymod{The decomposition system of \mainname.}{A greedy implementation of the decomposition system.}}
	\label{alg:deductive}
	\LinesNumbered
	\KwIn{A lifting problem $\mathsf{LP}(\targetname, \m{op})$.}
	\KwOut{A solution $(\m{aux}, \m{comb})$ to lifting problem $\mathsf {LP}(\targetname, \m{op})$.}
	\SetKwProg{Fn}{Function}{:}{}
	\Fn{\SPLP{$\targetname, \m{op}, \m{aval}$}}{
		$\m{subtask}_1 \gets$ the first subtask generated by variable elimination (Formula \ref{formula:spec-aux}); \\
		$\m{aux}_1 \gets \IAUX{$\m{subtask}_1$}$; \\ 
		$\m{subtask}_2 \gets$ the second subtask corresponding to $\m{aux}_1$ (Formula \ref{formula:spec-comb});\\
		\Return $(\m{aux}_1, \ICOMB{$\m{subtask}_2$})$; 
	}
	\Fn{\SGLP{$\targetname, \m{op}, \m{aval}$}}{
		$(\m{aux}_1, \m{comb}_1) \gets \SPLP{$\targetname, \m{op}, \m{aval}$}$; \\
		\lIf{$aux_1 = \textit{null}$}{\Return $(\textit{null}, \m{comb}_1 \spl \textit{null})$}
		$(\m{aux}_2, \m{comb}_2) \gets \SGLP{$\m{aux}_1, \m{op}, \m{aval} \spl \m{aux}_1$}$; \\
		\Return $(\m{aux}, \m{comb})$ constructed from $(\m{aux}_1, \m{comb}_1)$ and $(\m{aux}_2, \m{comb}_2)$; 
	}
	\Return \SGLP($\targetname, \m{op}, \targetname$); 
\end{algorithm}}

\begin{algorithm}[t]
	\small
	\caption{\jrymod{The decomposition system of \mainname.}{A greedy implementation of the decomposition system.}}
	\label{alg:deductive}
	\LinesNumbered
	\KwIn{A lifting problem $\mathsf{LP}(\targetname, \m{op})$.}
	\KwOut{A solution $(\m{aux}, \m{comb})$ to lifting problem $\mathsf {LP}(\targetname, \m{op})$.}
	\SetKwProg{Fn}{Function}{:}{}
	\Fn{\SPLP{a subtask in the form of Formula \ref{formula:ce-sub1}}}{
		$\m{subtask}_1 \gets$ the first subtask generated by variable elimination (Formula \ref{formula:spec-aux}); \\
		$\m{aux}_1 \gets \IAUX{$\m{subtask}_1$}$; \\ 
		$\m{subtask}_2 \gets$ the second subtask corresponding to $\m{aux}_1$ (Formula \ref{formula:spec-comb});\\
		\Return $(\m{aux}_1, \ICOMB{$\m{subtask}_2$})$; 
	}
	\Fn{\SGLP{\text{a lifting problem in the form of Formula \ref{formula:lifting-definition-force}}}}{
		$\m{subtask}_1 \gets$ the first subtask generated by component elimination (Formula \ref{formula:ce-sub1}); \\
		$(\m{aux}_1, \m{comb}_1) \gets \SPLP{$\m{subtask}_1$}$; \\
		\lIf{$aux_1 = \textit{null}$}{\Return $(\textit{null}, \m{comb}_1 \spl \textit{null})$}
		$\m{subtask}_2 \gets$ the second subtask corresponding to $(\m{aux}_1, \m{comb}_1)$ (Formula \ref{formula:ce-sub2});\\
		$(\m{aux}_2, \m{comb}_2) \gets \SGLP{$\m{subtask}_2$}$; \\
		\Return $(\m{aux}, \m{comb})$ constructed from $(\m{aux}_1, \m{comb}_1)$ and $(\m{aux}_2, \m{comb}_2)$; 
	}
	\Return \SGLP(the original lifting problem); 
\end{algorithm}

\subsubsection{Decomposition System}
\jrydel{The decomposition system of \mainname (Algorithm \ref{alg:deductive}) decomposes a lifting problem by repeatedly applying component elimination and variable elimination.}

\jryadd{

As mentioned before, two strategies exist for applying the decomposition methods, greedy and backtracking. Here we focus on the greedy strategy, the one used in our implementation. Algorithm \ref{alg:deductive} shows the pseudocode of a decomposition system using the greedy strategy.
}
\begin{enumerate}
    \item Algorithm \ref{alg:deductive} starts the decomposition by applying component elimination to the original lifting problem (Line 13). The first subtask is further decomposed by variable elimination (Lines 7-8), and the second one is solved by recursively applying component elimination (Lines 10-11). The recursion terminates when no new auxiliary value is found (Line 9).
    \item Algorithm \ref{alg:deductive} applies variable elimination to the first subtask generated by component elimination (Lines 1-5) and solves the two subtasks via inductive synthesis (Lines 3 and 5), which shall be discussed later (Section \ref{subsection:inductive}).
\end{enumerate}

\begin{figure}
   \includegraphics[width=0.65\linewidth]{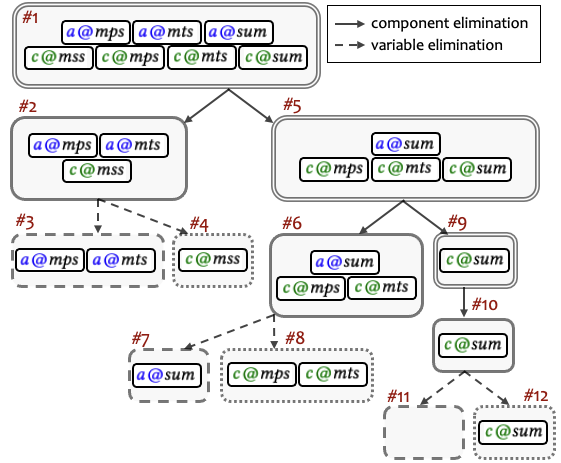}
 \captionof{figure}{The decomposition performed by the decomposition system to synthesize programs in Figure \ref{fig:mss-dac}.}
\label{fig:mss-synthesis-2}
\end{figure}
\begin{example}

Figure \ref{fig:mss-synthesis-2} illustrates how Algorithm \ref{alg:deductive} decomposes the \m{mss} task, where nodes are subtasks (\#1 denotes the original problem), arrows indicate task decomposition, tags within nodes indicate the sub-program synthesized from each subtask, and different line styles indicate different decomposition/subtask types. As we can see, the scale of the leaf subtasks (\#3, \#4, \#7, \#8, \#11, and \#12, each including at most $2$ components) is greatly reduced compared to the original lifting problem (\#1, including $7$ components). There are four types of subtasks: (1) lifting problems, including the original lifting problem (\#1) and the second subtasks of component elimination (\#5, \#9), (2) the first subtasks of component elimination (\#2, \#6, \#10), (3) the first subtasks of variable elimination (\#3, \#7, \#11), which involve only sub-programs $\m{aux}$, and (4) the second subtasks of variable elimination (\#4, \#8, \#12), which involve only sub-programs of $\m{comb}$. 

\ignore{\begin{figure*}
	  \begin{minipage}{.51\textwidth}
		\begin{table}[H]
			\caption{The parameters of generalized lifting problems in Figure \ref{fig:mss-synthesis-2}.} \label{table:mss-glp}
			\begin{tabular}{|c|c|c|}
				\Xhline{1pt}
				Subtask & $\targetname$  &  $\m{aval}$ \\
				\Xhline{1pt}
				\#1 & \m{sndmin} &  \m{sndmin}\\
				\hline
				\#5 & $\m{mps} \spl \m{mts}$  & $\m{sndmin} \spl (\m{mps} \spl \m{mts})$ \\
				\hline
				\multirow{2}{*}{\#9} & \multirow{2}{*}{$\m{sum}$} & $(\m{sndmin} \spl (\m{mps} \spl \m{mts}))$\\
				& & $\spl \m{sum}$ \\
				\Xhline{1pt}
			\end{tabular}
		\end{table}
	  \end{minipage}
	\hfill
	\begin{minipage}{0.44\textwidth}
		\begin{table}[H]
			\begin{spacing}{1}
			\caption{Subtasks in Figure \ref{fig:mss-synthesis-2} solved by each function in Algorithm \ref{alg:deductive}.} \label{table:solved}
					\begin{tabular}{|c|c|}
						\Xhline{1pt}
						Function & Subtasks   \\
						\Xhline{1pt}
						{\small \SGLP} & $\#1, \#5, \#9$ \\
						\hline
						{\small \SPLP} & $\#2, \#6, \#10$ \\
						\hline
						{\small \IAUX} & $\#3, \#7, \#11$ \\
						\hline
						{\small \ICOMB} & $\#4, \#8, \#12$ \\
						\Xhline{1pt}
					\end{tabular}
				\end{spacing}
		\end{table}
	  \end{minipage}
	  \end{figure*}}

Besides, the following points are worth noting about Figure \ref{fig:mss-synthesis-2}.
\begin{itemize}
  \item The tags in each node represent the synthesis result of the subtask. They are unavailable when the subtask is generated or further decomposed. 
  \item For both decomposition methods, the second subtask relies on the synthesis result of the first one so that it will not be generated until the first subtask is solved. The indices of nodes in Figure \ref{fig:mss-synthesis-2} reflect the generation order of subtasks.
  \item The decomposition terminates after solving subtask \#10 (a first subtask of component elimination) because the synthesis result shows that no new auxiliary value is required.
\end{itemize}
\end{example}

\subsection{Inductive Synthesis for Leaf Tasks} \label{subsection:inductive}
The decomposition system generates two types of leaf tasks, corresponding to the two subtasks of variable elimination (Formulas \ref{formula:spec-aux} and \ref{formula:spec-comb}), respectively. We apply the CEGIS framework~\cite{DBLP:conf/asplos/Solar-LezamaTBSS06} to convert both types of tasks into example-based synthesis tasks. 

CEGIS (Algorithm \ref{alg:cegis}) synthesizes by iteratively invoking an example-based synthesizer and a verifier. It records an example set that is initially empty (Line 1). In each iteration, the example-based synthesizer generates a candidate program $\m{prog}$ from existing examples (Line 3) and the verifier generates a counter-example $e$ under which $\m{prog}$ is incorrect, i.e., $\neg \phi(\m{prog}, e)$ is satisfied (Line 4). The candidate program will be returned if it is verified to be correct (i.e., no counter-example exists) (Line 5). Otherwise, the counter-example will be recorded for further synthesis (Line 6).

In this paper, we assume the existence of the verifier for both types of leaf tasks and focus on the example-based synthesis tasks. In practice, the verifier can be selected among off-the-shelf ones on demand. \jrymod{For example, bounded model checking~\cite{DBLP:journals/ac/BiereCCSZ03} can be used when all related programs are symbolically executable, and in our implementation, a probabilistic verifier based on random testing is used by default (Section \ref{section:implementation}).}{In our implementation, we use a combination of bounded model checking~\cite{DBLP:journals/ac/BiereCCSZ03} and random testing by default.}

\begin{figure*}
\begin{minipage}{.45\textwidth}
\SetKwFunction{SY}{Synthesis}
\SetKwFunction{CE}{CounterExample}
\vspace{-0.1em}
\begin{algorithm}[H]
	\small
	\caption{CEGIS framework}
	\label{alg:cegis}
	\LinesNumbered
	\KwIn{
	A specification $\Phi = \forall \quanti{\overline{x}}, \phi(\var{\m{prog}}, \quanti{\overline{x}})$.}
	\KwOut{A valid program.}
	\SetKwProg{Fn}{Function}{:}{}
	$\variable{examples} \gets \emptyset$;\\
	\While{true}{
		$\m{prog} \gets \SY{$\forall \quanti{\overline{x}} \in \textit{examples}, \phi(\var{\m{prog}}, \quanti{\overline{x}})$}$; \\
		$e \gets \CE{$\m{prog}, \Phi$}$; \\
		\lIf{$e = \bot$}{\Return $\m{prog}$}
		$\variable{examples} \gets \variable{examples} \cup \{e\}$;
	}
\end{algorithm} 
\end{minipage}
\hfill
\begin{minipage}{.5\textwidth}
\SetKwFunction{Poly}{SynthesisFromIOExamples}
\begin{algorithm}[H]
	\small
	\caption{Example-based solver of \m{comb}.}
	\label{alg:synthesis-comp}
	\LinesNumbered
	\KwIn{An example set $\m{examples}$ and programs $(\m{orig}, \m{op}, \m{aux}_1)$ specifying the task.}
	\KwOut{A valid combinator $\m{comb}_1^*$.}
	\SetKwProg{Fn}{Function}{:}{}
	$\variable{ioexamples} \gets \emptyset$;\\
	\ForEach{$(c, \overline{a}) \in \m{examples}$}{
		$\m{input} \gets (c, (\m{orig} \spl \m{aux}_1)^n\ \overline{a})$; \\
		$\m{output} \gets \m{orig}\ (\m{op}\ (c, \overline{a}))$; \\
		$\m{ioexamples} \gets \m{ioexamples} \cup \{(\m{input}, \m{output})\}$; 
	}
	\Return \Poly{$\m{ioexamples}$};
\end{algorithm}
\end{minipage}
\end{figure*}

\smallskip

\noindent \textbf{Example-based synthesizer for $\m{comb}$}. We begin with the leaf subtask of $\m{comb}$ (Formula \ref{formula:spec-comb}), the simpler case. An example of this task is an assignment to $(c, \overline{a})$ and is in the input-output form, requiring $\m{comb}_1$ to output $\m{orig}\ (\m{op}\ (c, \overline{a}))$ from input $(c, (\m{orig} \spl \m{aux}_1)^n\ \overline{a})$. This task can be solved by those existing synthesis algorithms relying on input-output examples. As shown in Algorithm \ref{alg:synthesis-comp}, the example-based synthesizer for $\m{comb}$ just converts the given examples into the input-output form (Lines 2-6) and then passes them to existing synthesizers.

\smallskip 

\noindent \textbf{Example-based synthesizer of $\m{aux}$}. For the leaf subtask of $\m{aux}$ (Formula \ref{formula:spec-aux}), an example is an assignment to $(c, \overline{a}, \overline{a'})$, with a constraint of $\var{\m{aux}_1}^n\ \overline{a} \neq \var{\m{aux}_1}^n\ \overline{a'}$. Note that the premise of this specification can be ignored in the example-based task because any example generated by CEGIS must be a counter-example of some candidate program, which is only possible when the premise of the specification is true. 

\begin{example}\label{example:aux-example} Table \ref{table:aux-example} shows two examples possibly generated from the subtask in Example \ref{example:decoupling}. The maximum prefix sum $\m{mps}$ satisfies example $\overline{\m{xs}}_1$ because $\m{mps}$ generates two different outputs $(1, 1)$ and $(1,0)$ on $([1], [1])$ and $([1], [-1, 1])$. Similarly, the maximum tail-segment sum $\m{mts}$ satisfies example $\overline{\m{xs}}_2$, and their pair $\m{mts} \spl \m{mps}$ satisfies both examples.
\begin{table*}
	\renewcommand\arraystretch{1.15}
	\caption{Two possible examples generated from the subtask in Example \ref{example:decoupling}.} \label{table:aux-example}
	\begin{spacing}{1}
		\small
		\begin{tabular}{|c|c|c|c|c|c|c|c|}
			\Xhline{1pt}
			Id  & $(\m{xs}_L, \m{xs}_R)$ & $(\m{xs}_L', \m{xs}_R')$ & premise & requirement \\
			\Xhline{1pt} 
			$\overline{\m{xs}}_1$ & $([1], [1])$ & $([1], [-1, 1])$ & $(1, 1) = (1, 1) \wedge 2 \neq 1$& $(\var{\m{aux}_1}\ [1], \var{\m{aux}_1}\ [1]) \neq (\var{\m{aux}_1}\ [1], \var{\m{aux}_1}\ [-1, 1])$\\
			\hline 
			$\overline{\m{xs}}_2$ & $([1], [1])$ & $([1, -1], [1])$ & $(1, 1) = (1, 1) \wedge 2 \neq 1$& $(\var{\m{aux}_1}\ [1], \var{\m{aux}_1}\ [1]) \neq (\var{\m{aux}_1}\ [1, -1], \var{\m{aux}_1}\ [1])$\\
			\Xhline{1pt}
		\end{tabular}
	\end{spacing}
\end{table*}
\end{example}

\begin{figure*} 

	\SetKwFunction{NC}{Extend}
	\SetKwFunction{IS}{Insert}
	\SetKwFunction{APP}{Append}
	\SetKwFunction{NE}{Next}
	\SetKwFunction{CHECK}{IsCovered}
	\SetKwFunction{OE}{OE}
	\SetKwFunction{OB}{ObservationalEquivalenceSolver}
	\begin{minipage}{0.75\textwidth}
	\begin{algorithm}[H]
		\small
		\caption{Example-based solver of \m{aux}.}
		\label{alg:synthesis-aux}
		\LinesNumbered
		\KwIn{An example set $\m{examples}$ and an integer $\m{lim}_c$ specifying the number of components considered by observational covering.}
		\KwOut{A valid auxiliary program $\m{aux}_1$.}
		\SetKwProg{Fn}{Function}{:}{}
		$\m{involvedInputs} \gets \{a\ |\ (c, \overline{a}, \overline{a'}) \in \m{examples} \wedge (a \in \overline{a} \vee a \in \overline{a'})\}$; \\
		$\m{oe} \gets \OB{$\m{involvedInputs}$}$; \\
		$\forall \m{size} \geq 0, \m{programs}[\m{size}] \gets []$; \ \ $\m{result} \gets \bot$; \\ 
		\Fn{\CHECK{$\m{prog}, \m{size}$}}{
			\Return $\exists \m{size'} \leq \m{size}, \exists \m{prog}' \in \m{programs}[\m{size}'], \m{prog}'$ satisfies all examples that are satisfied by $\m{prog}$;
		}
		\Fn{\IS{$\m{prog}, \m{size}$}}{
			\lIf{$\m{prog}$ satisfies all examples $\wedge \m{result} = \bot$}{$\m{result} \gets \m{prog}$}
			\lIf{$\neg \CHECK{$\m{prog}, \m{size}$}$}{$\m{programs}[\m{size}].\APP{$\m{prog}$}$}
		}
		\Fn{\NC{}}{ 
			$\m{component} \gets \m{oe}.\NE{}$;\ \ $\IS{$\m{component}, 1$}$\\
			$\m{prePrograms} \gets \m{programs};$\\
			\ForEach{$\m{size} \in [1, \dots, \m{lim}_c - 1]$ and $\m{prog} \in \m{prePrograms}[\m{size}]$}{
				\IS{$\m{prog} \spl \m{component}, \m{size} + 1$};
			}
		}
		\IS{$\m{null}, 0$}; \\
		\lWhile{$\m{result} = \bot$}{\NC{}}
		\Return $\m{result}$;
	\end{algorithm}
	\end{minipage}
\end{figure*}

The example-based synthesizer for $\m{aux}$ (Algorithm \ref{alg:synthesis-aux}) is built upon an existing enumerative synthesizer, namely \textit{observational equivalence (OE)}~\cite{DBLP:conf/pldi/UdupaRDMMA13} (Line 10). OE enumerates programs from small to large following a bottom-up manner. It constructs larger programs by combining those existing smaller programs via language constructs. OE uses an effective pruning strategy to avoid duplicated programs with the same input-output behaviors. This strategy is parameterized by an input set and will prune off those programs producing duplicated outputs on this input set (compared to existing programs).

\begin{example} Consider a synthesis task with a single input $x$. When the input set is $\{1\}$, OE will skip program $x \times 2$ if $x + 1$ has been visited before because both programs output $2$ from the input. Furthermore, those programs constructed from $x \times 2$, such as $(x \times 2) + 1$ and $(x \times 2) \times x$, will be implicitly skipped as well because $x \times 2$ will no longer be used to construct larger programs.
\end{example}

In the example-based synthesis task of $\m{aux}$, whether a program satisfies an example $(c, \overline{a}, \overline{a'})$ is determined by its outputs on those components inside $\overline{a}$ and $\overline{a'}$. Therefore, OE can be applied by including all inputs involved in examples into the input set (Lines 1-2, Algorithm \ref{alg:synthesis-aux}).

\begin{example} When the example set is $\{\overline{xs}_1, \overline{xs}_2\}$ (examples in Example \ref{example:aux-example}), OE can be invoked with input set $\{[1], [-1, 1], [1, -1]\}$. Any two programs outputting the same from these inputs must satisfy the same subset of examples.
\end{example}

\noindent \textbf{Optimization: observational covering}. The example-based synthesizer for $\m{aux}$ also includes a specialized optimization for better handling the case where multiple auxiliary values are required. Specifically, we observe that many practical tasks require multiple auxiliary values, for example, \m{mps} and \m{mts} are both required for calculating \m{mss} in D\&C. In these cases, the form of $\m{aux}_1$ can be assumed as a tuple of components (i.e., $\m{comp}_1\! \spl\! \dots\! \spl\! \m{comp}_k$), each in a smaller scale. To better synthesize such programs, we optimize the enumeration by invoking OE to generate only basic components and combining these components explicitly on the top level (Lines 6-14).

To implement the optimized enumeration, our example-based synthesizer for $\m{aux}$ maintains a program storage $\m{programs}$ during the enumeration, where $\m{programs}[\m{size}]$ stores existing programs formed by \m{size} components (Line 3). In each iteration, our synthesizer invokes OE to generate the next component (Line 10) and then combines it with existing programs to form larger tuples (Lines 11-14). To limit the combination space, our synthesizer is configured by an integer $\m{lim}_c$ and considers only combining at most $\m{lim}_c$ components at the top level (Line 12). Note that such a limitation would not affect the effectiveness: our synthesizer is still complete (i.e., never fails on a realizable task) even when $\m{lim}_c$ is set to $1$. A valid program of a realizable task will ultimately be found as a single component because OE directly enumerates the whole program space. 

\smallskip

To further speed up the combination, we propose an optimization method named \emph{observational covering}. This method follows the key idea of \m{OE}, that is, to prune off programs whose effect is covered by some other programs on a set of given examples. Recall that an example $(c, \overline{a}, \overline{a'})$ here requires the auxiliary program to return different results on $\overline{a}$ and $\overline{a'}$, which means, an example is satisfied by an auxiliary program when and only when it is satisfied by some components in the auxiliary program. Therefore, when the goal is to satisfy a set of given examples using at most $\m{lim}_c$ components, the effect of a program $\m{prog}$ is covered by another program $\m{prog}'$ if (1) $\m{prog}'$ uses fewer components than $\m{prog}$ and (2) $\m{prog}'$ satisfies all examples satisfied by $\m{prog}$. At this time, the covered program $\m{prog}$ can be safely skipped from the top-level combination.



\begin{example} When the example set is $\{\overline{\m{xs}}_1, \overline{\m{xs}}_2\}$ (Example \ref{example:aux-example}), the effect of $\m{max}$ is covered by $\m{null}$ (the empty auxiliary program) because both programs satisfy no example and $\m{max}$ uses one more component. Therefore, $\m{max}$ can be safely skipped from the combination: whenever there is a program combined from $\m{max}$ (assumed as $\m{prog} \spl \m{max}$) satisfying both examples, there must exist another valid program $\m{prog} \spl \m{null}$ (i.e., \m{prog}) using fewer components.
\end{example}

Using this property, our synthesizer skips all covered programs. Only programs that are not covered by existing ones will be inserted into the storage for further combination (Lines 4-5 and 8). 

\begin{example} Algorithm \ref{alg:synthesis-aux} runs as below when the example set is $\{\overline{\m{xs}}_1, \overline{\m{xs}}_2\}$ (Example \ref{example:aux-example}), the limit $\m{lim}_c$ is $2$, and the first three components returned by OE are $\m{max}, \m{mps}$, and $\m{mts}$ in order.
\begin{itemize}
	\item Before the first invocation of \NC, $\m{null}$ is inserted and the storage is $\m{programs}[0] = [\m{null}]$. 
	\item In the first invocation, OE generates $\m{max}$ and no other program is constructed. $\m{max}$ will not be inserted into the storage as it is covered by $\m{null}$. So the storage will remain unchanged.
	\item In the second invocation, OE generates $\m{mps}$ and no other program is constructed. $\m{mps}$ is not covered by $\m{null}$ as it satisfies $\overline{\m{xs}}_1$, an example violated by $\m{null}$. So $\m{mps}$ will be inserted, and the storage will become $\m{programs}[0] = [\m{null}], \m{programs}[1] = [\m{mps}]$. 
	\item In the third invocation, OE generates $\m{mts}$ and program $\m{mps}\! \spl\! \m{mts}$ is generated by combination. Both programs are not covered and will be inserted, and the storage will become as follows.
	$$\m{programs}[0] = [\m{null}] \quad \m{programs}[1] = [\m{mps}, \m{mts}] \quad \m{programs}[2] = [\m{mps}\! \spl\! \m{mts}]$$ 
	
	Then $\m{mps} \spl \m{mts}$ will be returned as the result as it already satisfies all given examples. 
\end{itemize}
\end{example}

\section{Properties} \label{subsection:properties}

\subsection{Soundness \label{subsection:soundness}} \mainname is sound when the verifiers of leaf subtasks are sound (Theorem \ref{theorem:soundness}). Specifically, when these verifiers are sound, the synthesis results of the leaf subtasks must satisfy their respective specifications. Recall that in every decomposition made by \mainname, the second subtask is always for completing the synthesis result of the first subtask into a valid solution for the original task. Therefore, the final results built up from correct results of leaf subtasks must also be correct, implying the soundness of \mainname. 

\begin{theorem}[Soundness] \label{theorem:soundness} The result of \mainname (Algorithm \ref{alg:deductive}) is valid for the original lifting problem if the verifiers of leaf subtasks accept only valid programs for respective subtasks.
\end{theorem}
\begin{proof}
	Proofs of the lemmas and theorems in this paper are available in Appendix \ref{appendix:proofs}.
	\end{proof}

\jrydel{
\noindent \textbf{Incompleteness}. In theory, \mainname is incomplete mainly because its decomposition system is incomplete. This system may generate unrealizable subtasks from a realizable lifting problem. In detail, for both decomposition methods, the specification derived for the first subtask is not precise and cannot ensure the realizability of the subsequent second subtask.
\begin{itemize}
	\item The first subtask of component elimination (Formula \ref{formula:ce-sub1}) only requires calculating the expected output (defined by $\m{orig}$). However, an $\m{aux}_1$ valid for this subtask may introduce auxiliary values that cannot be calculated using auxiliary programs in $\mathcal L_{\textit{aux}}$ and combinators in $\mathcal L_{\textit{comb}}$, making the second subtask (Formula \ref{formula:ce-sub2}) unrealizable.
	\item The first subtask of variable elimination (Formula \ref{formula:spec-aux}) only requires $\m{aux}_1$ to provide enough auxiliary values such that a combination function exists for calculating the expected output. However, the expressiveness of $\mathcal L_{\textit{comb}}$ may not be enough to implement such a function, making the second subtask (Formula \ref{formula:spec-comb}) unrealizable.
\end{itemize}
}

\jryadd{
\subsection{Cost of Approximation}\label{subsection:incompleteness} Recall that \mainname uses approximate specifications when decomposing lifting problems. Such a treatment, in theory, will have negative effects on performance. The form of such effects depends on the specific strategy used to implement the decomposition system.
\begin{itemize}
	\item For the greedy strategy, every unrealizable subtask will make \mainname fail in solving lifting problems, affecting the effectiveness.
	\item For the backtracking strategy, every unrealizable subtask will make \mainname roll back and search for other solutions, affecting the efficiency.
\end{itemize}
} 

\jrydel{Fortunately, such theoretical incompleteness of \mainname seldom exposes in practice. In our evaluation, \mainname can solve almost all realizable tasks within a short timeframe (Section \ref{section:evaluation}). The direct reason for this phenomenon is the excellent practical performance of $\mathcal S_{\textit{aux}}$, the synthesizer for the auxiliary program (Algorithm \ref{alg:synthesis-aux}). Specifically, the decomposition procedure of \mainname is fully determined by the result of $\mathcal S_{\textit{aux}}$ because each subtask depends only on the original lifting problem and the auxiliary programs synthesized previously. In our evaluation, $\mathcal S_{\textit{aux}}$ can always find the intended auxiliary program, making all subtasks generated by decomposition realizable.}

\jryadd{Fortunately, we have empirical results showing that the negative effects of our approximations are negligible. In our evaluation, \mainname never decomposes realizable tasks into unrealizable and can solve almost all realizable tasks within a short time (Section \ref{subsection:rq1}). 

The direct reason for this phenomenon is the excellent practical performance of our example-based synthesizer for $\m{aux}$ (Algorithm \ref{alg:synthesis-aux}). Specifically, each subtask generated by \mainname depends only on the original lifting problem and the sub-programs of $\m{aux}$ synthesized previously, making the decomposition procedure of \mainname fully determined by the result of the $\m{aux}$ synthesizer. Our $\m{aux}$ synthesizer works well on this aspect. In our evaluation, it can always find the intended auxiliary program and thus avoid \mainname from generating unrealizable subtasks.}

As discussed in the $\m{sndmin}$ example (Section \ref{subsection:moti-overview}), we ascribe the effectiveness of our \m{aux} synthesizer to two reasons: (1) the \textit{compressing property} of a practical lifting problem (i.e., both $\m{orig}$ and programs in $\mathcal L_{\textit{aux}}$ map a large input space to a small output space) makes the specification of $\m{aux}$ (Formula \ref{formula:spec-aux}) strong enough to exclude most candidates, and (2) the \textit{preference to simpler auxiliary programs} helps avoid those unnecessarily complex solutions. In the remainder of this section, we shall provide formal results corresponding to these two factors. \jryadd{
\begin{itemize}
	\item First, in Section~\ref{subsection:probUnrealizableSubtasks}, we analyze the probability for \mainname to generate unrealizable subtasks under a probabilistic model, where the semantics of $\mathcal L_{\textit{aux}}$ and $\mathcal L_{\textit{comb}}$ are modeled as random.  
	Then in Section~\ref{subsection:effecCompressing}, we prove that this probability is almost surely small under the compressing property (Theorem \ref{theorem:completeness}). 
	\item Second, in Section~\ref{subsection:preferenceForSimpler}, we consider the concrete semantics of $\mathcal L_{\textit{aux}}$ and $\mathcal L_{\textit{comb}}$ and prove that our \m{aux} synthesizer (Algorithm \ref{alg:synthesis-aux}) can always find a minimal solution (Theorem \ref{theorem:minimal}).
\end{itemize}}

\jryadd{
\smallskip 
\subsubsection{Probability to generate unrealizable subtasks \label{subsection:probUnrealizableSubtasks}} To study the effectiveness of \mainname, we aim to analyze the probability for \mainname to generate unrealizable subtasks. However, calculating this probability precisely is extremely difficult because the languages of candidate programs ($\mathcal L_{\textit{aux}}$ and $\mathcal L_{\textit{comb}}$) are usually complex. Both of these languages may include infinitely many candidate programs, each assigned with possibly complex semantics. Consequently, it is almost impossible to precisely predict the performance of \mainname in synthesis.

We overcome this challenge by introducing a probabilistic model and conducting an approximate analysis instead. Specifically, we assume the semantics of programs in $\mathcal L_{\textit{aux}}$ and $\mathcal L_{\textit{comb}}$ as random functions and then analyze the probability for \mainname to generate unrealizable subtasks.

\smallskip

The following are the details of our probabilistic analysis.

Given lifting problem $\mathsf{LP}(\m{orig}, \m{op})$, we construct a corresponding probabilistic model $\mathcal M[\m{orig}, \m{op}]$ by modeling the semantics of programs in $\mathcal L_{\textit{aux}}$ and $\mathcal L_{\textit{comb}}$ as uniformly random functions. The detailed construction of $\mathcal M[\m{orig}, \m{op}]$ is shown below. 
\begin{itemize}
	\item $\mathcal M[\m{orig}, \m{op}]$ is constructed on a set of given parameters, including (1) the types and semantics of $\m{orig}$ and $\m{op}$, and (2) the syntax and the type of every program in $\mathcal L_{\textit{aux}}$ and $\mathcal L_{\textit{comb}}$. The only thing this model does is to assign random semantics to programs in $\mathcal L_{\textit{aux}}$ and $\mathcal L_{\textit{comb}}$. 
	\item For simplicity, we make the following assumptions when constructing $\mathcal M[\m{orig}, \m{op}]$.
	\begin{enumerate}[leftmargin=1.5em]
	\item Programs in $\mathcal L_{\textit{aux}}$ are formed as tuples of components (i.e., $\m{comp}_1 \spl \dots \spl \m{comp}_k$), where each component $\m{comp}_i$ outputs only a single auxiliary value.
	\item There is a universal value type $V$ capturing the types of the output of $\m{orig}$, every auxiliary value, and the complementary input required by $\m{op}$.
	\item The numbers of different values in the input type $A$ (the input type of $\m{orig}$) and the value type $V$ are both finite\footnote{Although there exist types including infinitely many values (e.g., lists), they can be approximated in our model by taking a large enough finite subset (e.g., setting a large enough length limit for lists).}, denoted as $s_A$ and $s_V$, respectively. At this time, the compressing property can be modeled as a domain property that $s_A \gg s_V$.
	\end{enumerate}
	\item $\mathcal M[\m{orig}, \m{op}]$ generates a lifting problem by independently and uniformly sampling the semantics for every candidate program in $\mathcal L_{\textit{aux}}$ and $\mathcal L_{\textit{comb}}$. For example, suppose that there is a program in $\mathcal L_{\textit{aux}}$ with type $A \rightarrow V$. The semantics of this program will be uniformly drawn from functions mapping from $A$ to $V$ ($s_V^{s_A}$ possibilities in total). 
\end{itemize}

The main result of our analysis is to bound the \textit{size-limited unrealizable rate of \mainname under the probabilistic model} using the \textit{\misfactor} of the given lifting problem. Before going into the details of this result, we shall first introduce the two concepts used in the analysis, starting from the \textit{size-limited unrealizable rate} (Definition \ref{definition:incomplete-rate}).

\begin{definition}[Size-Limited Unrealizable Rate] \label{definition:incomplete-rate} Given a lifting problem $\mathsf{LP}(\m{orig}, \m{op})$ and an integer $\m{lim}_s$, the \textit{size-limited unrealizable rate} of \mainname, denoted as $\m{unreal}(\m{orig}, \m{op}, \m{lim}_s)$, is defined as the probability for \mainname to decompose a random lifting problem (sampled from $\mathcal M[\m{orig}, \m{op}]$) into unrealizable subtasks, under the condition that the random problem has a valid solution whose size is no larger than $\m{lim}_s$, i.e.,
\begin{align*}
&\Pr_{\varphi \sim \mathcal M[\textit{orig}, \textit{op}]}\big[\text{\mainname generates an unrealizable subtask from }\varphi\ \big | \\
&\hspace{7em}\exists (\m{aux}, \m{comb}), \big(\textit{size}(\m{aux}, \m{comb}) \leq \m{lim}_s \wedge (\m{aux}, \m{comb})\text{ is valid for }\varphi\big)\big],
\end{align*}
where $\varphi \sim \mathcal M[\m{orig}, \m{op}]$ represents that $\varphi$ is a random lifting problem sampled from $\mathcal M[\m{orig}, \m{op}]$ and $\textit{size}(\m{aux}, \m{comb})$ represents the total size of programs $\m{aux}$ and $\m{comb}$. 
\end{definition}


In this definition, we take the size of the smallest valid program of lifting problems into consideration by introducing the size limit $\m{lim}_s$. Such a treatment enables a more refined analysis that relies on the size of the smallest valid program.


\smallskip 

The second concept is the \textit{\misfactor} of a lifting problem $\mathsf{LP}(\m{orig}, \m{op})$ (Definition \ref{definition:mismatch-factor}). This factor counts the number of independent examples that $\m{aux}$ must satisfy to ensure a function exists for calculating the output of $\m{orig}$.

\begin{definition}[\Misfactor]\label{definition:mismatch-factor} Given lifting problem $\mathsf{LP}(\m{orig}, \m{op})$ and integer $t$, the \textit{\misfactor} of this lifting problem is at least $t$ if there exists $t$ pairs of values $(\overline{a_i}, \overline{a_i'}) \in A^n \times A^n$ satisfying the following two conditions.
\begin{itemize}
	\item For each pair $(\overline{a_i}, \overline{a_i'})$, the inputs of ${\m{comb}}$ when no auxiliary value is used (i.e., the output of $\m{orig}^n$) are the same but the outputs of ${\m{comb}}$ (i.e., the output of $\m{orig} \circ \m{op}$) are different. In other words, every pair $(\overline{a_i}, \overline{a_i'})$ must satisfy the formula below.
	\begin{align}
	\exists c \in C, \big(\bluec{\m{orig}}^n~\overline{a_i} = \bluec{\m{orig}}^n~\overline{a_i'} \wedge \bluec{\m{orig}}~(\bluec{\m{op}}~(c, \overline{a_i})) \neq \bluec{\m{orig}}~(\bluec{\m{op}}~(c, \overline{a_i'})) \big) \label{formula:mismatch-first-condition}
	\end{align}
	\item All components involved in these pairs ($2nt$ in total) are different.
\end{itemize}
Recall that $A$ denotes the input type of $\m{orig}$, $n$ denotes the arity of $\m{op}$, and $C$ denotes the type of the complementary input required by $\m{op}$.
\end{definition} 

The mismatch factor reflects the strength of the specification that \mainname uses to synthesize ${\m{aux}}$. To see this point, let us consider the first leaf subtask of $\m{aux}$ (Formula \ref{formula:spec-aux}) generated when solving lifting problem $\mathsf{LP}(\m{orig}, \m{op})$, as shown below. 
\begin{align*}
\big(\bluec{\m{orig}}^n~\quanti{\overline{a}} = \bluec{\m{orig}}^n~\quanti{\overline{a'}} \wedge \bluec{\m{orig}}~(\bluec{\m{op}}~(\quanti{c}, \quanti{\overline{a}})) \neq \bluec{\m{orig}}~(\bluec{\m{op}}~(\quanti{c}, \quanti{\overline{a'}}))\big) \rightarrow \var{\m{aux}_1}^n~\quanti{\overline{a}} \neq \var{\m{aux}_1}^n~\quanti{\overline{a'}}
\end{align*}

The premise of this specification is exactly the first condition in Definition \ref{definition:mismatch-factor}. Through this connection, a mismatch factor of at least $t$ implies that, in the approximate specification derived by \mainname, ${\m{aux}_1}$ (a sub-program of ${\m{aux}}$) needs to output differently on $t$ independent pairs of inputs. Intuitively, with a larger mismatch factor, an incorrect ${\m{aux}_1}$ will be less likely to satisfy this specification, and thus \mainname will be more likely to solve the lifting problem successfully.

\smallskip 

With the above two concepts, Theorem \ref{theorem:main-result} shows the main result of our probabilistic analysis. We prove that for any lifting problem, the size-limited unrealizable rate of \mainname is bounded by the \misfactor, the size limit, and the size of the output domain (i.e., $s_V$). 
\begin{theorem}[Upper Bound on the Unrealizable Rate] \label{theorem:main-result} Given a lifting problem $\mathsf{LP}(\m{orig}, \m{op})$ of which the \misfactor is at least $t$, the size-limited unrealizable rate of \mainname is bounded, as shown below.
$$
\m{unreal}(\m{orig}, \m{op}, \m{lim}_s) \leq 2^w \exp(-t/s_V^{n \cdot w}), \textbf{where } w \triangleq \m{lim}_c \cdot \m{lim}_s
$$
\end{theorem} 

\subsubsection{Effectiveness under the compressing property \label{subsection:effecCompressing}} Theorem \ref{theorem:main-result} shows that the unrealizable rate of \mainname is small when the \misfactor is far larger than the size of the output domain. Fortunately, this is the usual case when solving lifting problems. One important domain property here is the \m{compressing property} of lifting problems, that is, the input domain of \m{orig} and auxiliary programs in $\mathcal L_{\textit{aux}}$ are usually far larger than their output domains. In the following, we shall first introduce the reason why the compressing property generally exists in lifting problems and then discuss how the compressing property implies a large \misfactor.

To see why the compressing property exists, let us recall the meaning of lifting problems discussed in Section \ref{subsection:lifting-problem}. In the sense of optimization, the inputs of $\m{orig}$ and \m{aux} correspond to the intermediate data structures constructed in an inefficient program, and their outputs correspond to the values calculated after eliminating these intermediate data structures. To achieve optimization, these programs must summarize a small result (e.g., a scalar value) from a large data structure (e.g., an inductive data structure that can be arbitrarily large), leading to the compressing property.

\begin{example} \label{example:compress}In the two lifting problems discussed in Sections \ref{subsection:example1} and \ref{subsection:incre-example} (Formulas \ref{formula:dac-sndmin} and \ref{formula:incre}), both the original program \m{sndmin} and the auxiliary program \m{aux} take an integer list as the input and output a single integer. The ratio between the number of integer lists to the number of integers tends to $\infty$ when the length of lists tends to $\infty$ and every integer is bounded within a fixed range.
\end{example}

To see the relation between the compressing property and the \misfactor, let us review the first condition in the definition of the \misfactor (Formula \ref{formula:mismatch-first-condition}). For an input pair $(\overline{a}, \overline{a'})$ and a fixed complementary input $c$, this condition requires $\m{orig}^n$ (whose output domain is $V^n$) to output the same on $\overline{a}$ and $\overline{a'}$, and requires $\m{orig} \circ \m{op}$ (whose output domain is $V$) to output differently on $(c, \overline{a})$ and $(c, \overline{a'})$. The strength of this condition can be estimated using probabilities. The probability for a given $(\overline{a}, \overline{a'})$ and $c$ to satisfy this condition can be estimated as $s_V^{-n}(1-1/s_V)$ because (1) the probability for two random values in $V^n$ to be the same is $s_V^{-n}$ and (2) the probability for two random values in $V$ to be different is $1 - 1/s_V$. It means, in the random sense, there is one pair $(\overline{a}, \overline{a'})$ satisfying this condition in every $O(s_V^n)$ pairs\footnote{Here we assume that $s_V > 1$, that means, there are at least two different outputs.}. Since there are up to $S_A/(2n)$ independent pairs in $A^n \times A^n$ and $s_A$ is far larger than $s_V$ by the compressing property, this estimation tells that the \misfactor is far larger than $s_V$ with a high probability. A concrete example of analyzing the \misfactor for a given lifting problem can be found in Appendix \ref{appendix:misfactor-example}.

Theorem \ref{theorem:completeness} combines the above analysis with Theorem \ref{theorem:main-result}. It demonstrates that the unrealizable rate of \mainname is almost always small when the compressing property holds.

\begin{theorem}[Unrealizable Rate Under the Compressing Property] \label{theorem:completeness} Consider the size-limited unrealizable rate of \mainname on a random lifting problem. When there are at least two values (i.e., $s_V > 1$), for any constant $\epsilon > 0$, the probability for this rate to exceed $\epsilon$ tends to $0$ when $s_A / s_V^{w'}$ tends to $\infty$, where $w' \triangleq n \cdot \m{lim}_c \cdot \m{lim}_s + n + 1$.
\end{theorem} 
} 
\jryadd{


\smallskip
}

\subsubsection{Preference of \mainname for simpler auxiliary programs \label{subsection:preferenceForSimpler}} In the discussion above, we model the semantics of candidate programs in $\mathcal L_{\textit{aux}}$ and $\mathcal L_{\textit{comb}}$ as independent to ease the analysis. However, a semantical dependency between programs indeed exists in practice because the semantics is usually defined along with the syntax under domain theories. Such a dependency weakens the specification for $\m{aux}$ (Formula \ref{formula:spec-aux}) because it makes the realizability of an \m{aux} subtask (i.e., the existence of a valid auxiliary program) imply the validness of (infinitely) many other programs. 

\begin{example} In the \m{sndmin} example (Section \ref{subsection:moti-overview}), many valid programs for the \m{aux} subtask (Formula \ref{formula:ve-1}) can be constructed from the target auxiliary program $\m{min}\ \m{xs}$, for example, by including more auxiliary values (e.g., $(\m{min}\ \m{xs}, \m{max}\ \m{xs})$) or performing invertible arithmetic operations (e.g., $(\m{min}\ \m{xs}) + (\m{min}\ \m{xs})$). Many of them may lead to unrealizable subtasks, for example, the combination function corresponding to $(\m{min}\ \m{xs}) + (\m{min}\ \m{xs})$ cannot be implemented in $\mathcal L_{\textit{comb}}^{\textit{ex}}$ (Figure \ref{fig:comb-space}).
\end{example}

The key for \mainname to perform well under such a dependency is its preference for simpler auxiliary programs, following the principle of \textit{Occam's razor}. The \m{aux} synthesizer in \mainname (Algorithm \ref{alg:synthesis-aux}) is an enumeration-based synthesizer and always returns a minimal possible auxiliary program (Theorem \ref{theorem:minimal}). Therefore, this synthesizer can successfully avoid those unnecessarily complex programs derived from the expected one.

\begin{theorem}[Minimality] \label{theorem:minimal} Given an example-based synthesis task for the auxiliary program, the program $\m{aux}^*$ synthesized by our \m{aux} synthesizer must be a minimal valid program. In other words, any strict sub-program of $\m{aux}^*$ must not be valid for the given task.
\end{theorem}
\section{Applications of \mainname}\label{section:application}

We have seen how to reduce the applications of D\&C and incrementalization to lifting problems (Sections \ref{subsection:example1} and \ref{subsection:example2}). Through these reductions, \mainname can be instantiated to synthesizers for the corresponding paradigms. In this section, we shall supply more details on how the efficiency of the resulting program is ensured in these reductions (Section \ref{subsection:efficiency-guarantee}) and provide reductions for several other D\&C-like paradigms (Section \ref{subsection:more-application}).



\subsection{Efficiency Condition} \label{subsection:efficiency-guarantee}
When applying an algorithmic paradigm, the synthesizer needs to ensure that the resulting program must be efficient. We achieve this guarantee by limiting the domain-specific languages $\mathcal L_{\textit{aux}}$ and $\mathcal L_{\textit{comb}}$, following the SyGuS framework. Specifically, we require these languages to include only programs that can lead to an efficient result. When instantiating \mainname for a specific algorithmic paradigm, we can establish different efficiency guarantees by using different languages.

In this paper, we consider a specific condition (denoted as the \m{efficiency condition}) for the domain-specific languages. This condition requires that (1) every program in $\mathcal L_{\textit{aux}}$ runs in constant time on a constant-sized input, and (2) every program in $\mathcal L_{\textit{comb}}$ runs in constant time. 

\begin{example} The domain-specific languages $\mathcal L_{\textit{aux}}^{\textit{ex}}$ and $\mathcal L_{\textit{comb}}^{\textit{ex}}$ (Figures \ref{fig:aux-space} and \ref{fig:comb-space}, discussed in the \m{sndmin} example) satisfy the efficiency condition.
\end{example}

The efficiency condition implies the efficiency of the D\&C and the incremental programs synthesized by the corresponding instantiation of \mainname, as shown below.
\begin{itemize}
\item In the D\&C program synthesized through the reduction in Section \ref{subsection:example1}, each of $\m{orig}, \m{aux}$, and $\m{comb}$ is invoked $O(n)$ times, and the former two are only invoked on singleton lists. Therefore, this program is ensured to be $O(n/p)$ time in parallel on a list of length $n$ and $p \leq n / \log n$ processors when (1) the efficiency condition is satisfied and (2) $\m{orig}$ runs in constant time on a singleton list.
\item In the incremental program synthesized through the reduction in Section \ref{subsection:incre-example}, $\m{comb}$ will be invoked once after each change. Therefore, this program is ensured to be constant-time per change when the efficiency condition is satisfied.
\end{itemize}
\subsection{More Applications}  \label{subsection:more-application}

\begin{figure*}
\hfill
\begin{minipage}{0.31\textwidth}
  \begin{figure}[H]
    \begin{lstlisting}
res = (orig([]), aux([]))
for v in xs:
  res = comb(v, res)
return res[0]
    \end{lstlisting}
    \vspace{-1em}
  \caption{A template of single-pass.}
  \label{fig:single-pass}
  \end{figure}
\end{minipage}
\hfill 
\begin{minipage}{0.35\textwidth}
        \begin{lstlisting}
info = (p([]), aux([]))
res, l = 0, 0
for r in range(len(xs)):
  info = comb1(xs[r], info)
  while l <= r and not info[0]:
    info = comb2(xs[l], info)
    l += 1
  res = max(res, r - l + 1)
return res
        \end{lstlisting}
        \vspace{-1em}
      \caption{A template of sliding window.}
      \label{fig:sliding-window}
      \end{minipage}
\hfill\ 
\end{figure*}

\textbf{Single-Pass}. Single-pass~\cite{DBLP:reference/db/Schweikardt09b} is an algorithmic paradigm widely applied in various domains such as databases and networking. It is also the input format required by \m{Parsynt}~\cite{DBLP:conf/pldi/FarzanN17,toronto21} (Section \ref{subsection:example1}). A single-pass program (Figure \ref{fig:single-pass}) scans the input list once from the first element to the last and iteratively updates the result after visiting each element. To apply single-pass to an original program $\m{orig}$, an auxiliary program $\m{aux}$ and a combinator $\m{comb}$ satisfying the formula below are required.
$$
\targetname'\ (\quanti{\m{xs}} \cat [\quanti{v}]) = \var{\m{comb}}\ (\quanti{v}, \targetname'\ \quanti{\m{xs}}), \key{where}\ \targetname' \triangleq \bluec{\targetname} \spl \var{\m{aux}}
$$

This task is equivalent to that for incrementalization (Section \ref{subsection:example2}) and can be regarded as a lifting problem $\mathsf{LP}(\m{orig}, \m{op})$ with $\m{op}\ (v, (\m{xs})) \triangleq \m{xs} \cat [v]$. Under the efficiency condition, the resulting single-pass program runs in $O(n)$ time on a list of length $n$ because its bottleneck is the $O(n)$ invocations of $\m{comb}$ in the loop. 

\smallskip

\noindent\textbf{General Incrementalization}. In the previous incrementalization example (Section \ref{subsection:example2}), the only allowed change is to append an element to the back of the input list. In the general case, there can be different types of changes in a single task, captured by a change set $C$ denoting all possible changes and a change operator $\m{change}:C\times A \mapsto A$ applying a change to an $A$-element. To incrementally update the result of the original program $\m{orig}$ after each change, an auxiliary program $\m{aux}$ and a combinator $\m{comb}$ satisfying the formula below are required. 
$$
\targetname'\ (\bluec{\m{change}}\ (\quanti{c}, \quanti{a})) = \var{\m{comb}}\ (\quanti{c}, \targetname'\ \quanti{a}), \key{where}\ \targetname' \triangleq \bluec{\targetname} \spl \var{\m{aux}}
$$

This task can be regarded as $\texttt{LP}(\m{orig}, \m{op})$ with $\m{op}\ (c, (a)) \triangleq \m{change}\ (c, a)$. Under the efficiency condition, the resulting incremental program must run in $O(1)$ time per change because only $\m{comb}$ is invoked once after each change.

\begin{example} \label{example:merge-change} Continuing with the previous incrementalization task (Section \ref{subsection:incre-example}), let us consider a new change that pushes a new element to the front of the input list. In the new task, the change set $C$ can be defined as $\{\m{``front''},\m{``back''}\} \times \texttt{Int}$ and the corresponding $\m{change}$ is as follows.
    \begin{align*}
        \m{change}\ ((\m{tag}, v), \m{xs}) \triangleq \textbf{if }\m{tag} = \m{``front''}\textbf{ then }[v] \cat \m{xs}\textbf{ else }\m{xs} \cat [v] 
      \end{align*}
\end{example}

\smallskip

\noindent \textbf{Longest Segment Problem}. Given a predicate $p$ and an input list, there may be multiple segments of the input list satisfying $p$, and the longest segment problem asks for the maximum length of valid segments. \citet{DBLP:journals/scp/Zantema92} studies three different subclasses of longest-segment problems and proposes three algorithmic paradigms for them, respectively. Here, we select and introduce one typical paradigm among them, and the details on the others can be found in Appendix \ref{appendix:paradigms}.

This paradigm enumerates segments via a technique named \textit{sliding window} (Figure \ref{fig:sliding-window}), where $l$ and $r$ are the indices of the current segment (from the $l$-th to the $r$-th element in the input list $\m{xs}$) and $\m{info}$ records (1) whether $p$ is currently satisfied and (2) necessary auxiliary values. The outer loop appends every element in the input list \m{xs} to the back of the current segment one by one, and the inner loop repeatedly removes the first element of the current segment until $p$ is satisfied. This enumeration guarantees to visit the longest valid segment when $p$ is prefix-closed, that is, every prefix of a list satisfying $p$ satisfies $p$ as well.

To apply this paradigm to a longest segment problem, an auxiliary program $\m{aux}$ and two combinators $\m{comb}_1$ and $\m{comb}_2$ are required to correctly update $\m{info}$ during the enumeration. Concretely, they must satisfy the formulas below, where \m{head} returns the first element of a list and \m{tail} returns the result of removing the first element from a list.
\begin{gather*}
    (\bluec{p} \spl \var{\m{aux}})\ (\quanti{\m{xs}} \cat [\quanti{v}]) = \var{\m{comb}_1}\ (\quanti{v}, (\bluec{p} \spl \var{\m{aux}})\ \quanti{\m{xs}}) \\
    {\m{head}}\ \quanti{xs} = \quanti{v} \rightarrow (\bluec{p} \spl \var{\m{aux}})\ ({\m{tail}}\ \quanti{\m{xs}}) = \var{\m{comb}_2}\ (\quanti{v}, (\bluec{p} \spl \var{\m{aux}})\ \quanti{\m{xs}})
\end{gather*}
The condition of $\m{head}\ \m{xs} = \m{v}$ is involved to allow the second combinator $\m{comb}_2$ to access the element to be removed. When reducing the above task to a lifting problem, this condition can be eliminated by assigning a dummy output to $\m{op}$ when the condition is violated, and the two formulas can be merged through the construction in Example \ref{example:merge-change}. A possible operator $\m{op}$ of the resulting lifting problem is shown below, where the complementary input is in $\{\m{``append''}, \m{``remove''}\} \times \texttt{Int}$.
\begin{align*}
    \m{op}\ ((\m{tag}, v), (\m{xs}))\triangleq\left\{
    \begin{array}{ccc}
      \m{xs} \cat [v] & & \m{tag} = \m{``append''}  \\
      \m{tail}\ \m{xs} & & \m{tag} = \m{``remove''} \wedge \m{head}\ \m{xs} = v \\
      \ \ []\ \  & & \textit{otherwise} \\
    \end{array}
    \right.
  \end{align*}

Under the efficiency condition, the resulting program runs in $O(n)$ time on a list of length $n$ because both $\m{aux}_1$ and $\m{aux}_2$ are invoked at most $n$ times. 
\smallskip

\noindent\textbf{Segment Trees}. The segment tree is a type of classical data structure for answering queries on a specific property of a segment in a possibly long list~\cite{Bent1977}. Given an initial list, after a linear-time pre-processing, a segment tree can efficiently evaluate a pre-defined function \targetname on a segment (e.g., ``answer the second minimum of the segment from the 2nd to the 5,000th element'') or applies a pre-defined change operator \m{change} to a segment (e.g., ``add each element in the segment from the 2nd to the 5,000th element by $1$''), each in logarithmic time w.r.t. the list length. 

The detailed template of a segment tree can be found in Appendix \ref{appendix:paradigms}, and in brief, it uses D\&C to respond to queries and uses incrementalization to respond to changes. Therefore, implementing a segment tree is to find an auxiliary program $\m{aux}$ and two combinators $\m{comb}_1$ and $\m{comb}_2$ such that $(\m{aux}, \m{comb}_1)$ is a solution to the lifting problem of D\&C and $(\m{aux}, \m{comb}_2)$ is a solution to the lifting problem of incrementalization; in other words, the formulas below need to be satisfied.
\begin{gather*}
(\bluec{\m{orig}} \spl \var{\m{aux}})\ (\quanti{xs_L} \cat \quanti{xs_R}) = \var{\m{comb}_1}\ ((\bluec{\m{orig}} \spl \var{\m{aux}})\ \quanti{xs_L}, (\bluec{\m{orig}} \spl \var{\m{aux}})\ \quanti{xs_R}) \\
(\bluec{\m{orig}} \spl \var{\m{aux}})\ (\bluec{\m{change}}\ (\quanti{c}, \quanti{xs})) = \var{\m{comb}_2}\ (\quanti{c}, (\bluec{\m{orig}} \spl \var{\m{aux}})\ \quanti{xs}) 
\end{gather*}
Through the construction in Example \ref{example:merge-change}, these two formulas can be unified into a single lifting problem with the following \m{op}, where the complementary input is either an element in the change set of $\m{change}$ or a token \m{``d\&c''} never used before. 
$$
\m{op}\ (c, (\m{xs}_L, \m{xs}_R)) \triangleq \textbf{if } c = \m{``d\&c''}\textbf{ then } \m{xs}_L \cat \m{xs}_R \textbf{ else }\m{change}\ (c, \m{xs}_L)
$$


Let $n$ be the length of the initial list. A segment tree invokes $\m{comb}_1$ and $\m{comb}_2$ $O(\log n)$ times when processing each operation (either a query or a change). Therefore, the resulting program must be $O(\log n)$ time per operation under the efficiency condition. More details on this guarantee can be found in Appendix \ref{appendix:paradigms}.
\section{Implementation}\label{section:implementation}
Our implementation of \mainname focuses on lifting problems related to integer lists and integers. It can be generalized to other cases if the corresponding types, operators, and grammars are provided. 

\begin{figure*}
    {\small \centering
        \begin{tabular}{cccl} 
            \toprule 
            Start symbol & $S$ & $\rightarrow$ & $N_{\mathbb Z}\ |\ (S, S)$\\
            Integer expr & $N_{\mathbb Z}$& $\rightarrow$ & $\text{IntConst}\ |\ N_{\mathbb Z} \oplus  N_{\mathbb Z}\ | \  \textit{sum}\ N_{\mathbb L}\ |\ \textit{len}\ {N_{\mathbb L}}\ |\ \textit{head}\ {N_{\mathbb L}}$ \\
            & & $|$ & $\textit{last}\ {N_{\mathbb L}}\ | \ \textit{access}\ N_{\mathbb Z}\ N_{\mathbb L}\ |\ \textit{count}\ {F_{\mathbb B}\ N_{\mathbb L}}\ |\ \textit{min}\ {N_{\mathbb L}}$ \\
            & & $|$ & $\textit{max}\ {N_{\mathbb L}}\ |\ \textit{neg}\ {N_{\mathbb Z}}$ \\
            List expr & $N_{\mathbb L}$& $\rightarrow$ & $\text{Input} \ | \ \textit{take}\ {N_{\mathbb Z}\ N_{\mathbb L}}\ |\ \textit{drop}\ {N_{\mathbb Z}\ N_{\mathbb L}}\ |\ \textit{rev}\ {N_{\mathbb L}}$  \\
            & & $|$ & $\textit{map}\ {F_{\mathbb Z}\ N_{\mathbb L}}\ |\ \textit{filter}\ {F_{\mathbb B}\ N_{\mathbb L}}\ |\ \textit{zip}\ {\oplus\ N_{\mathbb L}\ N_{\mathbb L}} \ | \ \textit{sort}\ {N_{\mathbb L}} $  \\
            & & $|$ & $\textit{scanl}\ {\oplus\ N_{\mathbb L}}\ |\ \textit{scanr}\ {\oplus\ N_{\mathbb L}}$ \\
            Binary Operator & $\oplus$ & $\rightarrow$ & $+ \ | \ -\ |\ \times\ |\ \textit{min}\ |\ \textit{max}$ \\
            Integer Function & $F_{\mathbb Z}$ & $\rightarrow$ &  $(+\ \text{IntConst}) \ | \ (-\ \text{IntConst}) \ | \ \textit{neg}{}$\\
            Boolean Function & $F_{\mathbb B}$ & $\rightarrow$ & $(< 0)\ |\ (> 0)\ |\ \textit{odd}{}\ |\ \textit{even}{}$ \\
            \bottomrule
    \end{tabular}}
\caption{The grammar of $\mathcal L_{\textit{aux}}$.}
\label{figure:gf}
\end{figure*}

\begin{figure*}
    {\small \centering
        \begin{tabular}{cccl} 
            \toprule 
            Start symbol & $S$ & $\rightarrow$ & $N_{\mathbb Z}\ |\ (S, S)$ \\
            Integer expr & $N_{\mathbb Z}$& $\rightarrow$ & $\text{IntConst}\ |\ N_{\mathbb Z} \oplus  N_{\mathbb Z}\ |\ \textit{if}\ {N_{\mathbb B}\textit{ then } N_{\mathbb Z}\textit{ else } N_{\mathbb Z}}\ |\ N_{\mathbb T}.i$\\
            Bool expr & $N_{\mathbb B}$& $\rightarrow$ & $\neg N_{\mathbb B}\ |\ N_{\mathbb B} \wedge N_{\mathbb B}\ |\ N_{\mathbb B} \vee N_{\mathbb B }\ |\ N_{\mathbb Z} \leq N_{\mathbb Z}\ |\ N_{\mathbb Z} = N_{\mathbb Z}$ \\
            Tuple expr & $N_{\mathbb T}$ &$\rightarrow$ & $\text{Input}\ |\ N_{\mathbb T}.i $ \\
            Binary Operator & $\oplus$ & $\rightarrow$ & $+ \ | \ -\ |\ \times\ |\ \textit{div}$ \\
            \bottomrule
    \end{tabular}}
    \caption{The grammar of $\mathcal L_\textit{comb}$.}
    \label{figure:gc}
\end{figure*}

\smallskip
\noindent \textbf{Domain-specific languages}. A lifting problem requires two languages $\mathcal L_{\textit{aux}}$ and $\mathcal L_{\textit{comb}}$ to specify the spaces of candidate auxiliary programs and combinators, respectively (Definition \ref{def:lifting}). 

Language $\mathcal L_{\textit{aux}}$ in our implementation (Figure \ref{figure:gf}) is the language of \m{DeepCoder}~\cite{DBLP:conf/iclr/BalogGBNT17}. It includes 17 list-related operators, including common higher-order functions (e.g., \m{map} and \m{filter}) and several operators that perform branching and looping internally (e.g., \m{count} and \m{sort}). Because the language of \m{DeepCoder} does not support producing tuples, we add an operator for constructing tuples at the top level to cope with the cases where multiple auxiliary values are required.

\begin{example} \label{example:implement-mps}Operator $\m{scanl}$ in our $\mathcal L_{\textit{aux}}$ receives a binary operator $\oplus$ and a list. It constructs all non-empty prefixes of the list and reduces each of them to an integer via $\oplus$, as shown below.
$$
\m{scanl}\ (\oplus)\ [\m{xs}_1, \dots, \m{xs}_n] \triangleq [\m{xs}_1, \m{xs}_1 \oplus \m{xs}_2, \dots, \m{xs}_1 \oplus \dots \oplus \m{xs}_n]
$$
Using this operator, the maximum prefix sum of list $\m{xs}$ can be implemented as $\m{max}\ (\m{scanl}\ (+)\ \m{xs})$. 
\end{example}
 
Language $\mathcal L_{\textit{comb}}$ in our implementation (Figure \ref{figure:gc}) is the language of the conditional arithmetic domain in SyGuS-Comp~\cite{DBLP:journals/corr/abs-1904-07146}, a world-wide competition for program synthesis. \jryadd{It includes basic arithmetic operators (e.g., $+$ and $\times$), comparison operators (e.g., $=$ and $\leq$), Boolean operators (e.g., $\neg$ and $\wedge$), and the branch operator \m{if-then-else}. This language can express complex scalar calculations by using the branch operator in a nested manner.} Similar to the case of $\mathcal L_{\textit{aux}}$, we add operators for accessing and constructing tuples (i.e., $(S, S)$ and $N_{\mathbb T}.i$) to deal with the cases requiring multiple auxiliary values.

\smallskip

These languages match the assumption we used when analyzing the probability for \mainname to generate unrealizable subtasks (Section \ref{subsection:properties}, assuming \jrydel{the expressiveness of $\mathcal L_{\textit{comb}}$ and} the compressing property of $L_{\textit{aux}}$) and when analyzing the efficiency of the resulting program (Section \ref{subsection:efficiency-guarantee}, assuming the efficiency condition).
\begin{itemize}
    \jrydel{\item For expressiveness, $\mathcal L_{\textit{comb}}$ can express complex integer calculations via nested branch operators. Although it cannot provide the strict guarantee in Assumption \ref{pre-assumption:expressive}, we believe it can handle those easy tasks in practice where a combination function directly exists.}
    \item For the compressing property, programs in $\mathcal L_{\textit{aux}}$ all map an integer list (that can be arbitrarily large) to a constant-sized integer tuple. Therefore, their input domains are far larger than their output domains when large enough lists are considered and every integer is bounded within a fixed range\footnote{The actual case is relatively more complex because many programs in $\mathcal L_{\textit{aux}}$ may enlarge the range of integers. We shall discuss this point in Section \ref{section:discussion}.}.
    \item For the efficiency condition, one can verify that every program in $\mathcal L_{\textit{aux}}$ runs in constant time on a constant-sized list, and every program in $\mathcal L_{\textit{comb}}$ runs in constant time. Therefore, the efficiency condition is satisfied, and our implementation can provide all efficiency guarantees established in Section \ref{section:application}.
\end{itemize}

\jrydel{\smallskip
\noindent \textbf{Verification}. \mainname applies the CEGIS framework to solve leaf subtasks and requires corresponding verifiers to generate counter-examples for incorrect programs (Section \ref{subsection:inductive}). A probabilistic verifier is used by default in our implementation. It generates random examples from a pre-defined distribution and then tests the candidate program on these examples. Such a verifier does not access the source code of \m{orig} and thus keeps the generality of \mainname.

This verifier can provide a probabilistic correctness guarantee for the CEGIS result when the number of tested examples is large enough. In our implementation, $10^4 \times i$ random examples are used in the $i$-th CEGIS iteration. Under this configuration, the probability for the error rate of the synthesized program (on a random example) to be more than $10^{-3}$ is at most $4.55 \times 10^{-5}$. The details of this guarantee can be found in Appendix \ref{appendix:proof-probabilistic}.}

\jryadd{\smallskip
\noindent \textbf{Verification}. Since \mainname applies the CEGIS framework to solve leaf subtasks (Section \ref{subsection:inductive}), it requires corresponding verifiers to generate counter-examples for incorrect programs. We implement these verifiers using bounded model checking~\cite{DBLP:journals/ac/BiereCCSZ03}. Specifically, we assume the original program is implemented in C++. Given a candidate program written in our domain-specific languages (Figures \ref{figure:gc} and \ref{figure:gf}), our verifier will first translate it and its specification into C++ and then apply CBMC~\cite{DBLP:journals/corr/abs-2302-02384}, a popular bounded model checker, to verify whether the specification is satisfied. In this procedure, we consider only lists within a length limit (6 by default) to bound the depth of loops and recursions.

Besides, since it is time-consuming to perform bounded-model checking, in each CEGIS iteration, we will first test the candidate program using a set of random examples and will invoke the above verifier only when the candidate program satisfies all of these examples. Note that the random testing itself can provide a probabilistic correctness guarantee  when the number of tested examples is large enough. Details on this point can be found in Appendix \ref{appendix:proof-probabilistic}.
}

Our implementation generates random examples as follows.
\begin{itemize}
    \item Every tuple is generated by recursively generating its components.
    \item Every list is generated by (1) uniformly drawing its length from integers in $[0, 10]$, and then (2) recursively generating every element.
    \item Every integer is uniformly drawn from integers in $[-5, 5]$ by default. This range may change for tasks with specialized requirements. For example, some tasks in the dataset collected by \citet{toronto21} consider only $01$ lists. Correspondingly, the range of integers is set to $[0, 1]$ on these tasks. 
\end{itemize}

\smallskip 
\noindent \textbf{Other configurations}. \jryadd{We implement the decomposition system in \mainname using the greedy strategy (Algorithm \ref{alg:deductive}). In each decomposition, our implementation considers only the first solution synthesized from the first subtask and will fail once an unrealizable subtask is generated.}

A synthesizer based on input-output examples is required to solve the leaf subtasks for \m{comb} (Algorithm \ref{alg:synthesis-comp}, Section \ref{subsection:inductive}). Our implementation uses \m{PolyGen}~\cite{DBLP:journals/pacmpl/JiXXH21}, a state-of-the-art synthesizer on the conditional arithmetic domain.

The example-based synthesizer for \m{aux} (Algorithm \ref{alg:synthesis-comp}, Section \ref{subsection:inductive}) is configured by an integer $\m{lim}_c$, representing the maximum number of components in the top-level combination. We set $\m{lim}_c$ to $4$ by default because $4$ auxiliary values are already enough for most known lifting problems.

\section{Evaluation}\label{section:evaluation}
To evaluate \mainname, we report two experiments to answer the following research questions.
\begin{itemize}
\item \textbf{RQ1}: How effective does \mainname solve lifting problems?
\item \textbf{RQ2}: Does \mainname outperform existing synthesizers in applying D\&C?
\item \textbf{RQ3}: Does \mainname outperform existing synthesizers in applying single-pass?
\item \textbf{RQ4}: How does observational covering affect the performance of \mainname?

\end{itemize}

\subsection{Experimental Setup} \label{subsection:setup}
\textbf{Baseline Solvers}. In our evaluation, we compare \mainname with two general-purpose synthesizers, \textit{Enum}~\cite{DBLP:conf/fmcad/AlurBJMRSSSTU13} and \textit{Relish}~\cite{DBLP:journals/pacmpl/0001WD18}. Both of them can be applied to solve lifting problems and can be instantiated as synthesizers for D\&C-like paradigms as \mainname does. 
\begin{itemize}
    \item \textit{Enum}~\cite{DBLP:conf/fmcad/AlurBJMRSSSTU13} is an enumerative solver. Given a lifting problem, \textit{Enum} enumerates all possible $(\m{aux}, \m{comb})$ from small to large until a valid one is found.
    \item \textit{Relish}~\cite{DBLP:journals/pacmpl/0001WD18} is a state-of-the-art synthesizer for relational specifications. It first excludes many invalid programs via a data structure namely \textit{hierarchical finite tree automata} and then searches for a valid program among the automata.
\end{itemize} 
\jrymod{Both \m{Enum} and \m{Relish} are re-implemented to support the list-related operators used in our paper.}{In our evaluation, we re-implement both \m{Enum} and \m{Relish} because of the reasons below.
\begin{itemize}
\item The original implementation of \m{Enum} cannot solve lifting problems in our dataset. Specifically, this implementation requires the synthesis task to be specified in the SyGuS input format~\cite{sygus-input-format}. However, this format does not support list-related operators and thus cannot express the language $\mathcal L_{\textit{comb}}$ we use (Figure \ref{figure:gc}).
\item The input format of the original implementation of \m{Relish} does not support the languages we use, and it is difficult to update this implementation to match our demand.
\end{itemize}
We re-implement these two synthesizers using the same setting as our implementation of \mainname (Section \ref{section:implementation}). Specifically, we use the same domain-specific languages to specify the program spaces and use the same verifier to generate counter-examples for incorrect candidate programs.
}


\smallskip
Besides, we also compare \mainname with two state-of-the-art specialized synthesizers.
\begin{itemize}
    \item \textit{Parsynt}~\cite{toronto21, DBLP:conf/pldi/FarzanN17} is a specialized synthesizer for D\&C that relies on syntax-based program transformations. Specifically, it requires the original program to be single-pass, then extracts $\m{aux}$ by transforming the loop body using pre-defined transformations, and at last, synthesizes a corresponding $\m{comb}$ via an existing synthesizer. There are two versions of \textit{Parsynt} available, denoted as \textit{Parsynt17}~\cite{DBLP:conf/pldi/FarzanN17} and \textit{Parsynt21}~\cite{toronto21}, where different syntax-based program transformations are used. We consider both versions of \m{Parsynt} in our evaluation.
    \item \textit{DPASyn}~\cite{DBLP:conf/oopsla/PuBS11} is a specialized synthesizer for single-pass. It reduces the application task of single-pass to a Sketch problem~\cite{DBLP:journals/sttt/Solar-Lezama13}, solves this task using existing Sketch solvers, and also includes specialized optimizations for dynamic programming programs. We use a re-implementation based on \m{Grisette}~\cite{DBLP:journals/pacmpl/LuB23} (provided by the authors of \m{DPASyn}) in our evaluation. 
\end{itemize}

\jryadd{We list two worth-noting details on these two specialized base solvers as follows.
\begin{itemize}
    \item \m{Parsynt} and \m{DPASyn} cannot be applied to each other's tasks because they both rely on the domain-specific property of their respective paradigms. Specifically, the program transformations used by \m{Parsynt} rely on the specific relationship between D\&C and single-pass, and the reduction to the Sketch problem in \m{DPASyn} is specifically designed for single-pass.
    \item These two solvers provide the same efficiency guarantee on the resulting program as \mainname. Specifically, both \mainname and \m{Parsynt} ensure that the resulting D\&C program runs in $\Theta(n/p)$ time on a list of length $n$ and $p \leq n / \log n$ processors; both \mainname and \m{DPASyn} ensure that the resulting single-pass program runs in $\Theta(n)$ time on a list of length $n$.
\end{itemize}
}


\smallskip
\noindent \textbf{Dataset}. Our evaluation is conducted on a dataset of $96$ tasks of applying D\&C-like algorithmic paradigms (Table \ref{table:dataset-profile}). These tasks are related to the four algorithmic problems.
\begin{table*}
	\renewcommand\arraystretch{1.15}
	\caption{The profile of synthesis tasks considered in our evaluation.} \label{table:dataset-profile}
	\begin{spacing}{1}
		\small
		\begin{tabular}{|c|c|c|c|c|c|}
            \hline
			Problem & D\&C & Single-pass & Longest Segment & Segment Tree & Total\\
            \hline
			\#Task & 36 & 39 & 8 & 13 & 96\\
			\hline 
		\end{tabular}
	\end{spacing}
\end{table*}
\begin{itemize}
\item \textit{Problem 1: applying D\&C to a program}. We collect $36$ such tasks from the datasets of previous studies~\cite{toronto21, DBLP:conf/pldi/FarzanN17, note1989}\footnote{\label{footnote:bugs}The original dataset of \textit{Parsynt21} contains two bugs in task \textit{longest\_1(0*)2} and \textit{longest\_odd\_(0+1)} that were introduced while manually rewriting the original program into single-pass. These bugs were confirmed by the original authors, and we fixed them in our evaluation. This also demonstrates that writing a single-pass program is difficult and error-prone.}, including all tasks used by~\citet{DBLP:conf/pldi/FarzanN17} and~\citet{note1989} and $12$ out of $22$ tasks used by~\citet{toronto21}. The other $10$ tasks used by \citet{toronto21} are out of the scope of \mainname because they cannot be reduced to lifting problems. They require a more general form of D\&C where the divide operator is not determined, making our reduction inapplicable.

\item \textit{Problem 2: applying single-pass to a program}. We consider the tasks used in the evaluation of \textit{DPASyn}~\cite{DBLP:conf/oopsla/PuBS11} and include 4 out of 5 tasks into our dataset. The last task includes multiple input lists and is not supported by our current implementation. 

Besides, we construct a series of tasks from our D\&C tasks as a supplement. For each D\&C task, a single-pass task with the same original program is constructed\footnote{A duplicated task involved by both \textit{DPASyn} and our D\&C dataset is ignored in this construction.}. These tasks correspond to a possible application for removing the restriction on the input program from existing transformation-based synthesizers for D\&C: one can first apply \mainname to get a single-pass program and then use existing synthesizers to generate a D\&C program.

\item \textit{Problem 3: Longest Segment Problem}. \citet{DBLP:journals/scp/Zantema92} proposes three algorithmic paradigms for longest segment problems and discusses $3, 1,$ and $4$ example tasks, respectively. For each example task, we include the task of applying the respective paradigm in our dataset.

\item \textit{Problem 4: applying segment trees to answer queries on a specific property of a segment in a list}. We collect some such tasks online because no previous work on segment trees provides a dataset. Specifically, we collect $13$ tasks by searching on Codeforces~(\url{https://codeforces.com/}), a website for competitive programming, using keywords "segment tree" and "lazy propagation"\footnote{A common alias of segment trees.}, and include all these tasks in our dataset.
\end{itemize}

\jrydel{Similar to the previous studies on automatically applying algorithmic paradigms~\cite{DBLP:conf/pldi/MoritaMMHT07, DBLP:conf/sosp/RaychevMM15, toronto21,DBLP:conf/ijcai/LinML19, acar2005self}, we assume the paradigm to be applied is given and directly apply the corresponding instantiation of \mainname to each task in our evaluation. In practice, when there are multiple paradigms available, we can either invoke all corresponding synthesizers in parallel or design a selector to select among them. Such a selection is out of the scope of this paper and is a direction for future work.}

\jryadd{
In our evaluation, we assume the correct paradigm for each task is available and thus evaluate \mainname (as well as baseline solvers \m{Enum} and \m{Relish}) by directly applying the corresponding instantiation to each task. This setting is the same as the previous studies on automatically applying algorithmic paradigms~\cite{DBLP:conf/pldi/MoritaMMHT07, DBLP:conf/sosp/RaychevMM15, toronto21,DBLP:conf/ijcai/LinML19, acar2005self} and corresponds to the usage scenario that the users need to decide which paradigm to use by themselves. We shall discuss the selection of paradigms in Section \ref{section:discussion}.
}

\smallskip

\noindent \textbf{Configuration}. Our experiments are conducted on Intel Core i7-8700 3.2GHz 6-Core Processor. Every execution is under a time limit of 300 seconds and a memory limit of 8 GB.

\subsection{RQ1: Comparison of Synthesizers for Lifting Problems} \label{subsection:rq1}
\jrydel{
\textbf{Procedure}. We compare \mainname with two baseline solvers \m{Enum} and \m{Relish} on all tasks in our dataset with a time limit of $300$ seconds and a memory limit of $8$ GB. We record the time cost of each successful synthesis to measure the efficiency of the solvers. 
\smallskip}

\jryadd{
\noindent \textbf{Procedure}. In this experiment, we compare \mainname with \m{Enum} and \m{Relish}, the two baseline solvers that support solving general lifting problems, on all tasks in our dataset. We consider three different aspects when comparing these solvers.
\begin{itemize}
    \item The effectiveness of synthesis, measured by the number of solved tasks.
    \item The efficiency of synthesis, measured by the average time cost of successful synthesis.
    \item The efficiency of the resulting program, measured by the time cost of the resulting program on a randomly generated test suite. Specifically, we calculate this time cost in three steps.
    \begin{enumerate}[leftmargin=1.5em]
        \item For each paradigm considered in our dataset, we implement a template in C++ and an automatic translator to fill the synthesis result of lifting problems into this template.
        \item For each task in our dataset, we construct a test suite by randomly generating 5 different inputs. These inputs are generated in the same way as random examples (Section \ref{section:implementation}), except the length of lists is fixed to $10^7$. We believe such an input scale is large enough to reflect the efficiency difference between different programs.
        \item For each successful synthesis, we first complete the synthesis result into an executable program using the translator of the corresponding paradigm, then execute this program on every input in the test suite of the corresponding task, and at last record the average time cost on these inputs (with the IO cost excluded). 
    \end{enumerate}
\end{itemize}
}


\smallskip
\begin{table*}
\jryadd{
	\caption{The results of comparing \mainname with \m{Enum} and \m{Relish}.}
    \renewcommand\arraystretch{1.15} 
	\begin{spacing}{1}
		\small
		\begin{tabular}{|c|c|c|c|c|c|c|c|c|c|}
			\Xhline{1pt}
            \multirow{2}{*}{Solver} & \multicolumn{4}{c|}{D\&C} & \multicolumn{4}{c|}{Single-pass}  \\
            \cline{2-9}
            & \#Solved & $\text{Time}_{\texttt{Base}}$ & $\text{Time}_{\texttt{Ours}}$  & $\text{Time}_{\texttt{Res}}$ & \#Solved & $\text{Time}_{\texttt{Base}}$ & $\text{Time}_{\texttt{Ours}}$ &  $\text{Time}_{\texttt{Res}}$ \\
            \Xhline{1pt}
            \mainname & \textbf{29}/36 & \multicolumn{2}{c|}{20.01} & $\times$1.000 &\textbf{33}/39 & \multicolumn{2}{c|}{8.861} & $\times$\textbf{1.000} \\
            \hline
            \textit{Enum}  & 5/36 & 46.76 & \textbf{0.247} & $\times$\textbf{0.981} & 9/39 & 9.544 & \textbf{0.337} & $\times$1.090  \\
            \hline
            \textit{Relish} & 12/36 & 34.75 & \textbf{7.283} & $\times$0.986& 16/39 & 18.03 & \textbf{4.366} & $\times$1.018  \\
            \Xhline{1pt}
            \multirow{2}{*}{Solver} & \multicolumn{4}{c|}{Longest Segment} & \multicolumn{4}{c|}{Segment Tree}  \\
            \cline{2-9}
            & \#Solved & $\text{Time}_{\texttt{Base}}$ & $\text{Time}_{\texttt{Ours}}$  &  $\text{Time}_{\texttt{Res}}$ &\#Solved & $\text{Time}_{\texttt{Base}}$ & $\text{Time}_{\texttt{Ours}}$ &  $\text{Time}_{\texttt{Res}}$  \\
            \Xhline{1pt}
            \mainname & \textbf{7}/8 & \multicolumn{2}{c|}{22.87} &  $\times$1.000& \textbf{13}/13 & \multicolumn{2}{c|}{43.30} & $\times$\textbf{1.000} \\
            \hline
            \textit{Enum}  & 1/8 & 14.99 & \textbf{0.530} & $\times$0.981& 4/13 & 115.1 & \textbf{19.84} & $\times$1.001\\
            \hline
            \textit{Relish} &3/8 & 4.229 & \textbf{2.377} & $\times$\textbf{0.980} & 7/13 & 86.47 & \textbf{42.20} & $\times$1.002\\
            \Xhline{1pt}
            \multirow{2}{*}{Solver} & \multicolumn{4}{c|}{Total} & \multicolumn{4}{c}{} \\
            \cline{2-5}& \#Solved & $\text{Time}_{\texttt{Base}}$ & $\text{Time}_{\texttt{Ours}}$  &  $\text{Time}_{\texttt{Res}}$ & \multicolumn{4}{c}{} \\
            \Xcline{1-5}{1pt}
            \mainname & \textbf{82}/96 & \multicolumn{2}{c|}{20.17} & $\times$\textbf{1.000}& \multicolumn{4}{c}{} \\
            \cline{1-5}
            \m{Enum} & 19/96 & 41.84 & \textbf{4.430} & $\times$1.035 & \multicolumn{4}{c}{} \\
            \cline{1-5}
            \m{Relish} & 38/96 & 34.98 & \textbf{14.07} & $\times$1.002 & \multicolumn{4}{c}{} \\
            \Xcline{1-5}{1pt}
		\end{tabular}
    \label{table:exp1} 
	\end{spacing}
}
\end{table*}
\jrydel{
\noindent \textbf{Results}. The results of this experiment are summarized in Table \ref{table:exp1}. For each solver, we report the number of solved tasks in column \#Solved, its average time cost (seconds) on solved tasks in column $T_{\texttt{Base}}$, and the average time cost of \mainname on the same tasks in column $T_{\texttt{Ours}}$. We conduct the two manual analyses below on the synthesis results.
\begin{itemize}
    \item We manually verify all results and confirm that they are all \textbf{completely correct}, though the verifier in our implementation provides only a probabilistic correctness guarantee\footnote{A gold medal winner in international programming competitions helped us to verify the synthesized programs.}.
    \item We manually verify the applications of our decomposition methods in every execution of \mainname and confirm that no unrealizable subtask is generated from realizable lifting problems, which matches the conclusion of our probabilistic analysis (Theorem \ref{theorem:completeness}).
\end{itemize} 

The results show that \mainname significantly outperforms the baseline solvers. It not only solves many more tasks but also solves much faster on those jointly solved tasks. }

\jryadd{
\noindent \textbf{Result}. The results of this experiment are summarized in Table \ref{table:exp1}, organized as follows.
\begin{itemize}
    \item Column \#Solved reports the number of tasks solved by each solver.
    \item Columns $\text{Time}_{\texttt{Base}}$ and $\text{Time}_{\texttt{Ours}}$ report the average time costs of each baseline solver and \mainname, respectively. Only those tasks solved by both the baseline solver and \mainname will be considered when calculating these time costs.
    \item Column $\text{Time}_{\texttt{Res}}$ reports the relative time cost of the resulting programs of each baseline solver compared with the resulting programs of \mainname. Specifically, for each task solved by both the baseline solver and \mainname, we calculate the relative time cost as ratio $t_{\texttt{Base}}/t_{\texttt{Ours}}$, where $t_{\texttt{Base}}$ and $t_{\texttt{Ours}}$ denote the time costs of the resulting programs of the baseline solver and \mainname, respectively. Then, we report the geometric average of these ratios in Column $\text{Time}_{\texttt{Res}}$. Here, an average ratio larger than $1$ means that the resulting program of \mainname is more efficient than that of the baseline solver.
\end{itemize}

Besides, we also manually analyze the synthesis results from two aspects, as shown below.
\begin{itemize}
    \item Since the verifier in our implementation (which is a combination of bounded model checking and random testing, Section \ref{section:implementation}) does not ensure the full correctness, we manually verify all synthesis results and confirm that they are all \textbf{completely correct}\footnote{A gold medal winner in international programming competitions helped us to verify the synthesized programs.}.
    \item We manually verify the applications of our decomposition methods in every execution of \mainname and confirm that no unrealizable subtask is generated from realizable lifting problems, which matches our probabilistic completeness guarantee (Theorem \ref{theorem:completeness}).
\end{itemize}

\smallskip

The results in Table \ref{table:exp1} demonstrate that \mainname significantly outperforms the baseline solvers. In the sense of solving lifting problems, \mainname not only solves many more tasks but also uses a much smaller time cost; and in the sense of synthesizing efficient programs, there is no significant efficiency difference between the resulting programs of \mainname and those of the baseline solvers, where the relative difference never exceeds $2\%$. 

It is expected that there is no significant efficiency gap between the resulting programs. Specifically, the efficiency of a program is roughly determined by two factors, its time complexity and the constant factor in its time cost. Both of these factors must be the same (or close) when comparing the resulting programs of \mainname, \m{Enum}, and \m{Relish}.
\begin{itemize}
    \item The time complexity of these resulting programs must be the same because our efficiency condition (Section \ref{subsection:efficiency-guarantee}) ensures that, for any paradigm considered in our evaluation, the time complexity of the resulting program must be the same no matter which $\m{aux}$ and $\m{comb}$ are synthesized from the domain-specific languages we use (Figures \ref{figure:gc} and \ref{figure:gf}). 
    \item The constant factor of these resulting programs must be close because (1) the constant factor is majorly determined by the time cost of the synthesized \m{comb}, (2) the time cost of \m{comb} is closely related to its size, and (3) solvers \mainname, \m{Enum}, and \m{Relish} all ensure to synthesize a \m{comb} whose size is close to the smallest. Specifically, both \m{Enum} and \m{Relish} ensure that the total size of the synthesized \m{aux} and \m{comb} is minimized, and \m{PolyGen} (the client synthesizer in \mainname for synthesizing \m{comb}) ensures that the size of the synthesized program is close to the smallest under the theory of Occam learning~\cite{DBLP:journals/ipl/BlumerEHW87}.
\end{itemize}
}

\mainname fails on 14 out of 96 tasks in our dataset, all of which are unrealizable because of the limited expressiveness of the default languages used in our implementation. These tasks require specialized operators such as regex matching on an integer list and the power operator on integers. These operators are not included in the general-purpose languages we used since they are not common in the domains of lists and integer arithmetic.

After supplying missing operators, \mainname can solve 13 more tasks and find a valid auxiliary program for the last remaining task.  The last failed task is \textit{longest\_odd\_(0+1)} constructed by \citet{toronto21}, on which \mainname fails because \m{PolyGen} times out in finding a corresponding combinator. This result suggests that \mainname can be further improved if missing operators can be inferred automatically, for example, by incorporating those transformation-based approaches and extracting useful operators from the source code. This is a direction for future work.

\subsection{RQ2: Comparison with Synthesizers for Divide-and-Conquer} \label{section:rq2}
\jrydel{
\textbf{Procedure}. We compare \mainname with the two versions of \textit{Parsynt} on D\&C tasks in our dataset and provide a single-pass implementation for each task to invoke \m{Parsynt}. This comparison favors \m{Parsynt} because it can access those auxiliary values provided in the single-pass implementation.

We failed in installing \textit{Parsynt17} because of some dependency issue, which is confirmed by the authors of \m{Parsynt17} but has not been solved yet. Therefore, we compare with \textit{Parsynt17} only on its original dataset using the evaluation results reported by \citet{DBLP:conf/pldi/FarzanN17}. 

Similar to the previous experiment, we use a time limit of $300$ seconds and a memory limit of $8$ GB and record the time cost of each successful synthesis. Please note that there is no difference between the time complexity of the synthesized programs because both \mainname and \m{Parsynt} ensure the time complexity to be exactly $\Theta(n/p)$ when a parallel template is used. \smallskip}

\jryadd{
\noindent \textbf{Procedure}. In this experiment, we compare \mainname with \m{Parsynt}, the baseline solver specialized for D\&C, on all D\&C tasks in our dataset. The details of the experiment setup are shown below.
\begin{itemize}
    \item Since \m{Parsynt} requires the original program to be single-pass (Section \ref{subsection:setup}), we provide a single-pass implementation of the original program for each task to invoke \m{Parsynt}\footnote{For those tasks taken from \textit{Parsynt}, we use the program in its original evaluation and fix the two bugs we found.}. Note that this setting favors \m{Parsynt} because (1) many programs cannot be implemented as single-pass unless some auxiliary values are manually introduced (Section \ref{subsection:moti-limitation}), and (2) \m{Parsynt} can access those auxiliary values provided in the single-pass implementation. 
    \item Since there are two versions of \m{Parsynt} available (i.e., \m{Parsynt17} and \m{Parsynt21}, mentioned in Section \ref{subsection:setup}), we consider both of them in this experiment. 
    \item We failed in installing \m{Parsynt17} because of some dependency issue. This issue has been confirmed by the authors of \m{Parsynt17} but has not been solved yet. Therefore, we compare with \m{Parsynt} only on its original dataset (which is a subset of ours) using the evaluation results reported by its original paper~\cite{DBLP:conf/pldi/FarzanN17}.  
    \item Similar to the first experiment (Section \ref{subsection:rq1}), we consider three metrics when comparing \mainname with \m{Parsynt}, including the number of solved tasks, the time cost of successful synthesis, and the time cost of the resulting programs. When evaluating the time cost of the resulting programs, we will fill the synthesis result of \m{Parsynt} into our template of D\&C because the original implementation of \m{Parsynt} does not provide a default template.
\end{itemize}
}

\smallskip
\jrydel{
\begin{table*}[t]
    \jrydel{
    \caption{The results of comparing \mainname with \textit{Parsynt}.}
    \renewcommand\arraystretch{1.15}
    \begin{spacing}{1}
        \small
        \begin{tabular}{|c|c|c|c|c|c|c|}
            \Xhline{1pt}
            Solver & \#Tasks  & $\text{\#S}_{\texttt{Base}}$ & $\text{\#S}_{\texttt{Ours}}$ &$T_{\texttt{Base}}$ & $T_{\texttt{Ours}}$ & $\#\text{Aux}_{\texttt{SP}}$ \\
            \Xhline{1pt}
            \textit{Parsynt17} & 20 & \textbf{19} & \textbf{19} & 15.6 & \textbf{5.84} & 39.3\% \\
            \hline
            \textit{Parsynt21} & 36 & 24 & \textbf{29} & 6.86 & \textbf{4.19}& 56.9\% \\
            \Xhline{1pt}
        \end{tabular}
    \end{spacing}
    \label{pre-table:exp2} 
    }
\end{table*}
}

\begin{table*}[t]
    \jryadd{
    \caption{The results of comparing \mainname with \textit{Parsynt}.}
    \renewcommand\arraystretch{1.15}
    \begin{spacing}{1}
        \small
        \begin{tabular}{|c|c|c|c|c|c|c|}
            \Xhline{1pt}
            Solver & $\text{\#Solved}_{\texttt{Base}}$ & $\text{\#Solved}_{\texttt{Ours}}$ &$\text{Time}_{\texttt{Base}}$ & $\text{Time}_{\texttt{Ours}}$ & $\text{Time}_{\texttt{Res}}$& $\#\text{Aux}_{\texttt{SP}}$ \\
            \Xhline{1pt}
            \textit{Parsynt17} &  \textbf{19}/20 & \textbf{19}/20 & 15.59 & \textbf{8.552} & N/A & 58.62\% \\
            \hline
            \textit{Parsynt21} & 24/36 & \textbf{29}/36 & \textbf{6.856} & 7.315& $\times$1.201 & 40.54\%\\
            \Xhline{1pt}
        \end{tabular}
    \end{spacing}
    \label{table:exp2} 
    }
\end{table*}

\jryadd{
\noindent \textbf{Results}. The results of this experiment are summarized in Table \ref{table:exp2}, organized as follows.
\begin{itemize}
    \item Columns \#Solved$_{\texttt{Base}}$ and \#Solved$_{\texttt{Ours}}$ report the number of tasks solved by each version of \m{Parsynt} and \mainname, respectively.
    \item Columns \#Time$_{\texttt{Base}}$ and \#Time$_{\texttt{Ours}}$ report the average time costs of \m{Parsynt} and \mainname, respectively, and Column \#Time$_{\texttt{Res}}$ report the relative time cost of the resulting programs of \m{Parsynt} compared with the resulting programs of \mainname. Values in these columns are calculated in the same way as the corresponding columns in Table \ref{table:exp1}. 
    \item Column \#Aux$_{\texttt{SP}}$ report the ratio of the number of auxiliary values provided in the input single-pass program to the number of auxiliary values used in the D\&C program synthesized by \m{Parsynt}. A larger value here means that \m{Parsynt} requires more extra inputs.
\end{itemize}
Besides, the value of cell $(\m{Parsynt17}, \text{Time}_{\texttt{Res}})$ in this table is unavailable because the original paper of \m{Parsynt17} does not provide the full synthesis results.

\smallskip

The results in Table \ref{table:exp2} show that compared with \m{Parsynt}, \mainname can offer competitive performance on synthesizing D\&C programs while using significantly less information from the input. Specifically, when compared with \m{Parsynt17}, \mainname solves the same number of tasks with a smaller time cost; and when compared with \m{Parsynt21}, \mainname solves more tasks and synthesizes more efficient D\&C programs, though requiring slightly more time for synthesis. Please note that in this comparison, \m{Parsynt} takes much more input than \mainname, including $40.54\%$-$58.62\%$ of those necessary auxiliary values.

\smallskip 

Now, we would like to discuss more the efficiency of the resulting programs. Table \ref{table:exp2} shows that, although \mainname and \m{Parsynt21} provide the same guarantee on the time complexity of the resulting programs (Section \ref{subsection:setup}), the resulting programs of \mainname tend to be more efficient than those of \m{Parsyn21}, with a relative advantage of about $20\%$. One important reason for this result is that the resulting programs of \m{Parsynt21} tend to use more auxiliary values than those of \mainname, leading to an extra time cost for calculating auxiliary values. In this experiment, the resulting programs of \m{Parsyn21} never use fewer auxiliary values than those of \m{AutoLifter} and use more auxiliary values on 10 tasks.

\m{Parsynt21} tends to use more auxiliary values because its syntax-based program transformations may be misled by the source code of the original program. For example, the original program of task \textit{line\_sight} (abbreviated as \textit{ls}) checks whether the last element is the maximum of the list. It can be implemented as single-pass with an auxiliary program $\textit{max}$ returning the maximum of a list, because $\m{ls}\ (\quanti{\m{xs}} \cat [\quanti{v}]) = \quanti{v} \geq (\m{max}\ \quanti{\m{xs}})$. Given this program, \textit{Parsynt} will extract the last element of a list as an auxiliary value because the last visited element $v$ is used in the loop body. However, this value is not necessary because
$
\textit{ls}\ (l_1 \cat l_2)$ is always equal to $(\textit{ls}\ l_2) \wedge (\textit{max}\ l_1 \leq \textit{max}\ l_2) 
$. \mainname can generate this simpler solution because it finds auxiliary values by enumerating programs in $\mathcal L_{\textit{aux}}$ instead of transforming the source code of the original program.

\smallskip 

At last, we find that when applying D\&C, the issue of missing operators on \mainname (Section \ref{subsection:rq1}) can be alleviated by combining \mainname with \m{Parsynt21}. Although the default languages we use are not expressive enough for 7 D\&C tasks in our dataset, our languages are enough for applying single-pass to the original programs of 5 tasks among these tasks. \mainname can successfully synthesize single-pass programs for these 5 tasks, and then \textit{Parsynt21} can synthesize D\&C programs for 4 among them. In this way, the combination of \mainname and \m{Parsynt21} can solve 33 out of 36 tasks, outperforming both individual solvers.
}

\subsection{RQ3: Comparison with Synthesizers for Single-Pass} \label{subsection:rq3}

\jrydel{\noindent \textbf{Procedure}. We compare \mainname with \m{DPASyn} on all single-pass tasks in our dataset with a time limit of 300 seconds and a memory limit of 8 GB. The time cost of each successful synthesis is recorded to measure the efficiency of the solvers. Please note that there is no difference between the time complexity of the synthesized programs because both \mainname and \m{DPASyn} ensure the time complexity to be exactly $\Theta(n)$. 

Besides, we also consider an enhanced configuration of \m{DPASyn} (denoted as $\m{DPASyn}_+$) in this experiment, where more compact program spaces are used. As a Sketch-based synthesizer, the time cost of $\m{DPASyn}$ increases dramatically when the scale of the target program increases. However, the program space $\mathcal L_{\textit{comb}}$ we used (Figure \ref{figure:gc})\footnote{\m{DPASyn} does not synthesize the auxiliary program explicitly and thus never use the other program space $\mathcal L_{\textit{aux}}$.} is so basic that a large program may be required for some simple functions. For example, the maximum of two integers $\max(a, b)$ has to be implemented as $\m{ite}(a \leq b, a, b)$ in $\mathcal L_{\textit{comb}}$. Therefore, to better reveal the ability of \m{DPASyn}, we customize the program space for each task when evaluating $\m{DPASyn}_+$, where (1) operators $\max$ and $\min$ are available, and (2) only those necessary operators are included. Note that the comparison between \mainname and $\m{DPASyn}_+$ favors the latter because $\m{DPASyn}_+$ explores a much smaller program space.
\smallskip}

\jryadd {
\noindent \textbf{Procedure}. In this experiment, we compare \mainname with \m{DPASyn}, the baseline solver specialized for single-pass, on all single-pass tasks in our dataset. Similar to the previous experiments, we consider three metrics in this experiment, including the number of solved tasks, the time cost of successful synthesis, and the time cost of the resulting programs. When evaluating the time cost of the resulting programs, we will fill the synthesis result of \m{DPASyn} into our template of single-pass because \m{DPASyn} does not provide a default template.

We consider two different configurations of \m{DPASyn} in this comparison, a normal configuration (denoted as \m{DPASyn}$_{=}$) for establishing a fair comparison and an enhanced configuration (denoted as \m{DPASyn}$_{+}$) for better revealing the performance of \m{DPASyn}.
\begin{itemize}
    \item \m{DPASyn}$_{=}$ uses the same language as \mainname to specify the program space.
    \item \m{DPASyn}$_{+}$ uses a more compact program space. We consider this configuration because, as a Sketch-based synthesizer, the time cost of \m{DPASyn} will increase dramatically when the scale of the target program increases. However, the language $\mathcal L_{\textit{comb}}$ we use (Figure \ref{figure:gc}) is so basic that a large program may be necessary for even simple tasks. Therefore, to better evaluate \m{DPASyn}, we customize the program space of each task for \m{DPASyn}$_{+}$ by (1) adding operators $\max$ and $\min$ and (2) excluding those operators that are not used.
\end{itemize}
Note that the comparison between \mainname and \m{DPASyn}$_+$ favors the latter because $\m{DPASyn}_+$ needs only to explore a much smaller program space.
}

\smallskip

\jrydel{
\noindent \textbf{Results}. The result are summarized in the right-side table, organized similarly to the table of the first experiment (Table \ref{table:exp1}, Section \ref{subsection:rq1}). These results demonstrate that \mainname significantly outperforms both versions of \textit{DPASyn} on both the number of solved tasks and the efficiency. }

\begin{wrapfigure}[]{r}{0.55\textwidth}
    \vspace{-0.7em}
    \makeatletter\def\@captype{table}\makeatother
    \centering
    \renewcommand\arraystretch{1.15}
    \jryadd{
	\begin{spacing}{1}
		\small
		\begin{tabular}{|c|c|c|c|c|}
			\Xhline{1pt}
            Solver & \#Solved& $\text{Time}_{\texttt{Base}}$ & $\text{Time}_{\texttt{Ours}}$  &  $\text{Time}_{\texttt{Res}}$   \\
            \Xhline{1pt}
            \mainname & \textbf{33}/39 & \multicolumn{2}{c|}{8.861} & $\times$1.000 \\
            \hline 
            \textit{DPASyn}$_=$ & 15/39  & 10.32  & \textbf{4.685} & $\times$1.019\\
            \hline 
            $\textit{DPASyn}_+$ & 21/39  & 27.74  & \textbf{3.951} & $\times$1.044\\
            \Xhline{1pt}
		\end{tabular}
	\end{spacing}}
\end{wrapfigure}
\noindent \jryadd{\textbf{Results}. The results of this experiment are summarized in the right-side table, organized in the same way as the table of the first experiment (Table \ref{table:exp1}, Section \ref{subsection:rq1}). These results show that \mainname outperforms both versions of \m{DPASyn}. Specifically, \mainname solves more tasks with a smaller time cost, and there is no significant efficiency difference between the resulting programs of \mainname and \m{DPASyn}.}

The advantage of \mainname majorly comes from its decomposition system, which decomposes the original task into subtasks on sub-programs with much smaller scales. In contrast, \textit{DPASyn} directly searches for the whole target program, leading to a combinatorially larger search space.

\subsection{RQ4: Comparison with the Variant without Observational Covering}

\jrydel{
\noindent \textbf{Procedure}. \mainname involves a specialized optimization named observational covering when solving the leaf subtasks of \m{aux} (Section \ref{subsection:inductive}). To test the effect of this optimization, we consider a variant of \mainname where leaf subtasks of \m{aux} are directly solved by OE (denoted as $\mainname_{\text{OE}}$) and compare it with the default \mainname. Similar to previous experiments, we use a time limit of 300 seconds and a memory limit of 8 GB and record the time cost of each successful synthesis. \smallskip}

\jryadd{
\noindent \textbf{Procedure}. \mainname uses a specialized optimization named observational covering when synthesizing \m{aux} (Section \ref{subsection:inductive}). In this experiment, we conduct an ablation study to test the effect of this optimization. Specifically, we consider a variant of \mainname (denoted as $\mainname_{\texttt{OE}}$) where the leaf subtasks of \m{aux} are solved by the pure OE without using observational covering. Then, we compare this variant with the default \mainname on all tasks in our dataset. Similar to the previous experiments, we consider three metrics in this comparison, including the number of solved tasks, the time cost of successful synthesis, and the time cost of the resulting programs.

}
\smallskip

\begin{table*}
    \jryadd{
	\caption{The results of comparing \mainname with $\mainname_{\text{OE}}$.}
        \renewcommand\arraystretch{1.15} 
        \begin{spacing}{1}
            \small
            \begin{tabular}{|c|c|c|c|c|c|c|c|c|c|}
                \Xhline{1pt}
                \multirow{2}{*}{Solver} & \multicolumn{4}{c|}{D\&C} & \multicolumn{4}{c|}{Single-pass}  \\
                \cline{2-9}
                & \#Solved & $\text{Time}_{\texttt{Base}}$ & $\text{Time}_{\texttt{Ours}}$  & $\text{Time}_{\texttt{Res}}$ & \#Solved & $\text{Time}_{\texttt{Base}}$ & $\text{Time}_{\texttt{Ours}}$ &  $\text{Time}_{\texttt{Res}}$ \\
                \Xhline{1pt}
                \mainname & \textbf{29}/36 & \multicolumn{2}{c|}{30.79} & $\times$1.000 &\textbf{33}/39 & \multicolumn{2}{c|}{8.861} & $\times$1.000 \\
                \hline
                $\mainname_{\texttt{OE}}$  & 13/36 & 3.329 & \textbf{1.010} & $\times$\textbf{0.985} & 28/39 & 5.563 & \textbf{4.141} & $\times$\textbf{0.954} \\
                \Xhline{1pt}
                \multirow{2}{*}{Solver} & \multicolumn{4}{c|}{Longest Segment} & \multicolumn{4}{c|}{Segment Tree}  \\
                \cline{2-9}
                & \#Solved & $\text{Time}_{\texttt{Base}}$ & $\text{Time}_{\texttt{Ours}}$  &  $\text{Time}_{\texttt{Res}}$ &\#Solved & $\text{Time}_{\texttt{Base}}$ & $\text{Time}_{\texttt{Ours}}$ &  $\text{Time}_{\texttt{Res}}$  \\
                \Xhline{1pt}
                \mainname & \textbf{7}/8 & \multicolumn{2}{c|}{22.87} & $\times$1.000 & \textbf{13}/13 & \multicolumn{2}{c|}{43.30} & $\times$1.000 \\
                \hline
                $\mainname_{\texttt{OE}}$  & \textbf{7}/8 & 46.79 & \textbf{22.87} & $\times$\textbf{0.968}& 8/13 & 48.69 & \textbf{39.66} & $\times$1.011  \\
                \Xhline{1pt}
                \multirow{2}{*}{Solver} & \multicolumn{4}{c|}{Total} & \multicolumn{4}{c}{} \\
                \cline{2-5}& \#Solved & $\text{Time}_{\texttt{Base}}$ & $\text{Time}_{\texttt{Ours}}$  &  $\text{Time}_{\texttt{Res}}$ & \multicolumn{4}{c}{} \\
                \Xcline{1-5}{1pt}
                \mainname & \textbf{82}/96 & \multicolumn{2}{c|}{23.27} &$\times$1.000 & \multicolumn{4}{c}{} \\
                \cline{1-5}
                $\mainname_{\texttt{OE}}$ & 56/96 & 16.36 & \textbf{10.83} & $\times$\textbf{0.971}& \multicolumn{4}{c}{} \\
                \Xcline{1-5}{1pt}
            \end{tabular}
        \label{table:exp4} 
        \end{spacing}
    }
    \end{table*}

\noindent \textbf{Results}. The results of this experiment are summarized in Table \ref{table:exp4}, organized in the same way as the table of the first experiment (Table \ref{table:exp1}, Section \ref{subsection:rq1}). These results demonstrate that observational covering significantly improves the efficiency of \mainname. 

Note that even when observational covering is removed, $\mainname_{\text{OE}}$ still outperforms \textit{Enum} and \textit{Relish} (Table \ref{table:exp1}) and outperforms \textit{DPASyn} on synthesizing single-pass programs (Table \ref{table:exp4}). This result also demonstrates the effectiveness of our decomposition system.

\subsection{\jrymod{Case Study}{Qualitative Analysis of Selected Tasks}}\label{subsection:case-study}
\jrymod{We also conduct a case study on two tasks in our dataset, showing (1) the advantage of inductive synthesis and (2) the ability of \mainname to solve tasks difficult for human programmers.}{To further illustrate the effectiveness of \mainname, we shall discuss two tasks in our dataset and show that (1) \mainname can solve tasks that are difficult for previous transformation-based approaches, and (2) \mainname can solve algorithmic tasks that are different for human programmers.}

\smallskip 
\noindent \textbf{Maximum segment product}.
The first task is named as \textit{maximum segment product (msp)}~\cite{note1989}, which is an advanced version of \textit{mss} (Section \ref{subsection:example2}). Given list $xs[1 \dots n]$, the problem is to select a segment $s$ from $xs$ and maximize the product of values in $s$. 

It is not easy to calculate the maximum segment product by D\&C. According to the experience in solving the \textit{mss} task, one may choose the maximum prefix/suffix product as the auxiliary values. However, these two values are not enough. The counter-intuitive point here is that the maximum segment product is also related to the \textbf{minimum} prefix/suffix product. This is because both the minimum suffix product of the left half and the minimum prefix product of the right half can be negative integers with large absolute values. Their product will flip back the sign, resulting in a large positive number. For example, the segment with the maximum product of $[-1, -5] \cat [-3, 0]$ is $[-5, -3]$, formed by the suffix with the minimum product of the left half (i.e., $[-5]$) and the prefix with the minimum product of the right half (i.e., $[-3]$).

\textit{Parsynt} fails to solve this task as its transformation rules are not enough to extract these auxiliary values (related to the minimum) from the original program (related to the maximum). In contrast, \mainname successfully solves this task in $287.9$ seconds (where $113.9$ seconds are used by bounded model checking) and finds an auxiliary program as follows.
\begin{align*}
\m{aux}\ xs \triangleq \big(\textit{max}\ (\textit{scanl}\ (\times)\ \m{xs}), \textit{max}\ (\textit{scanr}\ (\times)\ \m{xs}), ~&\textit{min}\ (\textit{scanl}\ (\times)\ \m{xs}),  \\
&\textit{min}\ (\textit{scanr}\ (\times)\ \m{xs}), \textit{head}\ (\textit{scanr}\ (\times)\ \m{xs})\big)
\end{align*}
This program calculates five auxiliary values, corresponding to the maximum prefix product, the maximum suffix product, the minimum prefix product, the minimum suffix product, and the product of all elements, respectively. We omit the combinator synthesized by \mainname because it is large in scale but is straightforward from the synthesized auxiliary program.

\smallskip

\noindent \textbf{Longest segment problem 22-2}.
The second problem is proposed by \citet{DBLP:journals/scp/Zantema92}, which is used as the second example on Page 22 of that paper. This problem is to find a linear-time program for the length of the longest segment $s$ satisfying $\textit{min}\ s + \textit{max}\ s > \textit{length}\ s$ for a given list.

This problem is difficult even for professional programmers in competitive programming. It was set as a problem in 2020-2021 Winter Petrozavodsk Camp, a worldwide training camp representing the highest level of competitive programming. Only $26$ out of $243$ participating teams solved this problem within the 5-hour competition.

The third algorithmic paradigm proposed by \citet{DBLP:journals/scp/Zantema92} can be applied to solve this problem. The synthesis task is to find an auxiliary program $\m{aux}$ and a combinator $\m{comb}$ such that the formula below is satisfied for any lists $xs_L, xs_R$ and integer $v$ satisfying $v < \textit{min}\ xs_L \wedge v \leq \textit{min}\ xs_R$.
$$
(\bluec{\targetname} \spl \var{aux})\ (\quanti{\m{xs}_L} \cat [\quanti{v}] \cat \quanti{\m{xs}_R}) = \var{\m{comb}}\ \big(\quanti{v}, \big((\bluec{\targetname}\spl\var{\m{aux}})\ \quanti{\m{xs}_L}, (\bluec{\targetname} \spl \var{\m{aux}})\ \quanti{\m{xs}_R}\big)\big)
$$
where $\targetname$ is an arbitrary reference program returning the length of the longest valid segment. However, it is difficult to find proper $\m{aux}$ and $\m{comb}$ satisfying the above formula. We encourage the readers to try to solve this task before moving to the discussion below. 

\mainname can find an auxiliary program $\m{aux}\ xs \triangleq (\textit{length}\ xs, \textit{max}\ xs)$ and a correct combinator $\m{comb}$ in $100.0$ seconds. The synthesized $\m{comb}$ includes $152$ AST nodes and is formed by several components dealing with different cases. Here, we only explain the component for calculating the expected output under the condition that $\m{max}\ \m{xs}_L \geq \m{max}\ \m{xs}_R$, as shown below. 
\begin{align*}
    &\m{comb}\ (v, (\m{res}_L, \m{res}_R))\!
    \triangleq\! \left\{\begin{array}{ccc}
        \!\!\max(\textit{lsp}_R, \text{min}(\textit{len}_L\! +\! \textit{len}_R\! +\! 1, v\! +\! \textit{max}_L\! -\! 1)) & v\! +\! \textit{max}\ \m{xs}_L > \textit{length}\ \m{xs}_L\! +\! 1  \\
        \!\!\max(\m{lsp}_L, \m{lsp}_R) &  \textit{otherwise} \\
      \end{array}\right.& \\
    &\qquad \textbf{where } \m{res}_L \text{ is unfolded to } (\m{lsp}_L, (\m{len}_L, \m{max}_L)), \m{res}_R \text{ is unfolded to } (\m{lsp}_R, (\m{len}_R, \m{max}_R)) \\
    &\qquad \textbf{assuming }\m{max}\ \m{xs}_L \geq \m{max}\ \m{xs}_R
\end{align*}

\textbf{Case 1}: $(\m{max}\ \m{xs}_L \geq \m{max}\ \m{xs}_R) \wedge (v + \textit{max}\ \m{xs}_L > \textit{length}\ \m{xs}_L + 1)$. 
    There are only three possible cases for the longest valid segment: the longest valid segment $s_L$ in $\m{xs}_L$, the longest valid segment $s_R$ in $\m{xs}_R$, or the longest valid segment $s_v$ including element $v$. In this case, $s_L$ is no longer than $s_v$ because segment $\m{xs}_L \cat [v]$ is valid under the condition that $v + \textit{max}\ \m{xs}_L > \textit{length}\ \m{xs}_L + 1$. Therefore, the longest valid segment of the whole list must be the longer one between $s_R$ and $s_v$. Since the length of $s_R$ is known as $\m{lsp}_R$, the remaining task is to get the length of $s_v$.
    
    An observation is that segment $\m{xs}_L \cat [v]$ already achieves the maximum possible $\m{min}\ s + \m{max}\ s$ among segments including $v$ because (1) $\m{min}\ s$ must be $v$, the minimum of the whole list, and (2) $\m{max}\ s$ must be no larger than $\m{max}\ \m{xs}_L$, the maximum of the whole list under the condition that $\m{max}\ \m{xs}_L \geq \m{max}\ \m{xs}_R$. Therefore, $s_v$ is the longest segment expanded from $\m{xs}_L \cat [v]$ until the length limit (i.e., $v + \m{max}\ \m{xs}_L$) is reached or the whole list is used up, that means, the length of $s_v$ must be $\min(\textit{len}_L + \textit{len}_R + 1, v + \textit{max}_L - 1)$.

    \smallskip 
    
\textbf{Case 2}: $(\m{max}\ \m{xs}_L \geq \m{max}\ \m{xs}_R) \wedge (v + \textit{max}\ \m{xs}_L \leq \textit{length}\ \m{xs}_L + 1)$. In this case, $s_v$ is no longer than $s_L$, so the longest valid segment of the whole list is the longer one between $s_L$ and $s_R$, and the result is $\max(\m{lsp}_L, \m{lsp}_R)$. This property can be proved in two steps. First, $s_v$ must be no longer than $\m{xs}_L$, as shown by the derivation below. 
    $$
    \m{length}\ s_v \leq v + \m{max}\ s_v -1 \leq v + \m{max}\ \m{xs}_L -1 \leq \m{length}\ \m{xs}_L   
    $$
    The first inequality uses the fact that $v$ is the minimum of the whole list, the second inequality uses the fact that $\m{max}\ \m{xs}_L$ is the maximum of the whole list under the condition that $\m{max}\ \m{xs}_L \geq \m{max}\ \m{xs}_R$, and the third inequality uses the condition that $v + \textit{max}\ \m{xs}_L \leq \textit{length}\ \m{xs}_L + 1$.

    Second, since $s_v$ is no longer than $\m{xs}_L$, there exists another segment $s'$ that includes the maximum of $\m{xs}_L$ and has the same length as $s_v$. As shown by the derivation below, $s'$ must be valid as well. Therefore, $s'$ is no longer than $s_L$, implying that $s_v$ is no longer than $s_L$.
    $$
    \m{min}\ s' + \m{max}\ s' \geq v + \m{max}\ s_v > \m{length}\ s_v = \m{length}\ s'
    $$

\smallskip
As we can see, the correct $\m{comb}$ here relies on several tricky properties. Finding this program is challenging for a human user. In comparison, \mainname can solve this problem quickly.  

\section{Discussion}\label{section:discussion} 

\jryadd{
\noindent \textbf{Selecting algorithmic paradigms.} As discussed in Section \ref{section:application}, \mainname can be instantiated as a series of synthesizers, each for applying a specific D\&C-like paradigm. However, the presence of multiple instantiations brings another problem in usage, that is, how to select a proper instantiation for a practical task. Currently, this selection problem is not yet an issue because the number of available instantiations of \mainname is not large ($6$ in our implementation): we can simply try all available instantiations in order or in parallel until the task of interest is solved. In the future, when more instantiations are developed, it may be necessary to design an automated approach to select among all possible choices. This is a direction for future work.

Besides, we believe \mainname can still have a significant practical effect even if the selection problem is left to the user.
\begin{itemize}
    \item On the one hand, some paradigms are so important that even a specialized synthesizer is still valuable. For example, the problem of automatic parallelization has long been studied in literature~\cite{DBLP:conf/pldi/MoritaMMHT07,DBLP:conf/pldi/FarzanN17,DBLP:conf/pldi/FedyukovichAB17,DBLP:conf/sosp/RaychevMM15}. The instantiation of \mainname on the D\&C paradigm solves this problem, and compared with previous approaches, this instantiation of \mainname is the first one that does not require the original program to be single-pass.
    \item On the other hand, it is usually not difficult for the user to select among paradigms because the application scope of different paradigms is usually clear. For example, the D\&C paradigm is usually used for parallelization, where the goal is to achieve a sublinear time complexity on multiple processors; the incrementalization paradigm is only available for incremental tasks where the original program will be executed multiple times on a series of similar inputs.
\end{itemize}}

\smallskip

\jryadd {\noindent \textbf{Verification}.  In this paper, we focus on designing an effective synthesizer for applying algorithmic paradigms and do not consider designing specialized verifiers. Instead, our approach can be combined with any off-the-shelf verifier to provide a correctness guarantee on the synthesis result.

Although our current implementation (Section \ref{section:implementation}, where bounded model checking is used) cannot ensure full correctness, it works well in our evaluation. As discussed in Section \ref{subsection:rq1}, we manually verify all synthesis results and confirm that all of them are completely correct. This result shows that our current implementation can already provide reliable synthesis results in practice. 

Besides, even when the user decides to manually verify the synthesized program, we believe this task will still be easier than solving the algorithmic task by the users themselves. On the one hand, an algorithmic task can be extremely difficult (e.g., the second task discussed in Section \ref{subsection:case-study}), and \mainname can provide a candidate solution for the user to check, which is correct with a high probability. On the other hand, \mainname ensures to synthesize simple programs (Theorem \ref{theorem:minimal} and the property of \m{PolyGen}~\cite{DBLP:journals/pacmpl/JiXXH21}), making its result usually easy to understand. 

In the future, we believe the ability of verifiers will be continuously improved and will be able to verify more and more complex programs. At that time, by combining with those more advanced verifiers, \mainname can potentially provide a stronger correctness guarantee on its result.

\smallskip 
}

\jryadd{
\noindent \textbf{The greedy strategy v.s. the backtracking strategy.} The effectiveness of \mainname comes from its decomposition system, which uses approximate specifications to break the dependency among sub-programs. There are two possible strategies for implementing such a decomposition system (Section \ref{subsection:properties}). The greedy strategy used in our implementation considers only the first solution of each subtask and will fail once an unrealizable subtask is generated. In comparison, the backtracking strategy will roll back and try to find other solutions each time an unrealizable subtask is met. It will not fail but will be inefficient if too many unrealizable subtasks are generated. 

At first glance, the backtracking strategy may seem like a good choice because it can realize a complete synthesis, that is, the synthesis will never fail on a realizable lifting problem. However, it is not easy to implement this strategy because there are still two challenges remaining.
\begin{itemize}
    \item First, to decide whether to roll back the decomposition, the backtracking strategy requires checking whether the current subtask is unrealizable. Although some research progress has been made~\cite{DBLP:conf/pldi/HuCDR20, DBLP:journals/pacmpl/KimHDR21}, proving unrealizability for a program synthesis task is still time-consuming in practice. 
    \item Second, some efficiency synthesis techniques will be unavailable for synthesizing $\m{aux}$ if we want to achieve completeness using the backtracking strategy. Specifically, the synthesizer for $\m{aux}$ needs to ensure that every possible auxiliary program will be considered after backtracking enough times, otherwise, some valid solutions may be missed. However, most synthesis techniques (e.g., observational equivalence and observational covering we used) fail to satisfy this requirement. They are designed only for synthesizing a single program and may skip many non-optimal programs during the synthesis for efficiency. 
\end{itemize}

Because of the above challenges in implementing the backtracking strategy, we use the greedy strategy to implement \mainname. Although in theory, the greedy strategy may fail in solving realizable lifting problems, both our probabilistic analysis (Section \ref{subsection:properties}) and our evaluation results (Section \ref{section:evaluation}) suggest that this failure seldom happens in practice.
}

\smallskip 

\noindent \textbf{The compressing property.} When analyzing the effectiveness of \mainname (Section \ref{subsection:properties}), we utilize the compressing property of practical lifting problems, that is, the original program and auxiliary programs usually map from a large input domain to a small output domain. Our analysis does not put a strict restriction on how much the input domain should be larger than the output domain, instead, it shows how the effectiveness of \mainname is gradually affected by a larger input domain. Specifically, when the input domain becomes larger (compared to the output domain), the mismatch factor of a lifting problem will become larger, making the unrealizable rate of \mainname become smaller (Theorem \ref{theorem:main-result}). This unrealizable rate will finally converge to $0$ when the size of the input domain approaches infinity (Theorem \ref{theorem:completeness}).

Please note that the compressing property is never a sufficient condition for \mainname to succeed in synthesis. It is still possible to construct a realizable lifting problem where the input domain is much larger than the output domain but \mainname fails. We study the compressing property in this paper only to explain the effectiveness of \mainname and clarify the boundary of \mainname.

\smallskip 

In the previous discussions, we simply regard those programs mapping from integer lists to a constant number of integers as compressing (Example \ref{example:compress} and Section \ref{section:implementation}). This claim holds only when the range of integers is assumed fixed and finite, but the practical situation is more complex. Many programs mapping from integer lists to integers will enlarge the range of integers, for example, the program calculating the sum of a list can return a number as large as $n \times m$ from a list whose length is no larger than $n$ and element integers are inside range $[-m, m]$. If such enlargement is not limited, non-compressing programs may exist, for example, there exist injective functions from integer lists to a single integer when the output integer is not bounded.

Luckily, the extreme case seldom happens in practice. Table \ref{table:enlarge} lists the output ranges of several programs mentioned before. As we can see, most of these programs enlarge the range of integers only polynomially and thus are still compressing because the number of integer lists is exponential to the list length and the input range. Even for the last program \m{msp} (maximum segment product, Section \ref{subsection:case-study}) that may return an exponentially larger integer, the compressing property still holds when the range of input integers is small. For example, the result of $\m{msp}$ must be in the form of $2^a3^b5^c$ for $0 \leq a \leq 2n$ and $0 \leq b, c \leq n$ when only integers within $[-5, 5]$ are used in the input. Consequently, \m{msp} maps $11^n$ different input lists to only $O(n^3)$ different outputs, leading to the compressing property.

\begin{table*}
    \renewcommand\arraystretch{1.15}
    \caption{The largest outputs from an integer list whose length $\leq n$ and element integers $\in [-m, m]$. }
    \label{table:enlarge} 
    \begin{spacing}{1}
        \small
        \begin{tabular}{|c|c|c|c|c|c|c|c|}
            \Xhline{1pt}
            Program & $\m{length}$  & $\m{min}$ & $\m{sndmin}$ & $\m{sum}$ & $\m{mps}$ & $\m{mss}$ & $\m{msp}$ (Section \ref{subsection:case-study}) \\
            \Xhline{1pt}
            Max Output & $n$ & \multicolumn{2}{c|}{$m$} & \multicolumn{3}{c|}{$n \times m$} & $n^m$ \\
            \Xhline{1pt}
        \end{tabular}
    \end{spacing}
\end{table*}

\smallskip 

Besides the case of mapping data structures to scalar values, we observe that the compressing property also holds for many programs mapping between data structures. For example, the sorting program maps all permutations of length $n$ ($n!$ possibilities in total) to the same output $[1, 2, \dots, n]$. This observation suggests interesting future work of applying \mainname to those tasks where not only auxiliary values but also auxiliary data structures are required. 
\section{Related Work}\label{section:related}
\noindent\textbf{Automatic applications of D\&C-like paradigms}. 
There have been previous studies on applying individual D\&C-like algorithmic paradigms. 

First, several approaches~\cite{DBLP:conf/pldi/MoritaMMHT07,DBLP:conf/pldi/FarzanN17,DBLP:conf/pldi/FedyukovichAB17,DBLP:conf/sosp/RaychevMM15} have been proposed to apply D\&C. All of these approaches are based on syntax-based program transformations and require the input program to be implemented as single-pass. 
Compared with them, \mainname does not require single-pass implementations but can still offer competitive performance compared with the previous state-of-the-art (Section \ref{section:rq2}).


\jryadd{
When applying D\&C, we assume the list is always divided from the middle and thus focus on synthesizing the auxiliary program $\m{aux}$ and the combinator $\m{comb}$. In this sense, \citet{toronto21} study the application of a more general form of D\&C where the divide operator is to be synthesized as well. Their approach and ours are complementary because their approach requires a single-pass implementation of the original program while \mainname does not. A possible future direction is to combine these approaches with \mainname.
}

Second, \citet{acar2005self} proposes an approach for incrementalization. This approach records the execution trace of the original program as the auxiliary value and lets the combinator re-evaluate only those operations affected by the change. Consequently, this approach can generate an efficient program only when the execution trace of the original program is affected little by the change. However, it can be difficult to satisfy this requirement. For example, in the incrementalization task for \m{sndmin} (Section \ref{subsection:incre-example}), the resulting program generated from the natural implementation of \m{sndmin} (Figure \ref{fig:smin}) will trace into the sorting function and thus runs in $O(\log n)$ time per change, much slower than the expected solution that runs in constant time per change (Figure \ref{fig:auxcomb4incre}).  

Both \mainname and \citet{acar2005self}'s approach have their advantages in automatic incrementalization. \mainname does not rely on the source code of the original program and thus can generate efficient results regardless of the user-provided implementation. However, when a proper original program is given, \citet{acar2005self}'s approach can construct incremental programs for extremely difficult tasks such as generating dynamic data structures requiring hundreds of lines of code~\cite{DBLP:journals/toplas/AcarBBHT09}, where even the decomposed subtasks generated by \mainname are still out of the scope of existing synthesizers. Scaling up inductive synthesis to these complex programs is future work.

Third, \citet{DBLP:conf/oopsla/PuBS11} propose an approach named \textit{DPASyn} to apply single-pass. This approach reduces the synthesis task to a Sketch problem and solves it via existing Sketch solvers. Compared with this approach, \mainname involves a decomposition system to decompose the synthesis task into subtasks with much smaller scales and thus greatly reduces the search space. Our evaluation results demonstrate the effectiveness of \mainname (Section \ref{subsection:rq3}).

\jrymod{Fourth}{At last}, there exist multiple approaches that do not support finding necessary auxiliary values when the paradigm cannot be directly applied. The related paradigms include D\&C~\cite{DBLP:conf/sigmod/AhmadC18,DBLP:conf/oopsla/RadoiFRS14,DBLP:conf/pldi/SmithA16}, structural recursion~\cite{DBLP:conf/cav/FarzanN21, DBLP:conf/pldi/FarzanLN22}, and incrementalization~\cite{DBLP:journals/lisp/LiuS03}. These approaches will fail when the output of the original program cannot be directly calculated, for example, when applying D\&C to \m{sndmin} (where the first minimum is required as an auxiliary value). Compared with these approaches, \mainname supports finding necessary auxiliary values.

\jrydel{
At last, in a lifting problem, we assume the operator $\m{op}$ is given and thus focus on synthesizing the auxiliary program $\m{aux}$ and the combinator $\m{comb}$. In this sense, there are two related studies on a more general task where the operator is to be synthesized as well~\cite{toronto21, DBLP:journals/pacmpl/MiltnerNBCD22}. Their approaches and ours are complementary because \citet{toronto21}'s approach requires a single-pass implementation and \citet{DBLP:journals/pacmpl/MiltnerNBCD22}'s approach does not support synthesizing the auxiliary program. A possible future direction is to combine these
approaches with \mainname.}


\smallskip

\noindent\textbf{Type- and resource-aware synthesis}. There is another line of work for synthesizing efficient programs, namely \textit{type- and resource-aware synthesis} ~\cite{DBLP:conf/pldi/KnothWP019,DBLP:journals/corr/abs-2103-04188}. These approaches use a type system to represent a resource bound, such as the time complexity, and use \textit{type-driven program synthesis}~\cite{DBLP:conf/pldi/PolikarpovaKS16} to find programs satisfying the given bound. 

Compared with \mainname, these approaches can deal with more refined efficiency requirements via advanced type systems. However, they suffer from a more serious scalability challenge because they need to synthesize the whole resulting program from the start. As far as we are aware, so far none of these approaches can scale up to applying algorithmic paradigms as our approach can.

\smallskip

\noindent\textbf{Program synthesis}. Program synthesis is an active field and many synthesizers have been proposed. Here we only discuss the most related approaches. 

\jrydel{
The divide-and-conquer-style synthesis framework of \mainname is similar to \textit{DraydSynth}~\cite{DBLP:conf/pldi/HuangQSW20}, which synthesizes programs by (1) transforming the synthesis task into separate subtasks by pre-defined rule, and (2) solving each subtask by enumerative solvers. However, the rules used in \textit{DryadSynth} are specialized for Boolean and arithmetic operators and thus cannot be used for lifting problems, where these operators do not occur in the specification.

\mainname is also related to \textit{Enum}~\cite{DBLP:conf/fmcad/AlurBJMRSSSTU13} and \textit{Relish}~\cite{DBLP:journals/pacmpl/0001WD18} as they are applicable to lifting problems. We compare \mainname with both of them in our evaluation, and the results demonstrate the better effectiveness of \mainname. \smallskip}

\jryadd{
First, \mainname addresses the scalability challenge by decomposing lifting problems into simpler subtasks. This decomposition-based  framework is common in program synthesis and has been applied to various scenarios. We list some representative approaches in this category as follows.
\begin{itemize}
    \item \m{Natural Synthesis}~\cite{DBLP:journals/pacmpl/QiuS17} uses loop invariants to decompose a loop synthesis problem into subtasks for pre-loop, in-loop, and after-loop codes, respectively.
    \item \m{Myth}~\cite{DBLP:conf/pldi/OseraZ15} and \m{Synquid}~\cite{DBLP:conf/pldi/PolikarpovaKS16} use a top-down enumeration scheme to synthesize recursive programs and will utilize type information to decompose certain intermediate synthesis problems into independent subtasks. 
    
    Our decomposition method, component elimination, is related to the tuple-decomposition method in \m{Myth}. Both of them are proposed to decompose an unknown program with a tuple output. In comparison, the method in \m{Myth} requires the specification of the unknown program to be input-output examples, while component elimination considers a specification where the input and the output both depend on another unknown program. 
    
    \item \m{DryadSynth}~\cite{DBLP:conf/pldi/HuangQSW20} proposes a general framework for reconciling inductive and deductive program synthesis. This framework repeatedly applies deductive rules to decompose a synthesis task into subtasks and then solves these subtasks using inductive program synthesis. \m{DraydSynth} implements this framework for the domain of conditional integer arithmetic, and \mainname can be regarded as an implementation of this framework for solving lifting problems.
    \item \m{Toshokan}~\cite{DBLP:conf/sas/HuangQ22}  proposes a decomposition method for dealing with complex library calls in component-based program synthesis. It uses library models to decompose a synthesis problem involving library calls into subtasks of library verification and client-code synthesis. This decomposition method cannot be applied to our task because lifting problems do not involve any libraries.
\end{itemize}

Second, since many algorithms can be regarded as recursive programs (e.g., D\&C and single-pass programs), \mainname is also related to previous studies on recursive program synthesis~\cite{DBLP:journals/pacmpl/MiltnerNBCD22, DBLP:journals/pacmpl/YuanRS23, DBLP:journals/pacmpl/LeeC23, DBLP:conf/cav/AlbarghouthiGK13, DBLP:conf/cav/FarzanN21, DBLP:conf/pldi/FarzanLN22}. However, previous recursive synthesizers cannot be applied to our tasks because of two major differences between recursive program synthesis and the automatic application of algorithmic paradigms.
\begin{itemize}
    \item The two problems treat input-output examples (or the original program) differently. Recursive program synthesis typically treats input-output examples as a full specification. It requires the synthesized recursive procedure to produce \textbf{exactly} the same outputs as specified in the examples. However, when applying algorithmic paradigms, we often need to introduce auxiliary values as part of the output of the recursive procedure. Sticking to the same outputs would lead to synthesis failures. 
    
    
    \item The two problems put different restrictions on the recursion. In recursive program synthesis, the synthesizer can use any recursion that can implement the target functions. In contrast, when applying algorithmic paradigms, the recursion of programs is prescribed by the paradigm, and the problem is how to calculate using the given recursions.

    Both settings have their respective challenges. In recursive program synthesis, the challenge is to find a proper recursion; and when applying algorithmic paradigms, the given recursion may significantly increase the scale of the resulting program. For example, the D\&C program of \m{sndmin} (Figures \ref{fig:auxcomb} and \ref{fig:dac}) is much more complex than its single-pass program (Figure \ref{fig:smin-single-pass}), though both of them can be regarded as recursive programs.
\end{itemize}
Besides, our reduction from applying algorithmic paradigms to lifting problems shares the same idea with \m{trace completeness} in recursive program synthesis~\cite{DBLP:conf/cav/AlbarghouthiGK13}. They both utilize the interpretation of the original program and then reduce the problem of synthesizing recursions to the problem of synthesizing the body of recursions. 

At last, \mainname is also related to \textit{Enum}~\cite{DBLP:conf/fmcad/AlurBJMRSSSTU13} and \textit{Relish}~\cite{DBLP:journals/pacmpl/0001WD18} because they can also be applied to solve lifting problems. We compare \mainname with both of them in our evaluation, and the results demonstrate the better performance of \mainname.
}

\section{Conclusion}\label{section:conclusion}

\jrydel{In this paper, we study the problem of applying D\&C-like algorithmic paradigms from the aspect of inductive synthesis. We capture the application of various paradigms as a novel class of synthesis problems, namely lifting problems, and propose a synthesizer \mainname for lifting problems. To address the scalability challenge, we propose two decomposition methods, namely component elimination and variable elimination, to divide a lifting problem into simpler subtasks and derive specifications for different sub-programs of the synthesis target.}
 
\jryadd{In this paper, we study the problem of applying D\&C-like algorithmic paradigms and aim to address the limitation of previous transformation-based approaches which put strict restrictions on the original program. To achieve this goal, we propose a novel approach named \mainname that applies D\&C-like paradigms by decomposition instead of by syntax-based transformation. To achieve an effective synthesis, \mainname repeatedly applies two decomposition methods, namely component elimination and variable elimination, to decompose an application task into simpler subtasks and derive specifications for different sub-programs of the synthesis target.}

\jrydel{Although \mainname does not ensure completeness, we demonstrate its effectiveness from both theoretical and practical perspectives. In theory, we prove that \mainname seldom fails on a random lifting problem when the language of combinators is expressive enough and the compressing property holds. In practice, we evaluate \mainname on a dataset of 96 lifting problems and the results show that \mainname can solve most of these problems within a short timeframe.}

\jryadd{To break the dependency among sub-programs, both decomposition methods in \mainname use approximate specifications in their first subtasks. 
We demonstrate that these approximations do not affect the effectiveness of \mainname by conducting theoretical analysis and empirical evaluation. In theory, we prove that these approximations will seldom produce unrealizable subtasks when the compressing property holds; in practice, we evaluate \mainname on a dataset of 96 tasks and the results show that \mainname can solve most of these problems within a short time.
}

We believe many techniques in this paper are general and can be potentially applied to other tasks. For example, variable elimination may be used to separate the composition of two unknown programs in other relational synthesis problems. Exploring other applications is future work.

The source code of our implementation and the experimental data of our evaluation are available online~\cite{material}.


\section*{Acknowledgement}
We thank all anonymous reviewers of this paper for their valuable suggestions on this work. We also thank Sirui Lu and Rastislav Bodik for their insightful feedback on this work and generous help in our evaluation. This work is supported by the National Key Research and Development Program of China under Grant No. 2022YFB4501902.

\bibliography{ref}
\clearpage  
\appendix
\section{Appendix: Proofs and Guarantees}
This section provides the proofs for the theorems in this paper (Section \ref{subsection:properties}) and supplies the details on the probabilistic correctness guarantee provided by our verifier (Section \ref{section:implementation}). 

\subsection{Proofs for Theorems} \label{appendix:proofs}

\begin{theorem}[Theorem \ref{theorem:soundness}] The result of \mainname (Algorithm \ref{alg:deductive}) is valid for the original lifting program if the verifiers of leaf subtasks accept only valid programs for respective subtasks.
\end{theorem}
\begin{proof} The soundness of \mainname is directly implied by the two facts below.
\begin{itemize}
    \item The result of \mainname must be valid when the sub-programs synthesized from leaf subtasks are valid because both decomposition methods in \mainname are sound by definition (i.e., the merged result is valid for the original task when the sub-results are valid for the subtasks).
    \item Since leaf subtasks are solved using the CEGIS framework, their results must be valid when the verifiers accept only valid programs. 
\end{itemize}

\end{proof}

\begin{theorem}[Theorem \ref{theorem:main-result}] Given a lifting problem $\mathsf{LP}(\m{orig}, \m{op})$ of which the \misfactor is at least $t$, the size-limited unrealizable rate of \mainname is bounded, as shown below.
    $$
    \m{unreal}(\m{orig}, \m{op}, \m{lim}_s) \leq 2^w \exp(-t/s_V^{n \cdot w}), \textbf{where } w \triangleq \m{lim}_c \cdot \m{lim}_s
    $$
    \end{theorem} 
\begin{proof} For simplicity, we shall abbreviate our probabilistic model $\mathcal M[\m{orig}, \m{op}]$ as $\mathcal M$ and interchangeably use a synthesis task as a predicate, where $\varphi(\m{prog})$ represents that $\m{prog}$ is a valid program for task $\varphi$. Besides, we shall use the following two notations in our proof.
\begin{itemize}
\item Let $\tilde{\varphi}$ be the first \m{aux} subtask generated from $\varphi$, of which the specification is shown below.
    $$
    (\bluec{\m{orig}} \spl \var{\m{aux}_1})^n\ \quanti{\overline{a}} = (\bluec{\m{orig}} \spl \var{\m{aux}_1})^n\ \quanti{\overline{a'}} \rightarrow \bluec{\targetname}\ (\bluec{\m{op}}\ (\quanti{c},\quanti{\overline{a}})) = \bluec{\targetname}\ (\bluec{\m{op}}\ (\quanti{c},\quanti{\overline{a'}}))
    $$
\item Let $\mathbb A(\varphi)$ be the set of auxiliary programs that can lead to a valid solution of $\varphi$ with a size no larger than $\m{lim}_s$, defined as below. Using this notation, the condition in the size-limited unrealizable rate (Definition \ref{definition:incomplete-rate}) can be restated as $|\mathbb A(\varphi)| > 0$.
$$
\mathbb A(\varphi) \triangleq \{\m{aux}\ |\ \exists \m{comb}, \varphi(\m{aux}, \m{comb}) \wedge \text{size}(\m{aux}, \m{comb}) \leq \m{lim}_s \}
$$
\end{itemize}

\noindent \textbf{Step 1: a sufficient condition}. Any program in $\mathbb A(\varphi)$ is valid for $\tilde \varphi$ by definition. Furthermore, \mainname will not generate unrealizable subtasks when a program in $\mathbb A(\varphi)$ is synthesized from $\tilde \varphi$. Specifically, when such a program is synthesized, \mainname will only generate $3$ subtasks after $\tilde{\varphi}$, as shown below.
\begin{enumerate}
    \item The first $\m{comb}$ subtask targets at a combinator for the output of $\m{orig}$. It is realizable because there is always a valid combinator for any program in $\mathbb A(\varphi)$. 
    \item The second \m{aux} subtask targets at an auxiliary program for the output of the synthesis result of $\tilde{\varphi}$. The result of this subtask must be $\m{null}$ as (1) it is valid by the definition of $\mathbb A_{\textit{aux}}$, and (2) it is the first choice of the corresponding synthesizer $\mathcal S_{\textit{aux}}$ (Algorithm \ref{alg:synthesis-aux}).
    \item The second \m{comb} subtask targets at a combinator for the output of the auxiliary program, which is also realizable by the definition $\mathbb A_{\textit{aux}}$. 
\end{enumerate}
Therefore, the event that the synthesis result of $\tilde \varphi$ is in $\mathcal A(\varphi)$ is a sufficient condition for the success of \mainname. As a result, we know that the size-limited unrealizable rate of \mainname is bounded by the probability below.
\begin{align}
\Pr_{\varphi \sim \mathcal M}\big[\mathcal S_{\textit{aux}}(\tilde{\varphi}) \not \in \mathbb A(\varphi)\ \big|\ |\mathbb A(\varphi)| > 0\big] \label{formula:original-sufficient} 
\end{align}

This probability is difficult for direct analysis because it involves second-order quantification (in the definition of $\mathbb A(\varphi)$) and the concrete behavior of a synthesizer, which can be very complex. Therefore, we conduct a series of derivations to eliminate these difficult parts from the probability.

\smallskip

\noindent \textbf{Step 2: eliminating left-side $\mathbb A(\varphi)$}. The conditional probability can be transformed as follows.
\begin{align}
    & \hspace{0.9em}\Pr_{\varphi \sim \mathcal M}\big[\mathcal S_{\textit{aux}}(\tilde{\varphi}) \not \in \mathbb A(\varphi)\ \big|\ |\mathbb A(\varphi)| > 0\big] \nonumber\\
    &= \Pr_{\varphi \sim \mathcal M} \big[\mathcal S_{\textit{aux}}(\tilde{\varphi}) \not \in \mathbb A(\varphi) \wedge  |\mathbb A(\varphi)| > 0\big] \bigg / \Pr_{\varphi \sim \mathcal M} \big[|\mathbb A(\varphi)| > 0\big] \nonumber \\
    &= \sum_{P \subseteq \mathcal L_{\textit{aux}}} [|P| > 0] \Pr_{\varphi \sim \mathcal M} \big[\mathcal S_{\textit{aux}}(\tilde{\varphi}) \not \in P \wedge  \mathbb A(\varphi) = P\big] \bigg / \sum_{P \subseteq \mathcal L_{\textit{aux}}} [|P| > 0] \Pr_{\varphi \sim \mathcal M} \big[\mathbb A(\varphi) = P\big] \\
    &\leq \frac{\sum_{P \subseteq \mathcal L_{\textit{aux}}}[|P| > 0] \left(\Pr_{\varphi \sim \mathcal M}\big[\mathcal S_{\textit{aux}}(\tilde{\varphi}) \not \in P \wedge  \mathbb A(\varphi) = P\big] + (|P| - 1)\Pr_{\varphi \sim \mathcal M} \big[\mathbb A(\varphi) = P\big]\right)}{\sum_{P \subseteq \mathcal L_{\textit{aux}}} |P|\Pr_{\varphi \sim \mathcal M} \big[\mathbb A(\varphi) = P\big]} \label{formula:C1-2}
\end{align}
The inequality of $(a + c)/(b+c) \geq a / b$ (when $a, b> 0, c\geq 0, a < b$) is used in the last step.

\smallskip

Let us consider the claim below under the premise that $|P| > 0$.
\begin{align}
  \Pr_{\varphi \sim \mathcal M}\big[\mathcal S_{\textit{aux}}(\tilde{\varphi}) \not \in P \wedge  \mathbb A(\varphi) = P\big] + (|P| - 1)\Pr_{\varphi \sim \mathcal M} \big[\mathbb A(\varphi) = P\big]&
  \nonumber
  \\
  = \sum_{\textit{aux}^* \in P} \Pr_{\varphi \in \mathcal M}\bigl[\mathcal S_{\textit{aux}}(\tilde\varphi) \neq \m{aux}^* \wedge \mathbb A(\varphi) = P\bigr]  & \nonumber
\end{align}
To prove this claim, let $\varphi$ be any task satisfying $\mathbb A(\varphi) = P$. There are two cases on $\mathcal S_{\textit{aux}}(\tilde\varphi)$.
\begin{itemize}
\item When $\mathcal S_{\textit{aux}}(\tilde\varphi) \not \in P$, the probability of $\varphi$ contributes to both sides for $|P|$ times.
\item Otherwise, the probability of $\varphi$ contributes to both sides for $|P| - 1$ times.
\end{itemize}
Therefore, the two probabilities involved in this claim must be the same.

\smallskip

Note that the size of any program in $\mathbb A(\varphi)$ is no larger than $\m{lim}_s$ by the definition of $\mathbb A(\varphi)$. By applying the claim to Formula \ref{formula:C1-2}, we further perform the derivation below, where $\mathbb L_{\leq \textit{lim}_s}$ denotes the subspace of $\mathcal L_{\textit{aux}}$ including only those programs of which the size is no larger than $\m{lim}_s$.
\begin{align}
\text{Formula}\ \ref{formula:C1-2}& = \sum_{P \subseteq \mathcal L_{\textit{aux}}} \sum_{\textit{aux}^* \in P} \Pr_{\varphi \in \mathcal M}\bigl[\mathcal S_{\textit{aux}}(\tilde\varphi) \neq \m{aux}^* \wedge \mathbb A(\varphi) = P\bigr]\bigg /\sum_{P \subseteq \mathcal L_{\textit{aux}}} \sum_{\textit{aux}^* \in P}\Pr_{\varphi \sim \mathcal M} \left[\mathbb A(\varphi) = P \right] \nonumber \\
& = \sum_{\textit{aux}^* \in \mathbb L_{\leq \textit{lim}_s}} \Pr_{\varphi \sim \mathcal M}\bigl[\mathcal S_{\textit{aux}}(\tilde\varphi) \neq \m{aux}^* \wedge \m{aux}^* \in \mathbb A(\varphi)\bigr] \bigg / \sum_{\textit{aux}^* \in \mathbb L_{\leq \textit{lim}_s}} \Pr_{\varphi \sim {\mathcal M}}\left[\m{aux}^* \in \mathbb A(\varphi)\right] \nonumber \\
&\leq \max_{\textit{aux}^* \in \mathbb L_{\leq \textit{lim}_s}} \left(\Pr_{\varphi \sim \mathcal M}\bigl[\mathcal S_{\textit{aux}}(\tilde\varphi) \neq \m{aux}^* \wedge \m{aux}^* \in \mathbb A(\varphi)\bigr] \bigg /  \Pr_{\varphi \sim {\mathcal M}}\left[\m{aux}^* \in \mathbb A(\varphi)\right]\right) \label{formula:C1-4} \\
&= \max_{\textit{aux}^* \in \mathbb L_{\leq \textit{lim}_s}} \Pr_{\varphi \sim \mathcal M}\bigl[\mathcal S_{\textit{aux}}(\tilde\varphi) \neq \m{aux}^* \ \big | \ \m{aux}^* \in \mathbb A(\varphi)\bigr] \label{formula:C1-5}
\end{align}
Formula \ref{formula:C1-4} is obtained via the following inequality (where $0/0$ is defined as $0$).
$$
\forall a_i, b_i \geq 0, \sum_{i=1}^n a_i \big / \sum_{i=1}^n b_i \leq \max_{i=1}^n (a_i / b_i)
$$

So far, the left-side $\mathbb A(\varphi)$ in the sufficient condition (Formula \ref{formula:original-sufficient}) has been eliminated.

\smallskip

\noindent \textbf{Step 3: eliminating the output of $\mathcal S_{\textit{aux}}$}. A worth noting property of $\mathcal S_{\textit{aux}}$ (Algorithm \ref{alg:synthesis-aux}) is that, given a task that has a solution with a size no lager than $\m{lim}_s$, the size of its synthesis result is no larger than $\m{lim}_c\m{lim}_s$. First, the result of $\mathcal S_{\textit{aux}}$ includes at most $\m{lim}_c$ components. Second, the size of each component must be no larger than $\m{lim}_s$, otherwise, the smallest solution (which is no larger than $\m{lim}_s$) will be found by OE and be synthesized instead.

Therefore, ``any program in $\mathbb L_{\leq \textit{lim}_c\textit{lim}_s}$ is invalid for $\tilde \varphi$ except $\m{aux}^*$'' forms a sufficient condition of $\mathcal S_{\textit{aux}}(\tilde \varphi) = \m{aux}^*$. Then, the conditional probability in Formula \ref{formula:C1-5} can be bounded as follows.
\begin{align}
    &\hspace{1.05em}\Pr_{\varphi \sim \mathcal M}\bigl[\mathcal S_{\textit{aux}}(\tilde\varphi) \neq \m{aux}^* \ \big | \ \m{aux}^* \in \mathbb A(\varphi)\bigr] \nonumber \\ 
    &\leq  \Pr_{\varphi \sim \mathcal M}\bigl[\exists \m{aux} \in \mathbb L_{\leq\textit{lim}_c\textit{lim}_s}, \m{aux} \neq \m{aux}^* \wedge  \tilde \varphi(\m{aux}) \ \big | \ \m{aux}^* \in \mathbb A(\varphi)\bigr] \nonumber \\
    &\leq \sum_{\textit{aux} \in \mathbb L_{\leq\textit{lim}_c\textit{lim}_s}} [\m{aux} \neq \m{aux}^*]\Pr_{\varphi \sim \mathcal M}\bigl[ \tilde \varphi(\m{aux}) \ \big | \ \m{aux}^* \in \mathbb A(\varphi)\bigr] \nonumber \\
    & \leq 2^{\textit{lim}_s\textit{lim}_c} \max_{\textit{aux} \in \mathbb L_{\leq\textit{lim}_c\textit{lim}_s}} [\m{aux} \neq \m{aux}^*]\Pr_{\varphi \sim \mathcal M}\bigl[ \tilde \varphi(\m{aux}) \ \big | \ \m{aux}^* \in \mathbb A(\varphi)\bigr] \label{formula:C1-6}
\end{align}
The last step uses the fact that the number of programs whose size is no larger than $k$ is at most $2^k$. So far, the concrete output of $\mathcal S_{\textit{aux}}$ has been eliminated.
\smallskip

\noindent \textbf{Step 4: eliminating the condition in the probability}. The probability in Formula \ref{formula:C1-6} can be transformed as follows by unfolding the definition of $\mathbb A(\varphi)$.
\begin{align*}
\Pr_{\varphi \sim \mathcal M} \left[\tilde \varphi(\m{aux})\ \big | \ \m{aux}^* \in \mathbb A(\varphi)\right] = \Pr_{\varphi \sim \mathcal M} \left[\tilde \varphi(\m{aux}) \ \big| \ \exists \m{comb}, \varphi(\m{aux}^*, \m{comb}) \wedge \text{size}(\m{aux}^*, \m{comb}) \leq \m{lim}_s \right]
\end{align*}

Recall that our probabilistic model $\mathcal M$ is parameterized by the semantics of $\m{orig}$ and $\m{op}$, and its randomness comes only from the semantics of programs in $\mathcal L_{\textit{aux}}$ and $\mathcal L_{\textit{comb}}$. In the probability above, the outcome $\tilde{\varphi}(\m{aux})$ is determined only by $\m{aux}$, while the condition $\m{aux}^* \in \mathbb A(\varphi)$ is determined by $\m{aux*}$ and programs in $\mathcal L_{\textit{comb}}$. Therefore, the two events in this conditional probability are independent, making it safe to directly ignore the condition, as shown below.
\begin{align*}
\Pr_{\varphi \sim \mathcal M} \left[\tilde \varphi(\m{aux})\ \big | \ \m{aux}^* \in \mathbb A(\varphi)\right] &= \Pr_{\textit{aux}} \left[\tilde \varphi(\m{aux})\right]  
\end{align*} 

\noindent \textbf{Step 5: bounding the probability of $\tilde \varphi(\m{aux})$ using the \misfactor.} Let us first unfold and transform the above probability $\Pr[\tilde{\varphi}(\m{aux})]$ as follows.
\begin{align}
    &\Pr_{\textit{aux}} \left[\tilde{\varphi}(\m{aux})\right] \nonumber\\
=& \Pr_{\textit{aux}} \left[(\m{orig} \spl \m{aux})^n~\quanti{\overline{a}} = (\m{orig} \spl \m{aux})^n~\quanti{\overline{a'}} \rightarrow \m{orig}~(\m{op}~(\quanti{c}, \quanti{\overline{a}})) = \m{orig}~(\m{op}~(\quanti{c}, \quanti{\overline{a'}}))\right] \nonumber \\
=& \Pr_{\textit{aux}} \left[\big(\m{orig}^n~\quanti{\overline{a}} =  \m{orig}^n~\quanti{\overline{a'}} \wedge \m{orig}~(\m{op}~(\quanti{c}, \quanti{\overline{a}})) \neq \m{orig}~(\m{op}~(\quanti{c}, \quanti{\overline{a'}}))\big) \rightarrow \m{aux}^n~\quanti{\overline{a}} \neq \m{aux}^n~\quanti{\overline{a'}}\right] \label{formula:D-1}
\end{align}

Suppose the \misfactor of $(\m{orig}, \m{op})$ is at least $t$, that means, there are $t$ pairs of $\m{orig}$ inputs $(\overline{a_i}, \overline{a'_i})$ such that (1) every pair satisfies the premise of the event in Formula \ref{formula:D-1}, as shown below, 
$$
\exists c, \big(\m{orig}^n~\overline{a_i} = \m{orig}^n~\overline{a_i'} \wedge \m{orig}~(\m{op}~(c, \overline{a_i})) \neq \m{orig}~(\m{op}~(c, \overline{a_i'}))\big)
$$
and (2) all components involved in these pairs ($2tn$ in total) are different.

The event in Formula \ref{formula:D-1} is satisfied only when $\m{aux}^n$ outputs differently on all these $t$ pairs of inputs. Using this fact, the target probability can be bounded as follows.
\begin{align}
\Pr_{\textit{aux}} \left[\tilde{\varphi}(\m{aux})\right] &\leq \Pr_{\textit{aux}}\left[\bigwedge_{i=1}^t \m{aux}^n~\overline{a_i} \neq \m{aux}^n~\overline{a_i'}\right] \leq \prod_{i=1}^t \Pr_{\textit{aux}} \left[\textit{aux}^n~\overline{a_i} \neq \textit{aux}^n~\overline{a_i'}\right] \nonumber \\
& \leq \big(1 - \text{pow}(s_V, -\m{lim}_c \m{lim}_s)^n\big)^t \leq \exp\big(-t/\text{pow}(s_V, n\textit{lim}_c\textit{lim}_s)\big) \nonumber
\end{align}

We supply some details on the above derivation as follows.
\begin{itemize}
    \item The second step uses the premise that all components involved in $\overline{a_i}$ and $\overline{a_i'}$ are different. This fact implies that events $\m{aux}^n~\overline{a_i} \neq \m{aux}^n~\overline{a_i'}$ are totally independent. 
    \item The third step uses the fact that the size of $\m{aux}$ is no larger than $\m{lim}_s \cdot \m{lim}_s$. Under this condition, \m{aux} can only output at most $\m{lim}_c \cdot \m{lim}_s$ auxiliary values, and thus the size of the output domain of $\m{aux}$ is no larger than $\text{pow}(s_V, \m{lim}_c \cdot \m{lim}_s)$.
\end{itemize}

\smallskip

\noindent \textbf{Step 6: Summary}. Let us now sum up the previous sub-results and then prove the target theorem. First, the inequality below is obtained by combining Steps 4 and 5.
\begin{align*}
\Pr_{\varphi \sim \mathcal M}\left[\tilde{\varphi}(\m{aux})\ \big|\ \m{aux}^* \in \mathbb A(\varphi)\right] \leq \exp(-t/s_V^{n \cdot w}), \textbf{where }w \triangleq \m{lim}_c \cdot \m{lim}_s
\end{align*}

Second, the inequality below is obtained by further combining Step 3.
$$
\Pr_{\varphi \sim \mathcal M}\bigl[\mathcal S_{\textit{aux}}(\tilde\varphi) \neq \m{aux}^* \ \big | \ \m{aux}^* \in \mathbb A(\varphi)\bigr] \leq 2^w\exp(-t/s_V^{n \cdot w})
$$

At last, we know the size-limited unrealizable rate of \mainname is bounded by the formula below after further combining Steps 1 and 2, which is exactly the target theorem.
$$
2^w\exp(-t/s_V^{n \cdot w}), \textbf{ where }w\triangleq\m{lim}_c \cdot \m{lim}_s
$$
\smallskip 
\end{proof}

Before proving Theorem \ref{theorem:completeness}, we shall first introduce and prove the following lemma. It shows that the \misfactor of a random lifting problem is almost surely large.

\begin{lemma} \label{lemma:intermeidate-for-ciompleteness} When there are at least two values (i.e., $s_V > 1$), for any constant $\epsilon > 0$, there always exists a constant $\delta$ such that the probability for the \misfactor of a random lifting problem to be smaller than $\delta \cdot s_A / (n \cdot s_V^{n + 1})$ does not exceed $\epsilon$, as shown below. 
    $$
    \forall \epsilon > 0, \exists \delta, \forall n, A, V, \Pr_{\textit{orig}, \textit{op}}\left[\text{the \misfactor of }\mathsf{LP}(\m{orig}, \m{op}) < \lfloor \delta \cdot s_A / (n \cdot s_V^{n + 1}) \rfloor \right] \leq \epsilon
    $$
where $\m{orig}$ is a random function with type $A \rightarrow V$, $\m{op}$ is a random function with type $V \times A^n \rightarrow A$, $s_A$ and $s_V$ are the numbers of values in types $A$ and $V$, respectively, and it is assumed that $s_V > 1$.
\end{lemma}
\begin{proof} In this proof, we only need to consider the case where $s_A/(n \cdot s_V^{n + 1})$ is large enough. To see this point, let $t$ be an arbitrary threshold that may depend on $\epsilon$. By taking $\delta$ as a value smaller than $1 / t$, the target inequality will be satisfied in cases where $s_A/(n \cdot s_V^{n + 1}) \leq t$, as shown below.
\begin{align*}
& \Pr_{\textit{orig}, \textit{op}}\left[\text{the \misfactor of }\mathsf{LP}(\m{orig}, \m{op}) < \lfloor \delta \cdot s_A / (n \cdot s_V^{n + 1}) \rfloor  \right] \\
= & \Pr_{\textit{orig}, \textit{op}}\left[\text{the \misfactor of }\mathsf{LP}(\m{orig}, \m{op}) < 0 \right] \\
= & ~0 \leq \epsilon
\end{align*}

Therefore, in the remainder of this proof, we shall focus on the cases where $s_A/(n \cdot s_V^{n + 1})$ is larger than a threshold $t$. Note that in these cases, $s_A$ must also be large because $s_V$ is at least $2$.

\smallskip
As the first step, we divide values in type $A$ into subsets $A_1$ and $A_2$, whose sizes are $\alpha s_A$ and $(1 - \alpha)s_A$ values for some constant $\alpha$, respectively. This construction ensures that the outputs of $\m{orig}$ on $A_1$ and $A_2$ are independent.

\smallskip 

Second, we prove that with a high probability, $\m{orig}$ will not map too many values in $A_2$ to the same value. Let $\m{num}(v)$ for $v \in V$ be the number of values in $A_2$ on which the outputs of $\m{orig}$ is $v$. We bound the probability of $\exists v \in V, \m{num}(v) \geq 2|A_2|/3$ (denoted as event $\mathcal E$) as follows.
\begin{align*}
    \Pr[\mathcal E] &\leq \sum_{v \in V} \Pr\big [\m{num}(v) > 2|A_2|/3 \big] \\ 
    & \leq \sum_{v \in V}  \Pr \big[\m{num}(v) > (1 + 1/3)\mathbb E\big[\m{num}(v)\big ]\big] \\
    & \leq s_V \cdot \exp\left(- \frac{(1 - \alpha)s_A}{21s_V}\right)
\end{align*}
Here the first step uses the union bound, the second step uses the facts that $\mathbb E[\m{num}(v)] \leq |A_2|/s_v$ and $s_v \geq 2$, and the last step uses the Chernoff bound.

\smallskip 

Third, we assume that $\mathcal E$ does not happen and then construct a valid sequence of input pairs for the \misfactor using only values in $A_1$. Concretely, values in $A_1$ can be arranged into a sequence $S$ including $m = \lfloor |A_1| / (2n) \rfloor$ independent input pairs in $A^n \times A^n$. Then, we call an input pair $(\overline{a}, \overline{a'})$ as \m{valid} if it satisfies the conditions below for a given value $c \in C$.
$$ 
 \m{orig}^n~\overline{a} = \m{orig}^n~\overline{a'} \qquad \m{orig}~(\m{op}~(c, \overline{a})) \neq \m{orig}~(\m{op}~(c, \overline{a'})) \qquad \m{op}~(c, \overline{a}) \in A_2 \qquad \m{op}~(c, \overline{a'}) \in A_2
$$
Let $S'$ be the sub-sequence of $S$ that includes only valid pairs. The left two conditions above ensure that $S'$ is a valid sequence for the \misfactor. Therefore, $|S'|$ provides a lower bound on the \misfactor, and the task remaining is to prove that $|S'|$ is large with a high probability.

For an input pair $(\overline{a}, \overline{a'})$ in $S$, the probability for it to satisfy the first condition is $s_V^{-n}$, and the probability for it to satisfy the other three conditions is at least $(1 - \alpha)^2/3$ when event $\mathcal E$ does not happen. These two probabilities are independent because the outputs of $\m{op}$, the outputs of $\m{orig}$ on $A_1$, and the outputs of $\m{orig}$ on $A_2$ are all independent. Therefore, the probability for a pair in $S$ to be valid is at least $(1 - \alpha)^2/3 \cdot s_V^{-n}$.

Now let us bound the length of sub-sequence $S'$. On the one hand, the expectation of $|S'|$ is $(1 - \alpha)^2 / 3 \cdot s_V^{-n} \cdot m$, which is no smaller than $\gamma \cdot s_A/(n \cdot s_V^{n})$ for some constant $\gamma < (1 - \alpha)^2 / 6$. On the other hand, it is easy to verify that the probabilities for each pair in $S$ to be valid are independent when the outputs of $\m{orig}$ on $A_2$ are fixed. Therefore, the Chernoff bound can be applied to provide a probabilistic lower bound for $|S'|$, as shown below, where $\tau$ is an arbitrary constant in $(0, 1)$.
\begin{align*}
&\Pr\left[|S'| \leq (1 - \tau) \cdot \gamma \cdot s_A / (n \cdot s_V^{n})~\big|~\neg \mathcal E \right] \\
\leq &\Pr\big[|S'| \leq (1 - \tau)\mathbb E[|S'|] ~\big|~\neg \mathcal E\big] \\
\leq & \exp\big(-\tau^2/2 \cdot \mathbb E[|S'|] \big)  \\ 
\leq & \exp \left(-\frac{\tau^2}{2} \cdot \gamma \cdot \frac{s_A}{n \cdot s_V^{n}}\right)  
\end{align*}

\smallskip 

By combining the above results, we can bound the \misfactor as follows.
\begin{align*} 
&\Pr\left[\text{the \misfactor of }\mathsf{LP}(\m{orig}, \m{op}) \leq (1 - \tau) \cdot \gamma \cdot s_A / s_V^{ n}\right] \\
\leq & \Pr\left[|S'| \leq (1 - \tau) \cdot \gamma \cdot s_A / (n \cdot s_V^{-n})~\big|~\neg \mathcal E \right] + \Pr[\mathcal E] \\
\leq & \exp \left(-\frac{\tau^2}{2} \cdot \gamma \cdot \frac{s_A}{n \cdot s_V^{n}}\right)  + s_V \cdot \exp\left(- \frac{(1 - \alpha)s_A}{21s_V}\right)
\end{align*}
Since we have assumed that $s_A/(n \cdot s_V^{n + 1})$ is no smaller than some value $t$, the value of $s_A / (n \cdot s_V^n)$ is no smaller than $t$, and the value of $s_A / s_V$ is no smaller than $t \cdot s_V$. Therefore, the formula above can be further simplified as below, resulting in the target lemma.
\begin{align*} 
    \leq& \exp \left(-\frac{\tau^2}{2} \cdot \gamma \cdot t\right)  + s_V \cdot \exp\left(- \frac{(1 - \alpha)}{21} \cdot s_V \cdot t\right) \\
    \leq& \exp \left(-\frac{\tau^2}{2} \cdot \gamma \cdot t\right)  + \exp\left(- \frac{(1 - \alpha)}{21} \cdot t\right) \textbf{ when }t > \frac{210}{1 - \alpha} \\
    \leq& ~\epsilon \textbf{ when } t > k \cdot \ln(1 / \epsilon) \text{ for a large enough constant }k
\end{align*}
\end{proof}

\begin{theorem}[Theorem \ref{theorem:completeness}] Consider the size-limited unrealizable rate of \mainname on a random lifting problem. When there are at least two values (i.e., $s_V > 1$), for any constant $\epsilon > 0$, the probability for this rate to exceed $\epsilon$ tends to $0$ when $s_A / s_V^{w'}$ tends to $\infty$, where $w' \triangleq n \cdot \m{lim}_c \cdot \m{lim}_s + n + 1$.
\end{theorem} 
\begin{proof} To prove this theorem, we need to prove that for any constants $\epsilon, \epsilon' > 0$, the probability for the unrealizable rate to exceed $\epsilon$ is at most $\epsilon'$ when $s_A / s_V^{w'}$ is large enough, as shown below.
$$
\Pr_{\textit{orig}, \textit{op}} \big[ \m{unreal}(\m{orig}, \m{op}, \m{lim}_s) > \epsilon\big] \leq \epsilon'
$$

By Lemma \ref{lemma:intermeidate-for-ciompleteness}, there exists a constant $\delta$ such that with a probability of at least $1 - \epsilon'$, the \misfactor on a random lifting problem will be at least $\lfloor \delta \cdot s_A / (n \cdot s_V^{n + 1}) \rfloor$, as shown below.
$$
\Pr_{\textit{orig}, \textit{op}} \big[ \text{the \misfactor of }\mathsf{LP}(\m{orig}, \m{op}) \geq \lfloor \delta \cdot s_A / (n \cdot s_V^{n + 1}) \rfloor \big] \geq 1 - \epsilon'
$$ 

When the event in the above probability happens (denoted as event $\mathcal E$), the size-limited unrealizable rate is bounded by Theorem \ref{theorem:main-result}, as shown below.
$$
\mathcal E \rightarrow \m{unreal}(\m{orig}, \m{op}, \m{lim}_s) \leq 2^w \exp(-\lfloor \delta \cdot s_A / (n \cdot s_V^{n + 1}) \rfloor/s_V^{n \cdot w}), \textbf{where } w \triangleq \m{lim}_c \cdot \m{lim}_s
$$

When $s_A / s_V^{w'}$ is larger than a threshold $k$, the above formula can be simplified as follows.
\begin{align*}
2^w \exp(-\lfloor \delta \cdot s_A / (n \cdot s_V^{n + 1}) \rfloor/s_V^{n \cdot w}) \leq 2^w \exp( - \delta \cdot k / n) \leq \epsilon \textbf{ when }k\text{ is large enough} 
\end{align*}

Therefore, we know the following inequality holds when $s_A / s_V^{w'}$ is large enough. This result directly implies the target conclusion.
\begin{align*}
    &\Pr_{\textit{orig}, \textit{op}} \big[ \m{unreal}(\m{orig}, \m{op}, \m{lim}_s) > \epsilon\big]  \\
    \leq &\Pr_{\textit{orig}, \textit{op}} \big[ \m{unreal}(\m{orig}, \m{op}, \m{lim}_s) > \epsilon~\big | ~\mathcal E\big] + \Pr_{\textit{orig}, \textit{op}}[\neg \mathcal E] \\ 
    < & ~\epsilon' \textbf{ when }s_A / s_V^{w'}\text{ is large enough}
\end{align*}
\end{proof}

\begin{theorem}[Theorem \ref{theorem:minimal}] Given an example-based task for the auxiliary program, let $\m{aux}^*$ be the synthesis result of $\mathcal S_{\textit{aux}}$. Then, any sub-program of $\m{aux}^*$ must not be valid for the given task.
\end{theorem}
\begin{proof}
    Assume that there is a strict sub-program of $\m{aux}^*$ (denoted as $\m{aux}$) that is also valid for the example-based task. Then there are two possible cases.

    \textbf{Case 1}: $\m{aux}$ is strictly included in a component of $\m{aux}^*$. Let $\m{comp}$ be the corresponding component. Since OE enumerates programs strictly from small to large, it must return $\m{aux}$ as a component before $\m{comp}$. Therefore, $\m{aux}$ will be considered by $\mathcal S_{\textit{aux}}$ before $\m{aux}^*$, so $\m{aux}$ should be the synthesis result of $\mathcal S_{\textit{aux}}$ instead, leading to a conflict.

    \textbf{Case 2}: $\m{aux}$ is not strictly included in any component of $\m{aux}^*$. Since $\m{aux}^*$ is formed as a tuple of components (i.e., in the form of $\m{comp}_1 \spl \dots \spl \m{comp}_k$), as a sub-program of $\m{aux}^*$, $\m{aux}$ must be a tuple of several components in $\m{aux}^*$. So $\m{aux}$ must be considered by $\mathcal S_{\textit{aux}}$ before $\m{aux}^*$ since the top-level combination enumerates the number of components from small to large. Consequently, $\m{aux}$ should be the synthesis result of $\mathcal S_{\textit{aux}}$ instead, leading to a conflict. 

    In summary, the assumption never holds, and thus the target theorem is obtained. 
\end{proof}

\subsection{Example of Analyzing the \Misfactor} \label{appendix:misfactor-example}
In Section \ref{subsection:example1}, we have discussed a lifting problem about applying D\&C to \m{sndmin}, where the original program $\m{orig}$ calculates \m{sndmin} and the operator $\m{op}$ concatenates two given lists. Now, we are going to prove a lower bound for the \misfactor of this lifting problem. For simplicity, in the discussion below, we assume each input list includes up to $l$ integers in the range $[1, s_V]$. 

To get a lower bound of the \misfactor, we need to find a sequence of input pairs satisfying the two conditions in Definition \ref{definition:mismatch-factor}. Specifically, in this example, we need to find a sequence of $4$-list tuples such that (1) every tuple $\left((\m{xs}_L, \m{xs}_R), (\m{xs}_L', \m{xs}_R')\right)$ in this sequence satisfies the formula below, and (2) every list is used at most once in this sequence. 
\begin{equation}
	\begin{array}{l}
	(\m{sndmin}~\m{xs}_L, \m{sndmin}~\m{xs}_R) = (\m{sndmin}~\m{xs}_L', \m{sndmin}~\m{xs}_R') \\ \hspace{10.5em}\wedge~\m{sndmin}~(\m{xs}_L \cat \m{xs}_R) \neq \m{sndmin}~(\m{xs}_L' \cat \m{xs}_R') 
	\end{array}\label{formula:example-first-condition}
\end{equation}

We can construct such a sequence from those lists formed by $l-2$ integers in range $[3, s_V]$. Specifically, for every such list $\m{ys}$, we construct a tuple $\left((\m{xs}_L, \m{xs}_R), (\m{xs}_L', \m{xs}_R')\right)$ as shown below\footnote{In this construction, we assume the value of $s_V$ is at least $3$.}.
$$
\begin{array}{ccc}
	\m{xs}_L \triangleq [1, 2] \cat \m{ys} & & \m{xs}_R \triangleq [1, 3] \cat \m{ys} \\
	\m{xs}_L' \triangleq [2, 2] \cat \m{ys} & & \m{xs}_R' \triangleq [2, 3] \cat \m{ys} 
\end{array}
$$
This construction results in a sequence of length $(s_V - 2) ^{l - 2}$, the number of different $\m{ys}$. We can verify that this sequence indeed satisfies the two conditions in Definition \ref{definition:mismatch-factor}.
\begin{itemize}
	\item For the first condition, Formula \ref{formula:example-first-condition} is satisfied by all tuples because the outputs of $\m{sndmin}$ on $\m{xs}_L$ and $\m{xs}_L'$ are always $2$, the outputs on $\m{xs}_R$ and $\m{xs}_R'$ are always $3$, but the outputs on $\m{xs}_L \cat \m{xs}_R$ and $\m{xs}_L' \cat \m{xs}_R'$ are always different, which are $1$ and $2$, respectively. 
	\item For the second condition, no two lists in this sequence are the same because (1) no two lists in the same tuple can be the same as their first two elements must be different, and (2) no two lists in different tuples can be the same as their last $l - 2$ elements must be different.
\end{itemize}

Therefore, the \misfactor of the lifting problem in Section \ref{subsection:example1} is at least $(s_V - 2)^{l - 2}$. By combining this lower bound with Theorem \ref{theorem:main-result}, we can get the conclusion that the size-limited unrealizable rate of this lifting problem tends to $0$ when the length of lists (i.e., $l$) tends to $\infty$.

\subsection{Probabilistic Correctness Guarantee of Our Verifier}\label{appendix:proof-probabilistic}

Our implementation of \mainname includes a testing procedure as a part of the verifier (Section \ref{section:implementation}). In the $i$-th CEGIS iteration, it tests the candidate program using $10^4 \times i$ random examples generated from a pre-defined distribution. This testing procedure provides a guarantee that the probability for the error rate of the synthesis result to be more than $10^{-3}$ is at most $4.55 \times 10^{-5}$. 

To prove this guarantee, let us first introduce some necessary notations. Let $n$ be the number of examples used in the first CEGIS iteration ($10^4$ in our implementation) and $\delta$ be the tolerable error rate in the probabilistic guarantee ($10^{-3}$ in the above claim). Then, let $\mathcal E$ be the event that the error rate of the synthesis result is more than $\delta$, and let $\mathcal E_i$ be the event that a program with an error rate larger than $\delta$ is accepted by the verifier in the $i$-th CEGIS iteration.

To get a probabilistic guarantee, our target is to derive an upper bound for $\Pr[\mathcal E]$. By the definition, $\mathcal E$ happens only when some event $\mathcal E_i$ happens, and $\mathcal E_i$ happens only when a program with an error rate larger than $\delta$ passes $n \cdot i$ random examples, of which the probability is no larger than $(1 - \delta)^{n \cdot i}$. Therefore, the following inequality holds.
$$
\Pr[\mathcal E] \leq \sum_{i=1}^{+\infty} \Pr[\mathcal E_i] \leq \sum_{i=1}^{+\infty} (1-\delta)^{n \cdot i} \leq \sum_{i=1}^{+\infty} \exp(-\delta \cdot n \cdot i) = w / (1 - w) \textbf{ for } w \triangleq \exp(-\delta \cdot n)
$$

When $(n, \delta)$ is set to $(10^4, 10^{-3})$, the value of $w$ is $\exp(-10) \approx 4.54 \times 10^{-5}$, and thus the upper bound of $\Pr[\mathcal E]$ is $w/(1-w) < 4.55 \times 10^{-5}$. Therefore, the probability for the error rate of the synthesis result to be more than $10^{-3}$ is at most $4.55 \times 10^{-5}$.
\section{Appendix: Algorithmic Paradigms} \label{appendix:paradigms}
In this section, we supply the details on the remaining two paradigms of longest segment problems and the paradigm of segment trees. For each paradigm, we introduce three aspects in order: (1) its procedure, (2) its reduction to lifting problems, and (3) the time complexity under the efficiency condition (Section \ref{subsection:efficiency-guarantee}), i.e., the efficiency guarantee provided by \mainname.

\subsection{The first algorithmic paradigm for the longest segment problem}
Recall that a longest segment problem is specified by a predicate $p$ on lists. Given a list $\m{xs}$, this problem asks for the length of the longest segment in $\m{xs}$ satisfying the predicate. 

\smallskip
\noindent \textbf{Procedure}. The first paradigm proposed by \citet{DBLP:journals/scp/Zantema92} aims at the cases where predicate $p$ is $\textit{predict-closed}$ and $\textit{overlap-losed}$, defined as follows.
\begin{itemize} 
    \item \textit{prefix-closed} means that for any list satisfying the predicate, all its prefixes must also satisfy the predicate, i.e.,
    $
    p\ (\quanti{\m{xs}}\cat \quanti{\m{ys}}) \rightarrow p\ \quanti{\m{xs}}
    $.
    \item \textit{overlap-closed} means that for any two overlapped segments satisfying the predicate, their join must also satisfy the predicate, i.e., 
    $$
    (\m{length}\ \quanti{\m{ys}} > 0 \wedge p\ (\quanti{\m{xs}} \cat \quanti{\m{ys}}) \wedge p\ (\quanti{\m{ys}} \cat \quanti{\m{zs}})) \rightarrow p\ (\quanti{\m{xs}} \cat \quanti{\m{ys}} \cat \quanti{\m{zs}})
    $$
\end{itemize}

This paradigm considers all prefixes of the input list $\m{xs}$ in order and calculates the longest suffix satisfying $p$ for each of them. Let $ls(\m{pref})$ be the longest suffix of $\m{pref}$ satisfying predicate $p$. For two consecutive prefixes $\m{pref}_1$ and $\m{pref}_2 = \m{pref}_1 \cat [v]$, when $p$ is both prefix-closed and overlap-closed, $ls(\m{pref}_2)$ must be one of $ls(\m{pref}_1) \cat [v], [v]$ and $[]$.

Figure \ref{fig:template-al1} shows a template of this paradigm (in C-like syntax), where $A$ and $n$ denote the input list and the length of the input list respectively. This template calculates the longest valid suffix for each prefix of $A$, stores its length (Line 6), and auxiliary values on the longest valid suffix as \textit{info} (Line 9). When a new element is considered, \m{lsp} verifies whether $ls(A[0 \dots {i-1}]) \cat [A_i], [A_i], []$ are valid, and picks the first valid one among them (Lines 9-18). In this procedure, combinator $\m{comb}$ is used to quickly update \texttt{info} and verify whether $ls(A[0 \dots {i-1}]) \cat [A_i]$ is valid (Line 9).

\begin{figure}[t]
        \footnotesize
        \hspace{0.3cm}\begin{minipage}[c]{0.6\linewidth}        
\begin{lstlisting}[language=c++, numbers=left]
struct Info {
    bool is_valid;
    // Variables representing the output of aux.
};
int lsp(int* A, int n){
    int res = 0, len = 0;
    Info info = {/*p [], aux []*/};
    for (int i = 0; i < n; ++i) {
        info = /*comb (A[i], info)*/;
        if (!info.is_valid) {
            info = {/*p [A[i]], aux [A[i]]*/};
            if (info.is_valid) {
                len = 1;
            } else {
                len = 0, info = {/*p [], aux []*/};
        } else {
            len += 1;
        }
        res = max(res, len);
    }
    return res;
}
\end{lstlisting}
\end{minipage}
\caption{The template of the first paradigm for the longest segment problem.}
\label{fig:template-al1}
\end{figure}

\smallskip
\noindent \textbf{Reduction to the lifting problem}. To apply this paradigm, we need to find a combinator $\m{comb}$ and an auxiliary program $\m{aux}$ satisfying the specification below, where $\m{comb}$ updates whether a segment is valid after a new element is appended, and $\m{aux}$ provides necessary auxiliary values.  
\begin{equation*}
    (\bluec{p} \spl \var{\m{aux}})\ (\quanti{\m{xs}} \cat [\quanti{v}]) = \var{\textit{comb}}\ \big(\quanti{v}, (\bluec{p}\spl \var{\m{aux}})\ \quanti{\m{xs}}\big) 
\end{equation*}
This task can be regarded as a lifting problem $\mathsf{LP}(p, \m{op})$ where $\m{op}\ (v, (\m{xs})) \triangleq \m{xs} \cat [v]$.

\smallskip 
\noindent \textbf{Time complexity}. The bottleneck here is $O(n)$ invocations of $\m{comb}$ in the loop. Therefore, under the efficiency condition, the resulting program will run in $O(n)$ time on a list of length $n$.

\subsection{The third algorithmic paradigm for the longest segment problem}
\noindent \textbf{Procedure}. This paradigm does not have any requirement on the $p$ and is based on a technique named \textit{segment partition}. Given list $A[1 \dots n]$, its segment partition is a series of consecutive segments $(r_0 = 0, r_1], (r_1, r_2], \dots, (r_{k-1}, r_k = n]$ satisfying (1) $\forall i \in [1,k], \forall j \in (r_{i-1}, r_i), A_j > A_{r_i}$, and (2) $\forall i \in [2, k], A_{r_{i-1}} \leq  A_{r_i}$. This paradigm first generates the segment partition of the given list and then gets the result by merging the information of all segments in the partition. 

\begin{figure}[t]
        \footnotesize
        \hspace{1.4cm}\begin{minipage}[c]{0.81\linewidth}        
\begin{lstlisting}[language=c++, numbers=left]
struct Info{
    int res; // Variable representing the output of orig
    // Variables representing the output of aux
}info[N];
int rpos[N];
int solve(int *A, int n) {
    int num = 0;
    for (int i = 0; i < n; i++) {
        Info now = {/*orig [], aux []*/};
        while (num > 0 && A[rpos[num]] > A[i]) {
            now = /*comb (A[rpos[num]], (info[num], now))*/;
            --num;
        }
        num++; rpos[num]=i; info[num]=now;
    }
    Info now = {/*orig [], aux []*/};
    for (int i = num; i > 0; i--) {
        now = /*comb (A[rpos[i]], (info[i], now))*/;
    }
    return now.res;
}
\end{lstlisting}
\end{minipage}
\caption{The template of the third paradigm for the longest segment problem.}
\label{fig:template-al3}
\end{figure}

Figure \ref{fig:template-al3} shows a template of this paradigm, which runs in two steps. In the first step, it constructs a segment partition of the whole list (Lines 8-15). Starting from the empty list, it considers each element in the list, updates the segment partition, and then gets the partition of the whole list when all elements are considered. In this procedure, several variables are used.
\begin{itemize}
    \item \m{num} represents the number of segments in the current partition (Line 7).
    \item $\m{rpos}[i]$ represents the index of the right end the $i$th segment in the partition (Line 3).
    \item $\m{info}[i]$ records the function value of $\targetname$ (i.e., the length of the longest valid segment) and $\m{aux}$ on the content of the $i$th segment.  
\end{itemize}
When a new element is inserted, the template merges the last several segments via combinator $\m{comb}$ to ensure that the remaining segments form a partition of the current prefix (Lines 9-14). 

In this second step, after the whole segment partition is obtained, the template merges these segments (Lines 16-20) using $\m{comb}$ again and thus gets the result of the whole list (Line 21).

\smallskip
\noindent \textbf{Reduction to the lifting problem}. In this template, combinator $\m{comb}$ is invoked on an element $v$ and the cared information on two lists $\m{xs}_L, \m{xs}_R$, and its task is to return the cared information on $\m{xs}_L \cat [v] \cat \m{xs}_R$. By the definition of the segment partition, all these invocations ensure that $v$ is the leftmost maximum in $\m{xs}_L \cat [v] \cat \m{xs}_R$, i.e., $\m{min}\ \m{xs}_L > v$ and $\m{min}\ \m{xs}_R \geq v$. 

Therefore, to apply this paradigm, we need to find a combinator $\m{comb}$ and an auxiliary program $\m{aux}$ satisfying the following specification.
\begin{align*}
    \m{min}\ \quanti{\m{xs}_L} > \quanti{v}~\wedge&~\m{min}\ \quanti{\m{xs}_R} \geq \quanti{v} \rightarrow \\
    &(\bluec{\targetname} \spl \m{\var{aux}})\ (\quanti{\m{xs}_L} \cat [\quanti{v}] \cat \quanti{\m{xs}_R}) = \m{comb}\ (\quanti{v}, ((\bluec{\targetname} \spl \m{\var{aux}})\ \quanti{\m{xs}_L}, (\bluec{\targetname} \spl \m{\var{aux}})\ \quanti{\m{xs}_R}))
\end{align*}
This task can be reduced to a lifting problem $\mathsf {LP}(\targetname, \m{op})$, where $\m{op}$ is defined as follows and the premise above is eliminated by setting the corresponding outputs of $\m{op}$ to a dummy list.
\begin{align*}
    \m{op}\ (v, (\m{xs}_L, \m{xs}_R))\triangleq\left\{
    \begin{array}{ccc}
      \m{xs}_L \cat [v] \cat \m{xs}_R & & \m{min}\ \m{{xs}}_L > {v} \wedge \m{min}\ \m{{xs}}_R \geq {v} \\
      \ \ []\ \  & &  \textit{otherwise} \\
    \end{array}
    \right.
  \end{align*}
\smallskip

\noindent \textbf{Time complexity}. Under the efficiency condition, the synthesized program runs in $O(n)$ time on a list of length $n$ because its bottleneck is $O(n)$ invocations of $\m{comb}$.

\subsection{Segment Tree}

The segment tree aims at the problem of \textit{{Range Update and Range Query}}~\cite{0Solution}, a classical data structure problem. Given an initial list $\m{xs}$, a query program $h$, and an update program $u$, the task is to process a series of operations in order. There are two types of operations.
\begin{itemize}
    \item Range update $(\texttt{U}, a, l, r)$: set the value of $\m{xs}_i$ to $u\ (a, \m{xs}_i)$ for each $i \in [l, r]$.
    \item Range query $(\texttt{Q}, l, r)$: calculate and output the value of $h\ [\m{xs}_l, \dots, \m{xs}_r]$.
\end{itemize}

The segment tree requires the semantics of the update operator $u$ to form a monoid, that is, there exists a constant $a_0$ and an operator $\oplus$ satisfying the following conditions. 
$$
    \forall w , u\ (a_0, w) = w \qquad \forall a_1, \forall a_2, w, u\ (a_1,u\ (a_2, w)) = u\ (a_1 \otimes a_2, w)
$$

Here, we assume that $a_0$ and $\otimes$ are directly given for simplicity. In general, finding $a_0$ and $\otimes$ for a given update program is an isolated synthesis task and thus can be treated separately.

\noindent \textbf{Procedure}. A segment tree is a tree-like data structure where each vertex corresponds to a segment in the list. On each vertex, several values with respect to the corresponding segment are maintained. 
\begin{itemize}
    \item For each update operation, the segment tree distributes the updated range into several disjoint vertices and applies the update in batch via \textit{lazy tags}, which will be discussed later.
    \item For each query operation, the segment tree also distributes the updated range into disjoint vertices and merges the maintained values on these vertices together.  
\end{itemize}
    \begin{figure}[t]
        \footnotesize
        \hspace{0.3cm}\begin{minipage}[c]{0.75\linewidth}       
\begin{lstlisting}[language=c++, numbers=left]
struct Info {
    Int res; // Variable representing the output of h.
    // Variables representing the output of aux.
}info[N];
Int tag[N];
void apply(int pos, Int a){
    info[pos] = /*comb2 (a, info[pos])*/;
    tag[pos] = tag[pos] `$\otimes$` a;
}
void pushdown(int pos) {
    apply(pos * 2, tag[pos]);
    apply(pos * 2 + 1, tag[pos]);
    tag[pos] = a0;
}
void initialize(int pos, int *A, int l, int r) {
    if (l == r) {info[pos] = /*h [A[l]], aux [A[l]]*/; return;}
    int mid = l + r >> 1;
    initialize(pos * 2, A, l, mid);
    initialize(pos * 2 + 1, A, mid + 1, r);
    info[pos] = /*comb1 (info[pos * 2], info[pos * 2 + 1])*/;
}
void update(int pos, int l, int r, int ul, int ur, Int a) {
    if (l > ur || r < ul) return;
    if (l >= ul && r <= ur) {apply(pos, a); return;}
    int mid = l + r >> 1; pushdown(pos);
    update(pos * 2, l, mid, ul, ur, a);
    update(pos * 2 + 1, mid + 1, r, ul, ur, a);
    info[pos] = /*comb1 (info[pos * 2], info[pos * 2 + 1])*/;
}
Info query(int pos, int l, int r, int ul, int ur) {
    if (l > ur || r < ul) return {/*h [], aux []*/};
    if (l >= ul && r <= ur) return info[pos];
    int mid = l + r >> 1; pushdown(pos);
    return /*comb1 (query(pos * 2, l, mid, ul, ur), query(pos * 2 + 1, mid +1, r, ul, ur))*/;
}
void range(int n, int *A, int m, Operator* op) {
    initialize(1, A, 0, n - 1);
    for (int i = 0; i < m; ++i) {
        if (op[i].type == `\texttt{Update}`) {
            update(1, 0, n - 1, op[i].l, op[i].r, op[i].a);
        } else {
            print(query(1, 0, n - 1, op[i].l, op[i].r));
        }
    }
}
\end{lstlisting}
\end{minipage}
\caption{The template for the algorithmic paradigm of segment trees. }
\label{fig:template-ar}
\end{figure}

Figure \ref{fig:template-ar} shows the sketch of segment trees. For simplicity, we assume the element in the list, the output of the query program $h$, and the parameter $a$ of the update program $u$ are all integers. This template uses an array $\m{info}$ to implement the segment tree, where
\begin{itemize}
    \item $\m{info}[1]$ records the information on the root node, which corresponds to the whole list.
    \item $\m{info}[2k]$ and $\m{info}[2k+1]$ correspond to the left child and the right child of node $k$ respectively. 
    \item For each node $k$, $\m{info}[k]$ records the output of $h$ and the auxiliary values on the segment corresponding to node $k$ (Lines 1-4). 
    \item Array \m{tag} records the lazy tag on each node. $\m{tag}[k]$ represents that all elements inside the range corresponding to node $k$ should be updated via $\lambda w. u\ (\m{tag}[k], w)$, but such an update has not been applied to those nodes strictly in the subtree of node $k$ yet.
\end{itemize}

There are several functions used in this template:
\begin{itemize}
    \item {\tt apply} deals with an update on all elements in the segment corresponding to node \m{pos} by updating $\m{info}[pos]$ via combinator $\m{comb}_2$ (Line 7).
    \item {\tt pushdown} applies the tag on node \m{pos} to its children (Lines 11-12) and then clears it (Line 13).
    \item {\tt initialize} initializes the information for node \m{pos} which corresponds to range $[l, r]$. It recurses into two children (Lines 18-19) and merges the sub-results via $\m{comb}_1$ (Line 20).
    \item {\tt update} applies an update ($[ul, ur]$, $\lambda w. u\ (a, w)$) to node \m{pos} that corresponds to range $[l, r]$. If $[l,r]$ does not overlap with $[ul, ur]$, the update will be ignored (Line 23). If $[l,r]$ is contained by $[ul, ur]$, a lazy tag will be put (Line 24). Otherwise, {\tt update} recurses into two children (Lines 26-27) and merges the sub-results via $\m{comb}_1$ (Line 28).
    \item {\tt query} calculates a sub-result for query $[ul, ur]$ by considering elements in node {\tt pos} only. It is implemented similarly to {\tt update}.
\end{itemize}

To solve a task, the segment tree is first initialized via function {\tt initialize} (Line 37) and then responds to each operation by invoking the corresponding function (Lines 39-43).

\smallskip
\noindent \textbf{Reduction to the lifting problem}. To apply this paradigm, we need to find two combinators $\m{comb}_1$, $\m{comb}_2$ and an auxiliary program $\m{aux}$ satisfying the following condition, where $\m{comb}_1$ merges the sub-results on two sub-segments, $\m{comb}_2$ update the result after an update operation is applied to each element in the segment, and $\m{aux}$ provides necessary auxiliary values.
\begin{gather*}
    (\bluec{h} \spl \var{\m{aux}})\ (\quanti{\m{xs}_L} \cat \quanti{\m{xs}_R}) = \var{\m{comb}_1}\ ((\bluec{h} \spl \var{\m{aux}})\ \quanti{\m{xs}_L}, (\bluec{h} \spl \var{\m{aux}})\ \quanti{\m{xs}_R}) \\ 
    (\bluec{h} \spl \var{\m{aux}})\ (\m{map}\ (\lambda w. \bluec{u}\ (\quanti{a}, w))\ \quanti{\m{xs}}) = \var{\m{comb}_2}\ (\quanti{a}, (\bluec{h} \spl \var{\m{aux}})\ \quanti{\m{xs}})
\end{gather*}
Similar to the second paradigm of the longest segment problem (Section \ref{subsection:more-application}), the two formulas above can be unified into a single lifting problem $\mathsf{LP}(h, \m{op})$, where $\m{op}$ is defined as below and its complementary input is from $\{\m{``merge'', ``update''}\}\ \times\ $(the update set of $u$).
\begin{align*}
    \m{op}\ ((\m{tag}, a), (\m{xs}_L, \m{xs}_R))\triangleq\left\{
    \begin{array}{ccc}
        \m{xs}_L \cat \m{xs}_R & & \m{tag} = ``merge" \\
        \m{map}\ (\lambda w. u\ ({a}, w))\ {\m{xs}_L}  & &  \m{tag} = ``update" \\
    \end{array}
    \right.
  \end{align*}

\smallskip 

\noindent \textbf{Time complexity}. Let $n$ be the length of the initial list. Under the efficiency condition, one can verify that (1) the bottleneck of $\texttt{initialize}$ is $O(n)$ invocations of $\m{comb}_1$, and (2) the bottleneck of $\texttt{query}$ and $\texttt{update}$ are $O(\log n)$ invocations of $\m{comb}_1$ and $\m{comb}_2$. Therefore, under the efficiency condition, the resulting program will take linear time to perform pre-processing and will respond to each operation in logarithmic time. 

\end{document}